\documentclass[10pt,journal,compsoc]{IEEEtran}

\ifCLASSOPTIONcompsoc
  \usepackage[nocompress]{cite}
\else
  \usepackage{cite}
\fi
\ifCLASSINFOpdf
\else
\fi
\usepackage{amsmath}
\usepackage{amsthm}
\usepackage{amssymb}
\usepackage{dsfont}
\usepackage{xspace}
\usepackage{graphicx}
\usepackage{algorithmic}
\usepackage{algorithm}
\usepackage{subcaption}
\usepackage[usenames, dvipsnames]{color}
\usepackage{soul}
\usepackage{url}
\usepackage{epstopdf} 
\usepackage{enumitem}
\usepackage{mathtools} 
\usepackage{comment}
\usepackage{amsbsy}

\usepackage{balance}

\usepackage{hyperref}

\usepackage{tikz}
\usetikzlibrary{matrix,calc,shapes}
\tikzset{
  treenode/.style = {shape=rectangle, rounded corners,
                     draw, anchor=center,
                     text width=8em, align=center,
                     top color=white, font=\footnotesize, bottom color=blue!20,
                     inner sep=1ex},
  treenode2/.style = {shape=rectangle, rounded corners,
                     draw, anchor=center,
                     text width=10em, align=center,
                     top color=white, font=\footnotesize, bottom color=blue!20,
                     inner sep=1ex},
  decision/.style = {treenode2, diamond, inner sep=-8pt,  minimum height=2.7cm,minimum width=2.8cm},
  root/.style     = {treenode, bottom color=red!30, text width=9em},
  env/.style      = {treenode},
  finish/.style   = {root, bottom color=green!40, text width=9em},
  dummy/.style    = {circle,draw}
}
\newcommand{\yes}{edge node [above] {\small yes}}
\newcommand{\yyes}{-- node [above] {\small yes}}
\newcommand{\no}{edge  node [left]  {\small no}}
\newcommand{\nno}{--  node  [above] {\small no}}

\newcommand{\suchthat}{\, \mid \,}

\makeatletter
\long\def\@IEEEtitleabstractindextextbox#1{\parbox{0.922\textwidth}{#1}}
\makeatother

\setlist{leftmargin=15pt,labelindent=15pt}

\theoremstyle{definition}
\newtheorem{note}{Note}
\newcommand{\MATLAB}{\textsc{Matlab}\xspace}

\newcommand{\eag}[1]{\textcolor{black}{#1}}

\newcommand{\ea}[1]{\textcolor{blue}{#1}}
\newcommand{\ifc}[1]{\textcolor{black}{#1}}
\newcommand{\red}[1]{\textcolor{red}{#1}}
\newcommand{\redd}[1]{\textcolor{black}{#1}}
\newcommand{\ogreen}[1]{\textcolor{black}{#1}}
\newcommand{\blue}[1]{\textcolor{black}{#1}}
\newcommand{\sepia}[1]{\textcolor{black}{#1}}
\newcommand{\ceru}[1]{\textcolor{black}{#1}}

\newcommand{\aBS}{$a\!B\!S$\xspace}
\newcommand{\aBSs}{$a\!B\!S$s\xspace}
\newcommand{\gBS}{$g\!B\!S$\xspace}
\newcommand{\gBSs}{$g\!B\!S$s\xspace}
\newcommand{\SNR}{S\!N\!R\xspace}

\begin{document}

\title{Fair Throughput Optimization with \\ a Dynamic Network of Drone Relays$^{\star}$\thanks{${}^\star$A preliminary 6-page version of this manuscript appeared in the proceedings of IEEE INFOCOM WKSHPS -- MiSARN 2019~\cite{arribas2019fair}.\vspace{-0mm}}}

\author{
        Edgar~{Arribas},~
        Vincenzo~{Mancuso},~
        Vicent~{Cholvi}
\IEEEcompsocitemizethanks{\vspace{-1mm}
\IEEEcompsocthanksitem E.~{Arribas} is with Universidad CEU San Pablo, Madrid, Spain.
E-mail: edgar.arribasgimeno@ceu.es
\IEEEcompsocthanksitem V.~{Mancuso} is with IMDEA Networks Institute, Madrid, Spain.
E-mail: vincenzo.mancuso@imdea.org
\IEEEcompsocthanksitem V.~{Cholvi} is with Universitat Jaume I (UJI), Castell\'o, Spain.
E-mail: vcholvi@uji.es
}
\vspace{-1mm}
\thanks{The work of E. Arribas is partially supported by the FPU15/02051 grant from the Spanish Ministry of Education, Culture and Sports (MECD).
The work of V. Mancuso is supported by the RYC-2014-16285 grant from the Spanish Ministry of Economy and Competitiveness and
by the Spanish Ministry of Science and Innovation grant PID2019-109805RB-I00 (ECID).}
}

\markboth{}{Arribas \MakeLowercase{\textit{et al.}}: Bare Demo of IEEEtran.cls for Computer Society Journals}

\IEEEtitleabstractindextext{%

\begin{abstract}
Aiding the ground cellular network with aerial base stations carried by drones has experienced an intensive raise of interest in the past years.
Reconfigurable air-to-ground channels enable aerial stations to enhance users access links by means of seeking good line-of-sight connectivity while hovering in the air.
In this paper, we propose an analytical framework for the 3D placement of a fleet of coordinated drone relays.
This framework optimizes network performance in terms of user throughput fairness, expressed through the $\alpha$-fairness metric.
The optimization problem is formulated as a mixed-integer non-convex program, which is  intractable.
Hence, we propose an extremal-optimization-based algorithm, {\it Parallelized Alpha-fair Drone Deployment}, that solves the problem online, in low-degree polynomial time.
We evaluate our proposal by means of numerical simulations over the real topology of a dense city.
We discuss the advantages of integrating drone relay stations in current networks and test several resource scheduling approaches in both static and dynamic scenarios, including with progressively larger and denser crowds.
\end{abstract}

\vspace{-1mm}
}

\maketitle

\IEEEdisplaynontitleabstractindextext

\IEEEpeerreviewmaketitle

\renewcommand{\baselinestretch}{0.94}
\IEEEraisesectionheading{\section{Introduction}\label{sec:introduction}}


\IEEEPARstart{T}{he} infrastructure of cellular networks is evolving towards flexible and reconfigurable solutions, able to cope with the highly variable densities of users.
Specifically, the current generation of cellular communications, namely 5G~\cite{Mohr2016},  embeds new transmission techniques as well as novel communication paradigms, including smart and flexible relaying~\cite{5GPPP2016_use-cases-5G}.
Besides, wireless relaying with mobility of relays is possible thanks to precise beamforming and highly efficient cooperative transmission techniques, which makes it possible to operate broadband wireless backhaul links~\cite{CRAN}.
Without such mature technological tools, many attempts toward mobile relaying have failed in the past, since the advent of broadband wireless data networks~\cite{MMR}.

\begin{figure}[t]
  \centering
   \vspace{-10mm}
         \includegraphics[width=1.01\columnwidth]{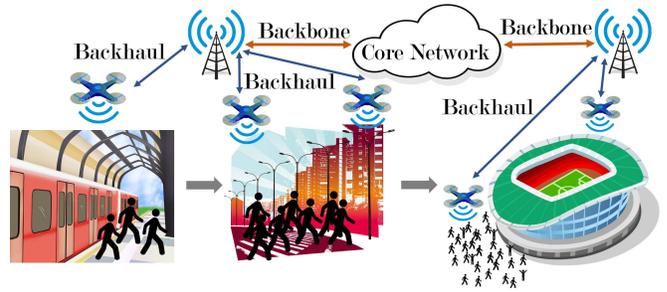}
         \vspace{-6mm}
        \caption{Reference scenario: multi-drone-aided network with people moving from a train station to a football stadium.
        }
        \vspace{-5mm}
        \label{fig:scenario}
\end{figure}

It is therefore currently possible to mount mobile relays on, e.g., transport vehicles and drones, which brings the possibility of moving the network with the users and position relays where the fixed infrastructure cannot sustain the user demand~\cite{drone-bs, guo2014performance}.
However, there exist serious concerns on the practicality of mobile relaying, due to interference management problems. For instance, Guo and O'Farrel~\cite{RelayCapacity}  have derived the capacity of OFDMA cellular networks (like LTE-A, but also in 5G) in the presence of relays reusing cellular frequencies, and showed that relays need to be operated onto orthogonal frequencies.
Besides, it is known that using over orthogonal frequencies gives additional advantages in terms of simplified resource allocation control~\cite{mozaffari2017wireless}.
Thus, we study the case of relay drones and base stations (BSs) transmitting on orthogonal frequencies.
Specifically, we tackle the optimization of 3D positions for a fleet of coordinated drone relays aiding a set of ground BSs, as depicted in Fig.~\ref{fig:scenario}, aiming at a fair throughput distribution among users.

We base the optimization on the throughput as evaluated by means of an $\alpha$-fairness metric~\cite{mo2000fair}, which is a high-level generalization of fairness metrics.
The parameter $\alpha$   can be tuned to analytically target, e.g., maximum throughput, proportional fairness or max-min fairness with a single framework. 
In the analysis, we account for the behavior of standard 3GPP networks and user access protocols, and model transmission technology details like the random variations of signal quality received by users over time, the interference caused by relays and BSs, the use of slotted time-frequency resources, the wireless backhaul attachment, and cell selection and resource allocation procedures.
Specifically, we adopt stochastic models for path-loss and availability of line-of-sight (LoS) and non-LoS (NLoS) links, and cast our problem into an OFDMA-like resource allocation scheme with several constraints.

The problem of finding the exact optimal drone positions is NP-complete.
In particular, our analysis unveils that the role of interference caused by drones and the stochastic characterization of LoS between drones and ground users make the optimization problem intractable.
However, is the problem really intractable in practice? To answer this question we analyze the drone-positioning problem under the perspective of a fair maximization of the network throughput provided to mobile users, and unveil its structure, which allows us to find efficient paths towards the computation of near-optimal solutions with low time complexity. Moreover, we show that the most critical part of the problem can be efficiently addressed by leveraging Extremal Optimization (EO) algorithms, which are a class of algorithms specifically designed for polynomial time optimization with intertwined variables~\cite{boettcher2001optimization}.
The EO operation is based on picking the ``least fit'' element of a finite set and change its configuration parameters to improve a global utility function.
EO offers a lightweight solution to problems otherwise intractable, and for which researchers proposed genetic algorithms. It is also more reliable than machine learning approaches--e.g., reinforcement learning schemes--since it does not require any training, which would instead be quite heavy and which does not scale with the number of drones to control (the state space in which to train the machine learning algorithm would quickly become unsustainable).
We therefore formulate a suitable utility function, targeting $\alpha$-fair user throughput across the network, and design PADD: a Parallelized Alfa-fair Drone Deployment Algorithm. By using an EO algorithm, PADD iteratively updates the position of the least fit drone, i.e., the drone relay that contributes the least to the utility function.
We validate our algorithm and evaluate its performance by means of simulations of realistic static and dynamic scenarios.
As an illustration of dynamic cases, we evaluate the performance of our algorithm when customers move towards a stadium before a sport event, so that their density grows over time.
Beside illustrating the advantages offered by PADD over state of the art algorithms, our numerical results show that optimizing network throughput without considering fairness is not beneficial in dense environments since drones serving the same area generate too much interference.

The contributions of this paper are summarized:
\begin{itemize}[wide=1.4pt, leftmargin=*]
  \item We propose an analytical framework for the 3D placement of a fleet of coordinated drone relays in cellular~networks.
  \item We propose an aerial cellular infrastructure that combines standard 3GPP tools, beamforming techniques, wireless backhauling and fair resource allocation.
  \item We show that the analyzed problem is NP-Complete.
  \item We formulate the 3D placement optimization problem in terms of the $\alpha$-fairness metric to guarantee fair distribution of resources and access link throughputs.
  \item We propose PADD as an extremal-optimization-based algorithm to solve the problem near-optimally.
  \item We deterministically solve optimally in polynomial time the $\alpha$-fair resource allocation problem for ground-plus-aerial cellular networks.
  \item We assess and validate our proposal by means of comprehensive numerical results.
\end{itemize}

The rest of the paper is structured as follows.
Section~\ref{s:related} discusses related work.
Section~\ref{s:model} presents the system model, while Section~\ref{s:NPCompleteness} derives the framework for optimizing drone positions under the $\alpha$-fairness metric. Section~\ref{s:extremal} describes the design of our optimization algorithm.
Section \ref{s:results} provides numerical results.
Section~\ref{s:lessonslearnt} discusses the findings of this paper and possible practical implementation issues.
Section~\ref{s:conclusions} summarizes and concludes the paper.

\vspace{-1mm}
\section{Related work}
\label{s:related}

\subsection{Relay Alternatives}

Mobile and non-terrestrial relays have been investigated in several forms and considering several technologies.
For instance, the usage of satellite networks~\cite{ferrus2016sdn} emerged several years ago. However, while satellites serve huge areas, users can only achieve relatively low rates.
Moreover, satellites cannot adjust to the user's topology, and service incurs high costs.
In contrast, drone relays may move dynamically at low altitudes and serve smaller target regions on demand, where the ground network cannot sustain the high demand from dense spots, so that a swarm of drones is able to rapidly act for aerial connectivity assistance.

Drone relays are also different from fixed relays and Device-to-Device (D2D)-based approaches~\cite{arribas2017multi,DBLP:journals/adhoc/ArribasM20}.
In fact, unlike those cases, drone communications are neither fixed nor opportunistic, and the channel propagation is impacted by the probability of communicating with LoS, 
as we account for in our analysis. That behavior is properly described in~\cite{DBLP:journals/tcom/BithasNKK20}, where the authors highlight a number of factors that affect the air-to-ground communications.

Thus, in general, satellites, balloons or terrestrial relays cannot face scenarios as the ones studied in this paper.
In fact, connectivity requirements used to design protocols for satellite, balloon and D2D communications, as well as technology constraints and signal propagation, are radically different from our case. An overview of the recent advances in drone communications, with an emphasis on integrating drones into the fifth-generation can be found in~\cite{DBLP:journals/comsur/MozaffariSBND19}.

\subsection{Drone Position Optimization}

In the recent years there have been various studies that optimize drone relay placement mainly focusing on coverage in static and oversimplified assumptions, as for instance neglecting inter-drone interference~\cite{strumberger2018moth} or ignoring fairness issues in resource allocation~\cite{mozaffari2016unmanned}.
With that, the resulting problem formulation is simple enough, typically quadratic, yet less realistic than what we derive in this paper. In our previous work, we have presented \texttt{OnDrone}, a lightweight multi-drone coverage framework that maximizes the number of users covered.
In there, we consider Quality-of-Service thresholds and show how to design flying paths for fast repositioning and augmented coverage over time~\cite{arribas2019coverage}. A similar approach has been followed in~\cite{DBLP:conf/infocom/KumarGS20}, where the authors propose the deployment of drones for video surveillance, obtaining a suitable topology (in terms of achieving zero blind spots) with minimum number of drones. However, they assume a region that lacks cellular infrastructure. A different approach is taken in~\cite{DBLP:journals/tvt/LiuLC19}, where a Q-learning solution is used to solve the deployment and movement of the drones that maximizes the Quality of Experience~\cite{DBLP:journals/twc/CuiLDFN18a}, although it requires explicit feedback from ground users and protocol changes in the way cells form. Galkin et al.~\cite{DBLP:journals/tvt/GalkinKD19} use a stochastic model to analyze the coverage when drones act as access points for users in urban areas. They demonstrate how the density of the drones determines whether drones should position closer to user hotspots to improve the received signal strength, or further away from one another to mitigate interference. By using stochastic and fractal geometry as a macromodel for both the  vehicular networks and for the fixed infrastructures, in~\cite{DBLP:conf/infocom/JacquetPM20} the authors derived  analytic  bounds in terms of connectivity. They discovered two regimes, in function of the existing infrastructure, and identified in which regime the deployment of nodes is desirable. In~\cite{alsharoa2019spatial}, the authors propose a communications framework for HetNets composed by macrocells, microcells and communication drones. Although the application scenario is similar to the one of this paper, we find remarkable and fundamental differences: they consider only a finite and small set of aerial locations for drones, ignore key communication features such as throughput or coverage and focus the decisions on minimizing the network energy consumption.

Although all these works provide valuable contribution and foundational results, they do not shed light on problems like capacity optimization with fairness targets when a fleet of drones is deployed to assist a cellular network, which is our goal in this paper. Neither they provide a realistic framework to integrate drone relay stations into current cellular networks, whereas we show that our practical approach accounts for integration. In the conference version of this paper~\cite{arribas2019fair}, we address this problem from a quite simple approach: in there, we relax critical constraints like wireless backhaul and backbone bottlenecks, and air-space constraints where drones can be located. Such simplifications allow for the design of a simpler system model and less complex algorithms and optimizations, yet bring the risk of being inaccurate. In this article, instead, we provide a fully general, realistic, precise and tractable model for an aerial network integrated in a cellular infrastructure.

Besides, here we use EO in a novel way.  EO algorithms took relevance on fields as biology or physics, and have been also applied to network science, e.g., to optimize transport in complex networks~\cite{danila2006optimal}. EO has been thought and used so far as a form of centralized static optimization. Instead, in this paper we use EO to design an algorithm that works in parallel threads  and dynamically, as the system evolves.

%
%

\begin{figure}[t]
  \centering
   \vspace{-3mm}
         \includegraphics[width=0.875\columnwidth]{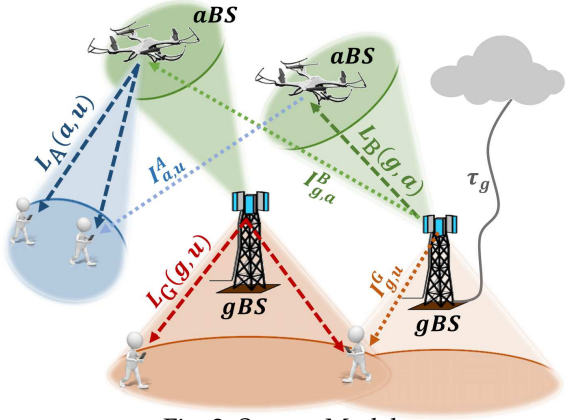}
         \vspace{-4mm}
        \caption{System Model}
        \vspace{-4mm}
        \label{fig:SystemModel}
\end{figure}

\section{System Model}
\label{s:model}

Our goal is to derive an analytical framework that finds optimal 3D locations of drone relays, given the position of users and ground stations, to optimize downlink communications.
We target $\alpha$-fair instantaneous user throughput.
To this aim, jointly with 
path-loss and interference in both access and backhaul, we address how users perform cell selection and get resources allocated.  The modelled system is a plain 3GPP cellular system comprising access and backhaul networks, with drones acting as standard 5G relays~\cite{BenMimoune2017}.

In order to illustrate the system model and parameters involved in the network, we show the main components in Fig.~\ref{fig:SystemModel} and describe the meaning of the parameters in Table~\ref{tab:symbols}.

\begin{table}[t]
\centering
     \renewcommand{\arraystretch}{1.2} 
     \setlength{\tabcolsep}{3pt} 

\caption{System model parameters} 
\label{tab:symbols}
\vspace{-1mm}
\begin{tabular}{|c|l|}
\hline
Parameter & Description   \\
\hline
\hline
$\alpha$ & Fairness level \\
\hline
$A_g$ & Maximum number of \aBSs that \gBS~$g$ can serve  \\
\hline
$\mathcal{A}$ & Set of \aBSs \\
\hline
$\mathcal{B}$ & Set of BSs, i.e., set of \aBSs and \gBSs jointly \\
\hline
$\mathcal{G}$ & Set of \gBSs \\
\hline
$\gamma_{a,u}^{\mathcal{A}}$ & SINR of aerial access link $(a,u)$\\
\hline
$\gamma_{g,a}^{\mathcal{B}}$ & SINR of aerial backhaul link $(g,a)$  \\
\hline
$\gamma_{g,u}^{\mathcal{G}}$ & SINR of ground access link $(g,u)$ \\
\hline
$I_{a,u}^{\mathcal{A}}$ & Interference from \aBS~$a$ to UE~$u$\\
\hline
$I_{g,a}^{\mathcal{B}}$ & Interference from \gBS~$g$ to \aBS~$a$ \\
\hline
$I_{g,u}^{\mathcal{G}}$ & Interference from \gBS~$g$ to UE~$u$ \\
\hline
$L_\mathcal{A}(a,u)$ & Loss of air-to-ground channel from \aBS~$a$ to UE~$u$ \\
\hline
$L_\mathcal{B}(g,a)$ & Loss of ground-to-air channel from \gBS~$g$ to \aBS~$a$ \\
\hline
$L_\mathcal{G}(g,u)$ & Loss of ground-to-ground channel from \gBS~$g$ to UE~$u$ \\
\hline
$\phi_{g,a}$ & Angle from main
lobe of \gBS~$g$ to \aBS~$a$\\
\hline
$P_{L\!o\!S}(a,u)$ & LoS-likelihood of link $(a,u)$\\
\hline
$U_{\max}$ & Maximum number of UEs a BS can serve  \\
\hline
$W_{\mathcal{A}}$ & Available bandwidth of aerial access channels\\
\hline
$W_{\mathcal{A}}^{\min}$ & Minimum bandwidth guaranteed to aerial access links \\
\hline
$W_{\mathcal{B}}$ & Available bandwidth of aerial backhaul channels \\
\hline
$W_{\mathcal{B}}^{\min}$ & Minimum bandwidth guaranteed to aerial backhaul links\\
\hline
$W_{\mathcal{G}}$ & Available bandwidth of ground access channels \\
\hline
$W_{\mathcal{G}}^{\min}$ & Minimum bandwidth guaranteed to ground access links \\
\hline
$\tau_g$& Backbone capacity of \gBS~$g$ \\
\hline
\end{tabular}
\vspace{-3mm}

\end{table}

\subsection{Reference Scenario}
\label{ss:reference}

We consider a flat ground surface $\mathcal{S}$ where a set $\mathcal{G}$ of $G$ ground base stations (\gBS) provide cellular service.
The position of each \gBS $g$ is denoted as $\Pi^g = (X^g, Y^g)$.
We assume that every \gBS $g$ is wired to the internet with a backbone capacity $\tau_g$.
We consider that a set $\mathcal{U}$ of $U$ user equipments (UEs) is on the ground, requesting cellular service.
We denote the position of each user $u$ as $\pi_u = (x_u, y_u)$.
The network disposes of a fleet $\mathcal{A}$ of $A$ aerial base stations (\aBS) that act as mobile relays.
Each \aBS is mounted on a drone (for ease of readability, we may refer just as ``drone'' to an \aBS).
We assume that the system operator disposes of two frequency bands to offer user access: the first band is used by \gBSs to provide both cellular access to ground users and backhaul connectivity to \aBSs; and the second band is used by \aBSs to provide cellular access to ground users.
This assumption mimics the common scenario in which an operator deploys different types of base stations and equip them with different frequencies to simplify network management. Indeed, it is common that an operator disposes of multiple licenses to use more than one cellular band. It is also common practice to coordinate transmissions from different base stations, especially in case of mixed deployments with macro and small cells. Frequency reuse is one of the possible and simplest schemes to use for coordination, which is what we adopt for simplicity. Nonetheless, please note that the analysis carried out in this article can be extended to the case of using a single frequency for all transmissions or any arbitrary number of frequencies (e.g., one per aerial and ground base station, in the most extreme case).

Under our assumptions, users connected to drones suffer no interference from \gBSs. Similarly, users connected to ground base stations are not interfered by drone transmissions.
Drones fly in the air space, so that we denote as $\Pi_a = (X_a, Y_a, h_a)$ the 3D position of drone $a$.
We denote as $\mathcal{B} = \mathcal{G}\cup\mathcal{A}$ the set of all the base stations (\gBSs jointly with \aBSs) that form the whole network.

We assume that \gBSs reuse the downlink spectrum used for the \gBS--UE access links in order to provide backhaul wireless service to \aBSs.
Usually, \gBSs dispose of three antenna sectors pointing mainly to the ground in order to provide cellular coverage to users. In addition to these antennas, we assume that \gBSs dispose of an additional full dimensional antenna array that performs 3D-beamforming over clear LoS links in order to establish \gBS--\aBS links for backhauling, as suggested and studied in~\cite{DBLP:journals/wc/ZengLZ19}.
%
%
%
%
 Hence, \gBS--UE and \gBS--\aBS links do not practically interfere.
Moreover, 3D antenna arrays allow \gBSs to relay traffic to several drones, by alternating transmissions every few milliseconds.
Therefore, we assume that each \gBS can set backhaul wireless links with more than one \aBS, and we also assume that a minimum backhaul bandwidth $W_\mathcal{B}^{\min}$ is guaranteed, which implies that only a limited number of drones can stay connected to a single ground base station.

In order to analyze a realistic network configuration, differently from what we assume for BSs talking to each other, we do not assume that transmissions to users can benefit from beamforming, therefore, we account for inter-cell interference for the groups of users connected to the same type of BS, be it \aBS or \gBS. We do so because beamforming to users requires non-negligible overheads to track the position of erratically moving users, and to fine tune the beam within potentially narrow spaces crowded with users.  Instead,   \aBSs and \gBS can
easily have precise information about their positions at any time, and do not risk to stay packed in small spaces. Having said that, our analysis can be
straightforwardly simplified to apply also to the ideal case in which all transmissions can be interference-free due to beamforming, should that possibility become~reality.


Another important aspect to consider in drone-assisted missions is the energy consumption management of drones, generally by means of a recharge strategy with additional drones. In this paper, we assume that $A$ drones are guaranteed to be operational in a desired position when the optimization is run. Hence, we assume that there is a parallel recharge scheduling strategy that uses a minimum number of extra drones to guarantee a persistent communications service with $A$ \aBSs at any moment. This parallel mechanism can certainly be achieved, as we have already proposed and demonstrated in~\cite{arribas2021optimal}.

\vspace{-2mm}
\subsection{Channel Modelling}
\label{ss:pathloss}

We assume that the network operator disposes of two orthogonal frequency bands. One band is assigned to \gBSs to provide access service to ground users as well as aerial backhaul service to \aBSs, with a fixed bandwidth $W_\mathcal{G}$ and $W_\mathcal{B}$, respectively. 
The other band is assigned to \aBSs for aerial user access and it has a fixed bandwidth $W_{\!\mathcal{A}}$.
Hence, we model three different kinds of channels: (i) air-to-ground and (ii) ground-to-ground channels in the access network, and (iii) ground-to-air channels in the backhaul network.

To optimize the network and compute link throughputs we need to compute the signal strength of each wireless channel, measured as the SINR. In order to compute the SINR, we study path-loss and fading of channels, as well as suffered interference. Then, throughputs will be measured with the well known Shannon formula.

The notation of each parameter of the channel model can be found in Table~\ref{tab:symbols}, and the
mathematical details of the modelling of each kind of channel can be found in separate sections of the online supplemental material, within Section~\ref{a:ss:pathloss}.


\textbf{Air-to-ground access channels.}
The loss of these channels differs notably depending on whether links are free of obstacles~\cite{al2014modeling}. Hence, this loss depends on the LoS-likelihood~\cite{al2014optimal} of links.
As \aBSs serve users in OFDMA channels, these users do not interfere among themselves. However, users suffer aggregated interference from other \aBSs serving other users.
We denote the loss from \aBS~$a$ to UE~$u$ as $L_\mathcal{A}(a,u)$.

%

\textbf{Ground-to-ground access channels.}
The loss of these channels follows
the well known path-loss model with slow fading~\cite{goldhirsh1998handbook}. As \gBSs serve users in OFDMA channels, there is no intra-cell interference. However, we account for inter-cell interference from neighbour ground cells.
We denote the loss from \gBS~$g$ to UE~$u$ as $L_\mathcal{G}(g,u)$.




\textbf{Ground-to-air backhaul channels.}
Since backhaul links 
follow directional beams in LoS, the loss of these channels also follows the well known path-loss model with slow fading~\cite{goldhirsh1998handbook}. Backhaul links reuse the spectrum from the ground access channels. However, \gBSs perform 3D beamforming for backhauling and hence there is no interference between ground access and aerial backhaul. Still, other \gBSs might serve other \aBSs in a beamforming direction such that served \aBSs suffer interference (attenuated depending on the radiating angle). We account for such interference.
We denote the loss from \gBS~$g$ to \aBS~$a$ as $L_\mathcal{B}(g,a)$.

\subsection{Cell Selection and Resource Allocation}
\label{ss:CellSelection}

BSs cannot provide service to unlimited numbers of users because $(i)$ available radio resources are limited and $(ii)$ it is necessary to guarantee a minimum set of radio resources to each connected user, to guarantee signaling exchange with the BS; this is needed to schedule data transmissions to and from the BS.
Of course, the number of devices is also limited by the minimum bandwidth that the system aims to guarantee to each user. 
Therefore, in general, the maximum number of users that can be simultaneously served is limited, and we denote by $U_{\max}$ such number.

\textbf{Cell Selection.}
We assume that users perform cell selection as in currently operational 3GPP networks: first, UEs select the BS with strongest Signal-to-Noise Ratio (SNR);  if the request is rejected because channel conditions deteriorate or the BS runs at maximum capacity, then the UE performs cell re-selection, and tries to attach to the BS with next strongest SNR, and so on until the user gets attached~\cite{sesia2011lte}.

\textbf{Resource Allocation.}
We assume that \gBSs and \aBSs schedule cellular users according to an OFDMA system. 
Today's BSs use an OFDMA system and dispose of a finite set of physical resource blocks organized in subframes, which repeat to form frames lasting a few milliseconds (1 to 10 ms in 3GPP-compliant networks).
A physical resource block is the smallest unit of time-frequency resources that can be allocated to a user.
Thus, we assume that the minimum bandwidth allocated to a user is the bandwidth corresponding to one resource block and the scheduler guarantees that each user receives, on average, at least one block per subframe in each OFDMA frame.
We denote as $W_\mathcal{G}^{\min}$ and $W_{\!\mathcal{A}}^{\min}$  the minimum bandwidth that a \gBS or an \aBS can allocate to a single user.

Backhaul links also use an OFDMA system, although \aBSs select a \gBS according a the global network optimization criterion rather than based on SNR.
Moreover, \gBSs can serve $A_g$ \aBSs at most and each backhaul link $(g, a)$ disposes of a minimum bandwidth $W_\mathcal{B}^{\min}$ to relay traffic.

\section{Optimization}
\label{s:NPCompleteness}

Here we derive an analytic framework for the 3D positions of \aBSs, to optimize throughputs based on $\alpha$-fairness~\cite{mo2000fair}. We seek optimal \aBS positions and BS--UE associations, and also optimal backhaul association and optimal allocation of physical resources. We formally present the optimization problem addressed in this paper:

\vspace{1mm}
\textsl{\textbf{Throughput Problem $\pmb{\mathcal{T}}$}:
Given a set $\mathcal{G}$ of $G$ fixed \gBSs, a fleet $\mathcal{A}$ of $A$ relay \aBSs hovering at heights in the range $[h_{\min}, h_{\max}]$, a set $\mathcal{U}$ of $U$ ground UEs that may connect to either a \gBS or an \aBSs, each of which can serve $U_{\max}$ UEs at most,
find the optimal position of each \aBS $a\!\in\!\mathcal{A}$, the optimal user association, the optimal backhaul association and the optimal user resource allocation so to maximize the $\alpha$-fair throughput utility function.}
\vspace{1mm}

We formulate the drone positioning problem as a Mixed-Integer Non-Convex Program (MINCP) in Fig.~\ref{fig:T}. The meaning of all the optimization variables is shown in Table~\ref{table:symbols}. 

On the access network side, we denote as $C_{a, u}\!\in\!\{0, 1\}$ the decision variable that tells whether $u$ connects to \aBS~$a$.
Similarly, $C_u^{g_u}\!\in\!\{0, 1\}$ tells whether $u$ connects to \gBS~$g_u$.
Decision variables $W_{b, u}$ and $T_{b, u}$ denote bandwidth and throughput allocated to link $(b, u)$.
Throughput is a decision variable and not directly computed with the Shannon formula, since, in addition to bandwidth limitations we must account for access, backhaul and backbone bottlenecks.

On the backhaul network side, we denote as $B^{g, a}\!\!\in\!\!\{0, 1\}$ the decision variable that tells whether \aBS $a$ is attached to \gBS $g$.
Variables $W^{g, a}$ and $T^{g, a}$ denote bandwidth and throughput of the backhaul link $(g, a)$, respectively.
%

\subsubsection*{Network utility and constraints}

The utility function of the optimization problem in Fig.~\ref{fig:T} corresponds to the $\alpha$-fairness metric. This utility function is additive in terms of utilities conveyed by single BSs. Depending on the value of $\alpha\!\geq\!0$, known as the $\alpha$-fairness level, the metric captures different fairness criteria such as weighted proportional fairness ($\alpha\!=\!1$), $\max$-$\min$ fairness ($\alpha\!\rightarrow\!+\!\infty$) or the maximum capacity ($\alpha = 0$).
Constraints in Fig.~\ref{fig:T} correspond to the following restrictions:

\begin{figure}[t]
\fbox{
\begin{minipage}{8.5cm}
\small
\begin{figure}[H]
\vspace{-7mm}
\begin{equation*}
  \begin{array}{ll}
    \!\quad \quad\quad  \quad\quad\quad \quad\qquad\qquad\quad\mathclap{\max\limits_{\substack{\Pi_a, C_{a,u}, W_{b,u}, \\ B^{g, a}, W^{g, a}}}  \quad\!\! U^\alpha_{thr} \!=\! \begin{cases}
    \sum\limits_{u\in\mathcal{U}} \!\! \left(\sum\limits_{b\in\mathcal{B}} T_{b, u}\!\right)^{1-\alpha} \!\!\!\cdot\! \frac{1}{1-\alpha}, & \alpha\!\neq\! 1 ;   \\
    \sum\limits_{u\in\mathcal{U}} \! \log\!\left(\sum\limits_{b\in\mathcal{B}} T_{b, u}\!\right), & \alpha \!=\! 1 ;
    \end{cases}} \\
    \text{s.t.:} &
    \\
    \textit{\gBS--\aBS association constraints:} & \\
    \sum\limits_{g\in\mathcal{G}} B^{g, a} =  1, \quad \sum\limits_{a\in\mathcal{A}}B^{g, a} \leq  A_g, & \forall g\in\mathcal{G}, \forall  a\in\mathcal{A}; \\
    \\
    \blue{\textit{\gBS--\aBS capacity constraints:}} & \\
    \blue{W_\mathcal{B}^{\min} \cdot  B^{g, a} \leq  W^{g, a} \leq  W_\mathcal{B} \cdot  B^{g, a},} & \blue{ \forall  g\in\mathcal{G}, \forall  a \in\mathcal{A} ;} \\
    \blue{\sum\limits_{a\in\mathcal{A}} W^{g, a} \leq  W_\mathcal{B},} & \blue{ \forall g\in\mathcal{G}  ;} \\
    \blue{T^{g, a} \leq  W^{g, a}   \log_2\left(1 + \gamma_{g, a}^\mathcal{B}\right),} & \blue{ \forall  g\in\mathcal{G}, \forall  a \in\mathcal{A}  ;} \\
    \\
    \blue{\textit{\gBS backbone constraint:}} & \\
    \blue{\sum\limits_{a\in\mathcal{A}} T^{g, a} + \sum\limits_{\substack{u\in\mathcal{U}: \\ g=g_u}} T_{g, u} \leq  \tau_g,} & \blue{\forall  g\in\mathcal{G};} \\
    \\
    \ogreen{\textit{BS--UE association constraints:}} & \\
    \ogreen{C_u^{g_u} + \sum\limits_{a\in\mathcal{A}} C_{a, u} = 1,} & \ogreen{\forall u\in\mathcal{U};} \\
    \ogreen{\sum\limits_{\substack{u\in\mathcal{U}: \\ g = g_u}} C_u^g \leq  U_{\max}, \quad \sum\limits_{u\in\mathcal{U}} C_{a, u} \leq  U_{\max},} & \ogreen{\forall g\in\mathcal{G}, \forall a\in\mathcal{A};} \\
    \\
    \sepia{\textit{\gBS--UE capacity costraints:}} & \\
    \sepia{W_\mathcal{G}^{\min} \cdot C^{g_u}_u \leq W_{g_u, u} \leq  W_\mathcal{G} \cdot C^{g_u}_u,}  &  \sepia{\forall u\in\mathcal{U};} \\
    \sepia{\sum\limits_{u\in\mathcal{U}} W_{g, u} \leq W_\mathcal{G},} & \sepia{\forall g\in\mathcal{G};}\\
    \sepia{T_{g_u, u}  \leq  W_{g_u, u} \log_2\left(1+\gamma^\mathcal{G}_{g_u,u}\right),}  & \sepia{\forall u\in\mathcal{U};}  \\
    \\
    \redd{\textit{\aBS--UE capacity constraints:}} & \\
    \redd{W_{\!\mathcal{A}}^{\min} \cdot C_{a,u} \leq W_{a, u} \leq  W_{\!\mathcal{A}}\cdot C_{a,u},}  & \redd{\forall a\in\mathcal{A}, \forall u\in\mathcal{U};} \\
    \redd{\sum\limits_{u\in\mathcal{U}} W_{a, u} \leq  W_{\!\mathcal{A}},} & \redd{\forall  a\in\mathcal{A};} \\
    \redd{T_{a, u} \leq  W_{a, u} \log_2\left(1+\gamma^\mathcal{A}_{a, u}\right),}  & \redd{\forall a\in\mathcal{A}, \forall u\in\mathcal{U};} \\
    \redd{\sum\limits_{u\in\mathcal{U}} T_{a, u} \leq \sum\limits_{g\in\mathcal{G}} T^{g, a},} & \redd{  \forall  a\in\mathcal{A};} \\
    \\
    \ceru{\textit{Air space constraint:} \,\,\,\, \ceru{\Pi_a \in \mathcal{S}_a,}} & \ceru{\forall  a\in\mathcal{A}.}
  \end{array}
\end{equation*}
\end{figure}
\normalsize
\vspace{-6mm}
\end{minipage}
}
\vspace{-1mm}
\caption{Fair Throughput Optimization Program.}
\label{fig:T}
\vspace{-4.5mm}
\end{figure}

\textit{\bf \pmb{\gBS--\aBS} association constraints} state that every \aBS must associate with one \gBS to set a wireless backhaul link, and that every \gBS can serve at most $A_g$ \aBSs.

\textit{\bf \pmb{\gBS--\aBS} capacity constraints} impose guarantees on bandwidth and throughput allocation on backhaul links.

The \textit{\bf \pmb{\gBS} backbone constraint} restricts every \gBS to provide connected \aBSs and connected users with an aggregated throughput not higher than the capacity $\tau_g$ of the backbone link that serves that \gBS.

\textit{\bf BS--UE association constraints} state that each user can associate only to one BS, either a ground station or a drone, and that the maximum allowed number of UEs served by each BS is limited to $U_{\max}$.

\textit{\bf \pmb{\gBS}--UE capacity constraints} impose guarantees on bandwidth and throughput allocation on \gBS--UE links.

\textit{\bf \pmb{\aBS}--UE capacity constraints} impose guarantees on bandwidth and throughput allocation on \aBS--UE links, while at the same time the throughput of a user cannot exceed the backhaul link capacity and the aggregate user throughput cannot exceed the backhaul throughput.

The \textit{\bf Air space constraint} delimits the 3D air space in within which an \aBS can be moved, $\mathcal{S}_a$, and which is a ball centered in the current position of the drone with a radius equal to the distance that the drone can fly within a fixed time (i.e., time itself is the real constraint).

Modeling air-to-ground connections brings unavoidable non-convex functions, so that the formulated problem is not tractable with currently available optimizers, which are able to deal {\sl only} with problems that are convex.

\subsubsection*{Feasibility and NP-Completeness}

Here we remark that the formulated optimization problem in Fig.~\ref{fig:T} \textsl{always} admits a feasible solution. We just have to place each \aBS $a\!\in\!\mathcal{A}$ randomly in each $\mathcal{S}_a$, attach each UE and \aBS to a random \gBS, allocate to each UE and \aBS the minimum allocatable bandwidth and assign null throughputs to every link in the network. This naive network setting is always a feasible solution of the problem.

Moreover, the
problem of finding the exact optimal drone positions is NP-Complete. Indeed, the NP-Complete Minimum-Geometric Disk-Cover (MGDC) problem~\cite{das2012discrete} can be reduced, in polynomial time, to a special instance of the problem where users get 1 bps if a drone serves them and 0 bps otherwise.
This result is a direct consequence of the NP-Completeness proof of the Coverage Problem that we have presented in~\cite{arribas2019coverage}, because coverage can be seen as a particularly simplified throughput problem using an on/off, SINR-threshold-based, throughput function.

\begin{table}[t]
\centering
     \renewcommand{\arraystretch}{1.2} 
\caption{Optimization variables \& parameters.} 
\label{table:symbols}
\vspace{-0.75mm}
\begin{tabular}{|c|l|}
\hline
Variable & Description   \\
\hline
\hline
$B^{g,a}$ & Binary variable for the existence of backhaul link $(g,a)$ \\
\hline
$C_{a,u}$ & Binary variable for the existence of access link $(a,u)$ \\
\hline
$C_u^{g_u}$ & Binary variable for the existence of access link $(g_u, u)$ \\
\hline
$\Pi_a$ & Aerial position of \aBS~$a$ \\
\hline
$\mathcal{S}_a$ & Air space restricted to \aBS~$a$  \\
\hline
$T_{b,u}$ & Throughput allocated to access link $(b,u)$ \\
\hline
$T^{g,a}$ & Throughput allocated to backhaul link $(g,a)$ \\
\hline
$W_{b,u}$ & Bandwidth allocated to access link $(b,u)$ \\
\hline
$W^{g,a}$ & Bandwidth allocated to backhaul link $(g,a)$ \\
\hline
\end{tabular}
\vspace{-3mm}

\end{table}

\vspace{-1mm}
\section{Extremal Optimization}
\label{s:extremal}

The optimization framework proposed in Section \ref{s:NPCompleteness} is non-convex and mixed-integer, hence not solvable with any off-the-shelf optimizer~\cite{burer2012non}. 
The problem is hard to solve because any change in a decision variable (e.g., a position of a drone) affects backhaul and user association, as well as interference and resource allocation for all users. Due to the intertwined nature of the decision variables of the problem, 
this is the kind of problems EO has been thought for.

To find time-efficient and near-optimal solutions, we propose a Parallelized Alfa-fair Drone Deployment (PADD) algorithm, based on a typical EO operation.
We base the design of PADD on decoupling the problem in the four main decisions that the optimization framework must make:
$(i)$ the 3D positions of the fleet of \aBSs;
$(ii)$ the sets of users attached to each \gBS and \aBS;
$(iii)$ \gBS--\aBS backhaul association;
$(iv)$ bandwidth allocation to backhaul links \gBS--\aBS as well as to access links from each \gBS or \aBS to their attached users.

\begin{figure}[t]
\vspace{-2mm}
\hspace{-3mm}
\begin{tikzpicture}[-latex]
  \matrix (chart)
    [
      matrix of nodes,
      column sep      = 2em,
      row sep         = 2ex,
    ]
    {
      &  & |[root]| Initial system setting. Performance: $\mathcal{U}_{\mathrm{thr},0}^\alpha$                       \\
      &  &|[treenode2]|  Select $a_0\!\!\in\!\!\mathcal{A}$ as the least fit \aBS (see \S\ref{ss:leastfit})                  \\
       & |[treenode]| Set:  $\mathcal{U}_{\mathrm{thr},0}^\alpha\!\leftarrow\!\mathcal{U}_{\mathrm{thr} \!, \mathrm{new}}^\alpha$ & |[treenode2]| Move \aBS $a_0$ locally to a non-search position                        \\
       & & |[treenode2]| BS--UE association: best-signal policy (see \S\ref{ss:userassociation}) \\
       & |[decision]| $\frac{\mathcal{U}_{\mathrm{thr}, \mathrm{new}}^\alpha \!-\! \mathcal{U}_{\mathrm{thr},0}^\alpha}{\mathcal{U}_{\mathrm{thr},0}^\alpha} \!\!<\!\! \varepsilon$ & |[treenode2]| \gBS--\aBS backhaul association: \texttt{knap} heuristic (see \S\ref{ss:backhaulassociation})     \\
        &   & |[treenode2]| Optimal bandwidth allocation (CP), $\forall  g\!\in\!\mathcal{G}$. Output: $\mathcal{U}_{\mathrm{thr}, \mathrm{new}}^\alpha$  (see \S\ref{ss:bandwidthallocation})     \\
      & |[treenode]| Allocate to $a_0$ this new position & |[decision]| $\frac{\mathcal{U}_{\mathrm{thr}, \mathrm{new}}^\alpha \!-\! \mathcal{U}_{\mathrm{thr},0}^\alpha}{\mathcal{U}_{\mathrm{thr},0}^\alpha} \!\!>\!\! \delta$ \vspace{-5mm}\\
      & |[finish]| Decided system setting. Performance: $\mathcal{U}_{\mathrm{thr}}^\alpha$ \\
    };
  \draw
        (chart-1-3) edge (chart-2-3)
        (chart-2-3) edge (chart-3-3)
        (chart-3-3) edge (chart-4-3)
        (chart-4-3) edge (chart-5-3)
        (chart-5-3) edge (chart-6-3)
        (chart-6-3) edge (chart-7-3);
  \draw
        (chart-7-3) \yes (chart-7-2)
        (chart-7-3) \nno +(2.3,0) |- (chart-3-3);
  \draw
        (chart-7-2) edge (chart-5-2)
        (chart-5-2) \no (chart-3-2)
        (chart-3-2) |- (chart-2-3);
  \draw
        (chart-5-2) \yyes +(-2.0,0) |- (chart-8-2);
\end{tikzpicture}
\vspace{-1mm}
\caption{Flow diagram of PADD operation.}
\label{fig:eoa}
\vspace{-4mm}
\end{figure}
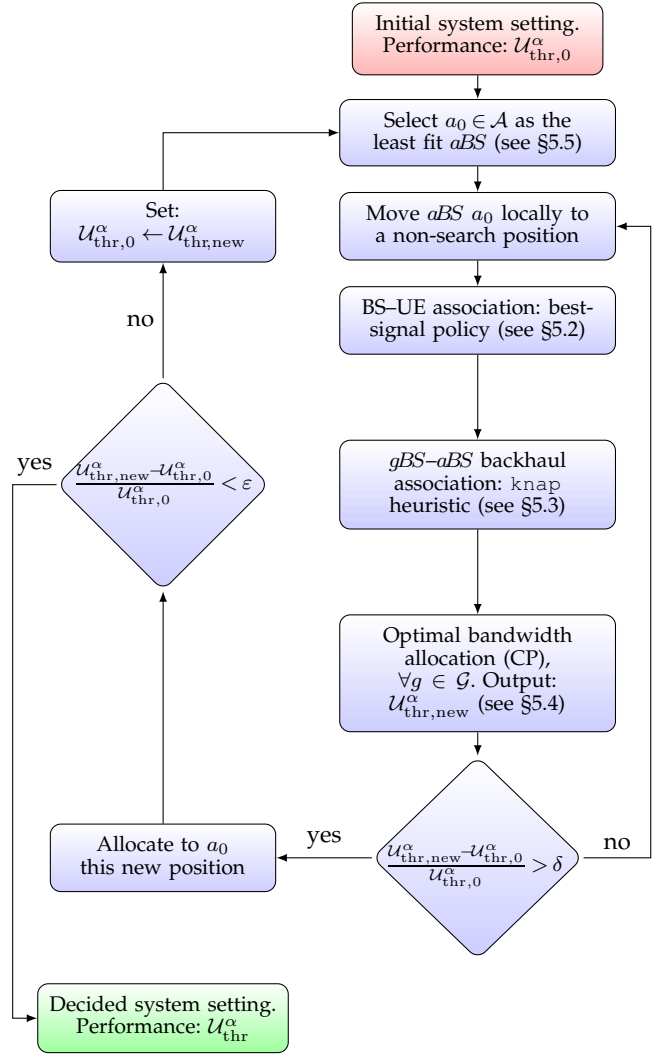

In what follows, we formally describe the PADD operation and provide details of each step that PADD takes.
The algorithm iteratively solves four steps, as pictured in Fig.~\ref{fig:eoa} as a flow-chart:
after deriving an initial feasible system setting, the least fit \aBS $a_0$ is selected in order to locally probe non-searched positions that improve the current network performance.
While probed positions do not provide a relative improvement of $\delta\geq 0$ over the current network performance, new local non-searched positions are probed.
In case a probed position improves performance, \aBS $a_0$ is {\sl moved} to such position and the current system performance is updated.
In case this new performance provides a relative improvement higher than $\varepsilon \geq \delta$, the new least fit \aBS is selected and the process begins again.
Otherwise, the algorithm converges and outputs the decided system setting.
We next formalize the operation of PADD more in detail.

\vspace{-2mm}
\subsection{Initial System Setting}
\label{ss:initial}

Initially, we consider a na\"ive drone positioning.
For instance, any random placement of drones provides a feasible solution that could be iteratively improved.
However, we more efficiently select those locations that are closer to \gBSs--so to guarantee good backhaul links--and those locations that are above densely populated regions--which are regions that potentially need drone relay assistance.
This initial drone setting yields utility $U_{\mathrm{thr}, 0}^\alpha$.
However, in order to compute $U_{\mathrm{thr}, 0}^\alpha$, we need to know also user association, backhaul association and resource allocation.
These decisions are made as described in the following subsections.

\vspace{-1mm}
\subsection{BS--UE Association: Best-Signal Policy}
\label{ss:userassociation}

For fixed positions of  the drones, the BS--UE association is performed individually by each user according to the best-signal policy described in Section~\ref{ss:CellSelection}, which is at most linear in the number of base stations
$|\mathcal{B}|$, with $\mathcal{B}  = \mathcal{G}  \cup \mathcal{A}$.
However, before association, we need to compute $|\mathcal{B}|$ SNR values for each UEs, and sort them in decreasing order, which goes with $| \mathcal{B} | \log | \mathcal{B} |$.
The complexity of user association is therefore $\mathcal{O}(| \mathcal{B} | \log | \mathcal{B} |)$ comparisons, for each user.
\\
\indent In the following, the
set of users attached to a \gBS $g\in\mathcal{G}$ and the set of users attached to an \aBS $a\in\mathcal{A}$ are denoted as $\mathcal{U}_g$ and $\mathcal{U}_a$, respectively.

\subsection{\!\!\pmb{\gBS--\aBS} Backhaul Association: \texttt{GAP-Knap} Heuristic}
\label{ss:backhaulassociation}
For fixed drone positions and given BS--UE association, the backhaul association is solved by assuming that the bandwidth
$W_{\mathcal{G}}$ is proportionally shared among users connected to a \gBS and to the \aBSs to be connected to that \gBS.

Specifically, the backhaul throughput is computed with the Shannon formula, using a fraction of bandwidth proportional to the number of users attached to the drone, and with the SINR resulting from current position of drone and the fixed position of \gBS.
Maximizing the $\alpha$-fairness of such throughput values for all drones, translates into a generalized assignment problem (GAP)~\cite{fisher1986multiplier}.
With the notation used in Fig.~\ref{fig:T}, in which $\gamma_{g, a}^\mathcal{B}$ is the backhaul SINR of link~$(b, a)$ and $B^{g, a}$ denotes the binary decision variable that tells wether \gBS~$g$ associates with \aBS $a$, we show the backhaul association optimization in Fig.~\ref{fig:GAP}.

\begin{figure}[t]
\fbox{
\begin{minipage}{8.5cm}
\small
\begin{figure}[H]
\vspace{-7mm}
\begin{equation*}
\hspace{-1.5mm}
\eag{
  \begin{array}{ll}
    \!\max\limits_{B^{g, a}} \, U_{\mathrm{bhl}}^\alpha \!=\! \!\begin{cases}
      \!\sum\limits_{g\in\mathcal{G}}\sum\limits_{a\in\!\mathcal{A}}\!\!\!\left(\!\frac{\left|\mathcal{U}_a\right|}{\left|\mathcal{U}_g\bigcup \mathcal{U}_a\right|}\! \log_2\!\!\left(1\!+\!\gamma_{g, a}^\mathcal{B}\!\right)\!\!\right)^{\!\!1-\alpha}\!\!\!\!\cdot\!  \frac{B^{g, a}}{1\!-\!\alpha}, & \!\! \alpha\!\neq\! 1; \\
      \!\sum\limits_{g\in\mathcal{G}}\sum\limits_{a\in\!\mathcal{A}} \!\log\!\left(\!\frac{\left|\mathcal{U}_a\right|}{\left|\mathcal{U}_g\bigcup \mathcal{U}_a\right|}\!\log_2\!\!\left(1\!+\!\gamma_{g, a}^\mathcal{B}\right)\!\!\right) \!\!\cdot\!\!  B^{g, a}\!\!, & \!\! \alpha \!=\! 1; \\
      \vspace{-3mm} &
    \end{cases} \\
    \text{s.t.:} \\
    \blue{\sum\limits_{a\in\mathcal{A}} B^{g, a} \leq  A_g, \,\quad\quad \forall g\!\in\!\mathcal{G};} \\
    \blue{\sum\limits_{g\in\mathcal{G}} B^{g, a} = 1,  \qquad\quad \forall a\!\in\!\mathcal{A}.}
  \end{array}
  }
\end{equation*}
\end{figure}
\normalsize
\vspace{-6mm}
\end{minipage}
}
\vspace{-1mm}
\caption{Backhaul Association Optimization Program.}
\label{fig:GAP}
\vspace{-3mm}
\end{figure}

The first constraint states that each \gBS can provide backhaul service to at most $A_g$ \aBSs.
The second constraint states that each \aBS $a$ must be associated with one \gBS.


GAP is an NP-hard MILP~\cite{cohen2006efficient}.
Hence, although the GAPs that PADD needs to solve have a small size and could be optimally solved by means of standard methods as a Branch\&Bound search~\cite{sun2006optimization}, such an approach would not lead to a polynomial-time algorithm.
Hence, multiple heuristics have been proposed in literature to find near-optimal solutions to the GAP~\cite{cattrysse1992survey, cohen2006efficient}.
In particular, we perform a simple heuristic based on an approximation to the $0$-$1$ knapsack problem~\cite{freville2004multidimensional} by means of dynamic programming~\cite{andonov2000unbounded}.
We name this heuristic as \texttt{GAP-knap}, which has a polynomial complexity of $\mathcal{O}(|\mathcal{G}|\!\cdot\! |\mathcal{A}|\!\cdot\! A_g)$.
The complexity of \texttt{GAP-knap} is linear with the sizes of $\mathcal{G}$ and $\mathcal{A}$ and the maximum number of \aBSs allowed to attached to \gBSs, $A_g$.

\subsection{Optimal Bandwidth Allocation: Convex Program}
\label{ss:bandwidthallocation}

For fixed drone positions, BS--UE association and \gBS--\aBS backhaul association, the optimal bandwidth allocation is solved by each \gBS in parallel, with a convex~program.

Each \gBS must perform bandwidth allocation to split backhaul resources among the served \aBSs, split \gBS--UE access resources among the attached users $\mathcal{U}_g$, and the attached \aBSs $a$ must split \aBS--UE access resources among the users attached to each \aBS $a$, $\mathcal{U}_a$.
Since all these bandwidth allocations are intertwined (\aBS--UE allocation depends on the backhaul bottleneck, \gBS--\aBS  allocation depends on the number of attached \aBSs to the same \gBS and the number of final users of each \aBS, and \gBS--UE access resources depend also on the backbone service that the \gBS gets from the internet), we formulate the bandwidth allocation optimization as a convex program 
in Fig.~\ref{fig:CP}.

\begin{figure}[t]
\fbox{
\begin{minipage}{8.5cm}
\small
\begin{figure}[H]
\vspace{-7.25mm}
\begin{equation*}
\hspace{-1.75mm}
  \eag{
  \begin{array}{ll}
    \,\,\quad\qquad\qquad\qquad\qquad\qquad\mathclap{\max\limits_{\blue{w^a,} \sepia{w_u,} \redd{T_u}} U_{\mathrm{cvx}, g}^\alpha =  \begin{cases}\sum\limits_{u\in\!\!\bigcup\limits_{b\in\mathcal{B}_g}\!\!\!\mathcal{U}_b} \left( T_u \right)^{1-\alpha} \cdot  \frac{1}{1-\alpha}, & \alpha \neq 1; \\
    \sum\limits_{u\in\!\!\bigcup\limits_{b\in\mathcal{B}_g}\!\!\!\mathcal{U}_b} \log\left( T_u \right), &  \alpha = 1;
    \end{cases}}\\
    \text{s.t.:} \\
    \blue{w^a \geq W^{\min}_\mathcal{B},} & \blue{\!\!\forall  a\!\in\!\mathcal{A}_g;}\\
    \blue{\sum\limits_{a\in\mathcal{A}_g} \! w^a = W_\mathcal{B};} & \!\!\\
    \blue{T^a \!\leq\! w^a  \log_2\!\left(1 \!+\! \gamma_{g, a}^\mathcal{B}\right)\!,} & \blue{\!\!\forall a\in\mathcal{A}_g;} \\
    \sepia{w_u \geq W^{\min}_\mathcal{G},} & \sepia{\!\!\forall  u\!\in\!\mathcal{U}_g;} \\
    \sepia{\sum\limits_{u\in\mathcal{U}_g} \! w_u = W_\mathcal{G};} & \!\!\\
    \redd{w_u \geq W^{\min}_\mathcal{A},} & \redd{\!\!\forall  u\!\in\!\bigcup\limits_{a\in\mathcal{A}_g} \!\! \mathcal{U}_a;} \\
    \redd{\sum\limits_{u\in\mathcal{U}_a} \! w_u = W_{\!\mathcal{A}},} & \redd{\!\!\forall  a\!\in\!\mathcal{A}_g;} \\
    \redd{T_u \!\leq\!  w_u \log_2\!\left(1 \!+\! \gamma_{b, u}\right)\!,} & \redd{\!\!\forall (b, u)\!\in\!\mathcal{B}_g \!\times \!\! \bigcup\limits_{b'\in\mathcal{B}_g} \!\!\mathcal{U}_{b'} \!\suchthat\! u \!\in\! \mathcal{U}_b;} \\
    \redd{\sum\limits_{u\in\mathcal{U}_a} \! T_u \!\leq  T^a,} & \redd{\!\!\forall  a\!\in\!\mathcal{A}_g;} \\
    \blue{\sum\limits_{u\in\mathcal{U}_g} \! T_u \!+ \! \sum\limits_{a\in\mathcal{A}_g} \! T^a \leq  \tau_g.} &\!\!
  \vspace{-0.15mm}
  \end{array}
  }
  \label{eq:CP}
\end{equation*}
\end{figure}
\normalsize
\vspace{-6mm}
\end{minipage}
}
\vspace{-1.5mm}
\caption{Bandwidth Allocation Optimization Program.}
\label{fig:CP}
\vspace{-4.4mm}
\end{figure}

In the program, $w^a$ is the share of total bandwidth that \gBS~$g$ allocates to \aBS $a$, $T^a$ is the backhaul throughput for \aBS $a$, $w_u$ is the share bandwidth that BS $b$ allocates to user $u$, and $T_u$ is the access service throughput for user $u$.
$\mathcal{A}_g$ is the set of \aBSs attached to \gBS~$g$, and $\mathcal{B}_g=\{g\}\!\cup\!\mathcal{A}_g$ is the set of \gBS~$g$ jointly with $\mathcal{A}_g$. $U_{\mathrm{cvx}, g}^\alpha$ is the utility function of \gBS~$g$, which is based on the $\alpha$-fairness metric.

The problem is convex, hence it is optimally solvable in polynomial time by means of standard interior-point methods~\cite{nesterov1994interior}.
However, these methods run in cubic time with respect to the number of users, i.e., the complexity is $\mathcal{O}(|\mathcal{U}|^3)$~\cite{bubeck2015convex}.
This cubic complexity might soon become prohibitive for real-time applications such as drone-aiding and fast repositioning in wireless networks (specially for big populations), as addressed in this paper.
However, we have derived KKT conditions~\cite{kuhn2014nonlinear} for problem 
of Fig.~\ref{fig:CP} that allow us to find  the exact solution analytically (see Section~\ref{a:ss:pathloss} in the online supplemental material).
The complexity of finding the exact solution for each \gBS~$g$ is linear with respect to the number of users and number of \aBSs attached to
\gBS~$g$, i.e., $\mathcal{O}(U_{\max} \cdot A_g)$, in the worst case.

\vspace{-2mm}
\subsection{Least Fit Drone Selection}
\label{ss:leastfit}

PADD iteratively improves the utility until convergence.
The algorithm uses the idea behind EO algorithms. It selects the {\it least fit} element and re-sets its parameters in order to improve system performance.
In our case, an \aBS is selected and a new location is {\it probed}.

The choice of which \aBS is the least fit is made based on the consideration that there are two factors that cause sub-optimality of \aBS positions:
(i) the \aBS has a bad backhaul connectivity, hence it provides a worse service to users than what the access channel conditions allow, i.e., access resources are wasted;
or (ii) the \aBS offers bad access connectivity to users due to inter-drone interference, even though it has a good backhaul service, so that backhaul resources are wasted.
Accordingly, we derive two indicators of sub-optimality as the relative difference between the aggregate utility from \aBS-served users (namely $U_{\mathrm{thr}, a}^\alpha$) and the following quantities:
(i) the utility of the \aBS assuming infinite backhaul capacity, namely $U_{\mathrm{thr}, a}^{\alpha, B_\infty}$;
and (ii) the utility of the \aBS assuming no inter-drone interference, namely $U_{\mathrm{thr}, a}^{\alpha, \SNR}\!$.
Eventually, we pick as least fit \aBS the drone~$a_0$ that provides the higher value of all sub-optimality indicators:

\vspace{-3mm}
\small
\begin{eqnarray}
  a_0 \!=\! \arg\max\limits_{a\in\mathcal{A}}\left(\!\!\max\!\left(\left|\frac{U_{\mathrm{thr}, a}^{\alpha, B_{\!\infty}} \!-\! U_{\mathrm{thr}, a}^\alpha}{U_{\mathrm{thr}, a}^{\alpha, B_{\!\infty}}}\right|\!, \left|\frac{U_{\mathrm{thr}, a}^{\alpha, \SNR} \!-\! U_{\mathrm{thr}, a}^\alpha}{U_{\mathrm{thr}, a}^{\alpha, \SNR}} \right|\right)\!\!\right)\!.
\label{eq:leastfit}
\end{eqnarray}
\normalsize

Computing utilities for \aBS $a$ needs $\mathcal{O}(U_{\max})$ sums and powers (logarithms for $\alpha\!=\!1$).
Hence, finding the maximum shown in Eq.~\eqref{eq:leastfit} has complexity $\mathcal{O}(|\mathcal{A}| \!\cdot\! U_{\max})$ powers (or logarithms), the sums incurring negligible extra complexity.

Having identified $a_0$, PADD selects a new random position within the allowed 3D ball space
around the current position of the drone.


\subsection{Complexity of PADD}
In this section we provide the complexity of PADD. The algorithm consists of sequential steps, involving: user association with BSs; the backhaul association solved with the GAP problem of Fig.~\ref{fig:GAP}; the resource allocation of each \gBS solved with the convex optimization of Fig.~\ref{fig:CP}; the system utility evaluation with the $\alpha$-fairness metric; and the least fit drone selection of the EO operation.
The overall PADD's complexity is $\mathcal{O} \left( N \left(\frac{1}{2\log\!2}|\mathcal{B}| \log |\mathcal{B}| + 2A_g |\mathcal{G}| |\mathcal{A}| + 4A_g  U_{\max} + | \mathcal{U} | | \mathcal{B} |\right) \right) $ simple operations (e.g., sums or comparisons) and $\mathcal{O} \left(N | \mathcal{B}|  + 4|\mathcal{A}| U_{\max}\right)$ complex operations (e.g., powers or logarithms). 
This means that PADD's complexity is low-order polynomial, linear with respect to the number of mobile users and log-linear with respect to the number of~BSs.
More details are available in
Section~\ref{a:Complexity} in the online supplemental material.

Moreover, PADD's iterations converge very fast (we have observed no more than a hundred iterations) and our algorithm implemented on \MATLAB took at most about 800 billion floating-point operations to optimize drone positions, which requires, e.g., about 10~s on a linux machine with an Intel Xeon E5-2670 processor at $2.6$~GHz with a processor performing 32 flops/cycle with 64-bit operations.

\vspace{-1mm}
\section{Numerical Simulations}
\label{s:results}

Here we present the numerical evaluation of the \ifc{$\alpha$-fair cellular capacity optimization} in both static and dynamic scenarios. All our simulations are performed over the real topology of an operational network deployed in a dense city: 150,000 inhabitants in a 10~km$^2$ area. (Legan\'es, south of Madrid, Spain). The area is covered by 10 \gBSs using the same LTE band, as shown in Fig.~\ref{fig:leganes}.

As there are several operators in Spain providing LTE service over multiple bands, and we only consider one operator in one band, we use a number of 1000 UEs that request service at the same time, unless otherwise specified.

First, we numerically validate our proposed PADD in comparison with optimal results approximated by means of Monte-Carlo (\texttt{MC}) simulations in static networks with limited population: We perform $10^7$ MC runs per instance where \aBSs are placed in feasible random positions. Then, we optimize the network for each run and save the best setting as the approximated optima.
Second, we analyze the robustness of the system model in order to prove that assuming perfect knowledge about user positions has limited impact on performance.
Eventually, we study three significative static and dynamic scenarios:
\begin{itemize}
\item {\sl Poisson Point Process (PPP)}: we statically place UEs on the city map according to a Poisson point process~\cite{last2017lectures}.
\item {\sl Stadium}: we statically place 60\% of the UEs in the surroundings of a stadium, and the rest like in {\sl PPP}.
\item {\sl Event}: 40\% of the population of UEs moves according to the random way-point model,
whereas other UEs keep arriving at an official scheduled rate of a train station and move towards the stadium.
\end{itemize}
For each scenario, we study drone placement, utility and throughput achieved, and system fairness measured with the Jain's index~\cite{jain1984quantitative}.

Table~\ref{t:parameters} reports the evaluation parameters used in our simulations.
We take a carrier bandwidth of $20$~MHz for both \gBS and \aBS channels (out of which, we consider that $10\%$ is for guard bands, so we only use $18$~MHz~\cite{sesia2011lte}). The carrier frequency of cellular links is $1815.1$~MHz, while for air-to-ground links we use $2630$~MHz. These are two commonly used LTE bands. The transmission power of \aBSs is 25~dBm, notably lower than the 44~dBm power transmission of the ground \gBSs. In addition, probed aerial positions arise from a lattice that spans equal-volume subspaces.

To position our proposal, we consider the performance without drones as baseline ({\tt Ground} in the figures).
Moreover, to compare PADD to state of the art drone-position optimization frameworks, we have implemented and tested the {\it Repulsion-Attraction} scheme (\texttt{RA})~\cite{andryeyev2017increasing}.

\begin{figure}[t]
  \centering
   \vspace{-0mm}
         \includegraphics[width=0.9\columnwidth]{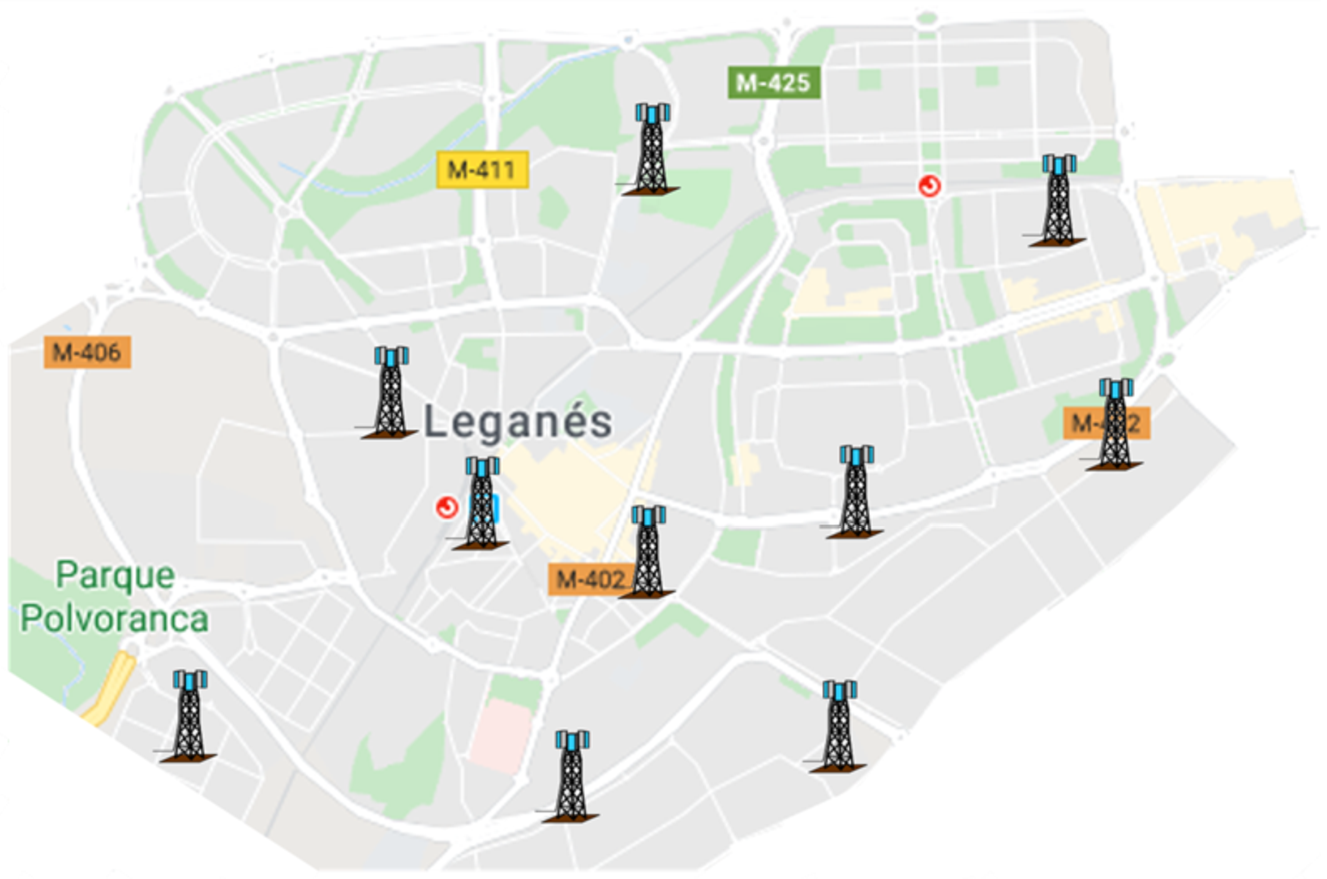}
         \vspace{-2mm}
        \caption{\eag{Topology of Legan\'es (Spain) and \gBSs placement.}}
        \vspace{-4mm}
        \label{fig:leganes}
\end{figure}

Although our analysis holds for generic values of $\alpha$, we present results for three specific and most interesting cases:
\begin{enumerate}[label={(\roman*)},, wide = 0.5em, leftmargin = 2.2em,  rightmargin = 0em]
	\item For $\alpha = 0$, we obtain the maximum throughput achievable ({\tt MaxThr} 	in the figures).
	\item For $\alpha = 1$, we optimize the proportional fairness network metric ({\tt 		PropFair} in the figures).
	\item For $\alpha \rightarrow \infty$, we optimize the max-min fairness network 		metric ({\tt MaxMin} in the figures).
\end{enumerate}
We study these three metrics for the following reasons.
The \texttt{MaxThr} metric provides the maximum achievable network capacity, without fairness. 
The \texttt{PropFair} metric takes into account the network capacity but it also does not let fairness decay. The mathematical design of this metric, which optimizes the aggregation of logarithms of user throughputs, allows for finding a trade-off between a high system capacity and high fairness values.
The \texttt{MaxMin} metric only targets fairness of the weakest customer, which comes at the cost of providing lower aggregate throughput.
Both the \texttt{PropFair} and \texttt{MaxMin} have been adopted in the implementation of real telecommunication systems~\cite{miao2016fundamentals} as well as in many research works~\cite{mosleh2016proportional, li2019max}.
Complementary, in
Section~\ref{a:OptimalCP} in the online supplemental material,
we provide more theoretical results for the PADD scheme for a generic value of $\alpha$.

We have simulated every analyzed use-case $1000$ times using \MATLAB \texttt{R2020a} and show average results.
Error bars in the figures are 95\% confidence intervals.

\begin{table} [t]
\vspace{2mm}
\caption{Evaluation Parameters}
\vspace{-2mm}
    \label{Parameters}
    \centering
    \begin{tabular}{|c|c|} 
    \hline
       \textbf{\textsl{Parameter}} & \textbf{\textsl{Value} }  \\ \hline \hline
       $\xi_{LoS}$, $\xi_{NLoS}$, $\beta_1$, $\beta_2$ & 1.6 dB, 23 dB, 12.08, 0.11 \\ \hline
       Carrier frequencies, $f_\mathcal{G}$, $f_\mathcal{A}$ & $1815.1$ MHz, $2.63$ GHz \\ \hline
       Bandwidths, $W_\mathcal{G}$, $W_{\!\mathcal{A}}$& $18$ MHz, $18$ MHz  \\ \hline
       Tx power, $P_{Tx}^g$, $P_{Tx}^a$ & $44$ dBm, $25$ dBm \\ \hline
       Thermal Noise Power & -174 dBm/Hz  \\ \hline
       Ground path loss exponent, $\eta_{BS}$ & $3$ \\ \hline
       Height range, $[h_{\min}, h_{\max}]$ & $[40, 300]$ m \\ \hline
       Urban area, $|\mathcal{S}|$ & $10$ km$^2$ \\ \hline
       Average walking speed & $2.5$ m/s \\ \hline
       Monte-Carlo runs per instance & $10^7$ \\ \hline
       Instances of simulations & $1000$ \\ \hline
    \end{tabular}
    \vspace{-3mm}
    \label{t:parameters}
\end{table}

\begin{figure*}[t]
\centering
\hspace{-6mm}
\vspace{-1mm}
\begin{minipage}[T]{.3\textwidth}
\hspace{-3mm}
    \includegraphics[width=6.15cm]{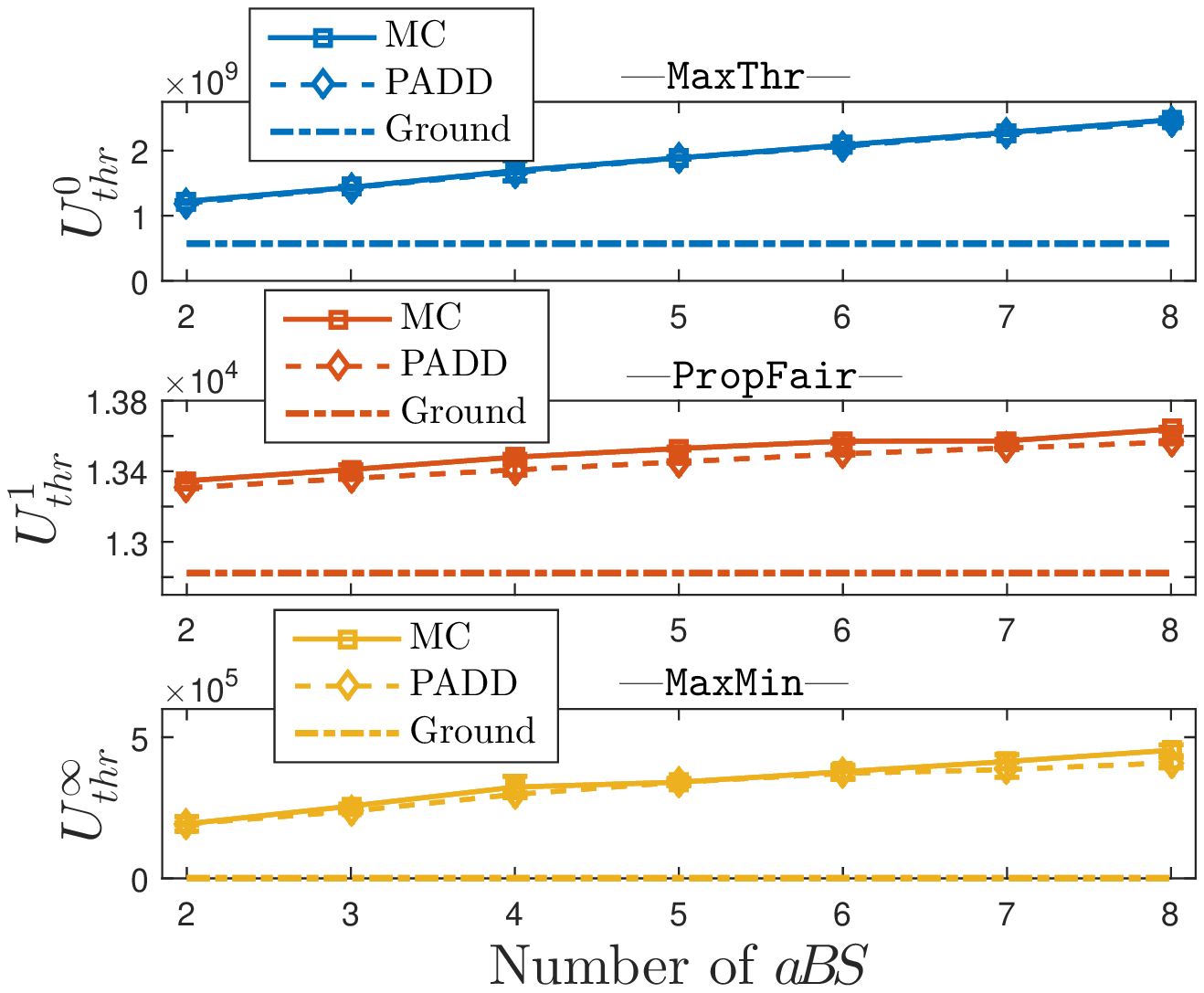}
    \caption{\eag{Utility validation for $\alpha~\in~\{0, 1, \infty\}$. $G = 10$, $U=1000$. Scenario: {\sl PPP}.}}
\label{fig:MC_metric}
\end{minipage}
\hspace{5mm}
\begin{minipage}[T]{.3\textwidth}
\hspace{-3mm}
    \includegraphics[width=6.15cm]{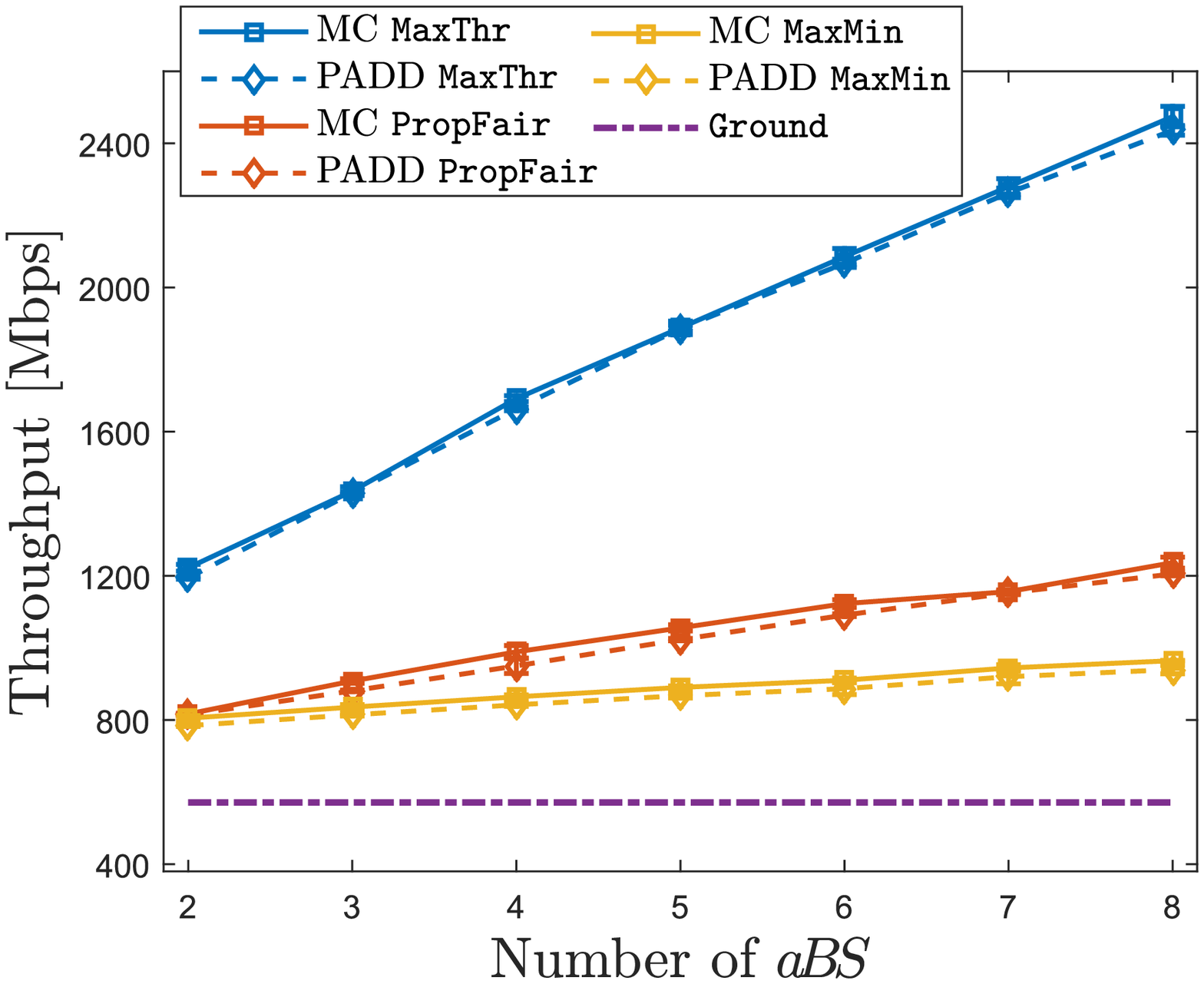}
    \caption{\eag{Network capacity validation for $\alpha\in\{0, 1, \infty\}$. $G = 10$, $U=1000$. Scenario: {\sl PPP}.}}
\label{fig:MC_thr}
\end{minipage}
%
\hspace{5mm}
\begin{minipage}[T]{0.3\linewidth}
\hspace{-5mm}
\includegraphics[width=6.15cm]{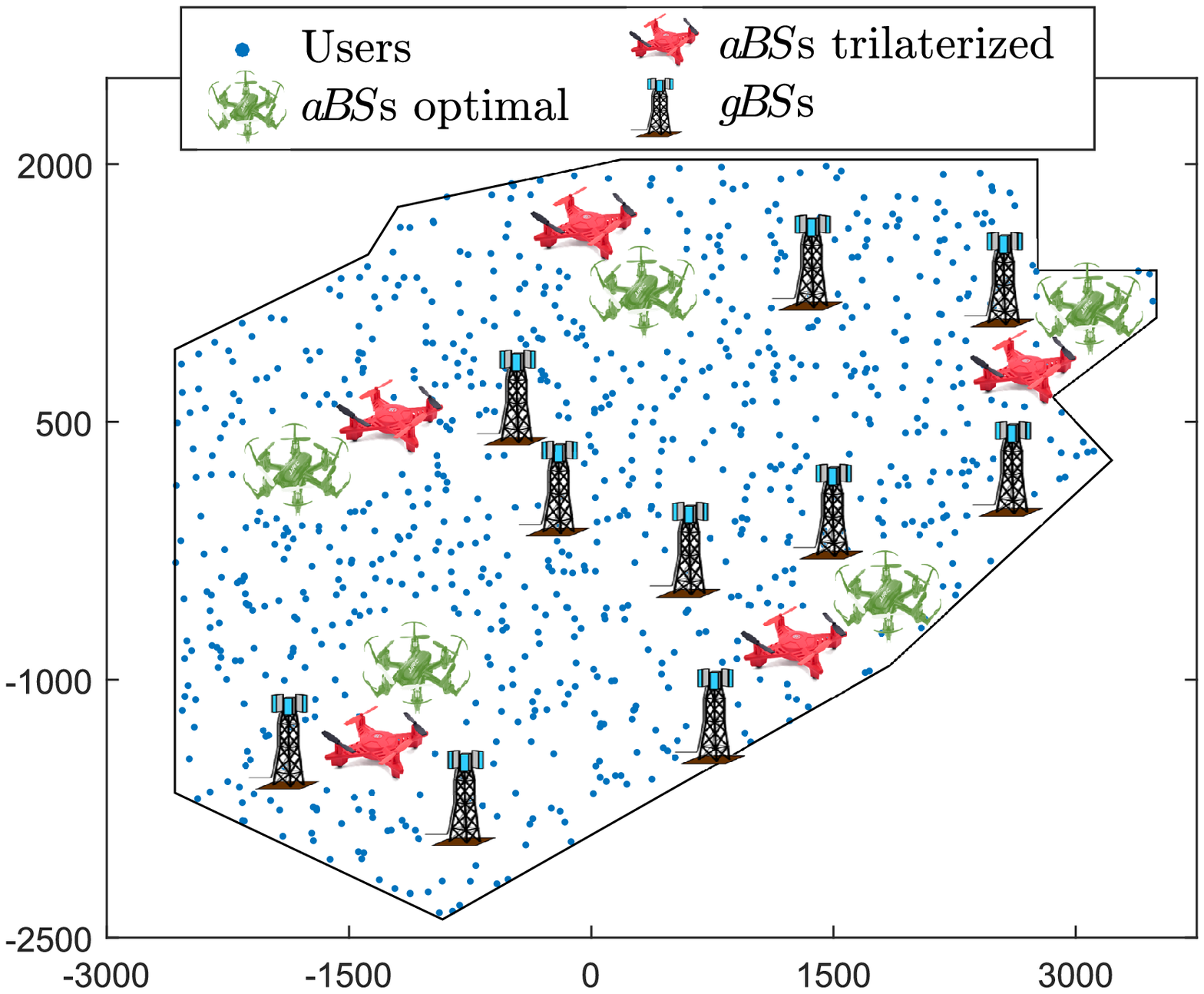}
    \caption{\ifc{Robustness validation for $\alpha=1$. \aBSs placement error. $G = 10$, $A = 5$, $U = 1000$. Scenario: {\sl PPP}.}}
\label{fig:Rob_pos}
\end{minipage}
\vspace{-5mm}
\end{figure*}

\subsection{Validation of PADD Operation}


Here we compare PADD results and optima \ifc{(approximated by means of \texttt{MC}, since there are no computationally feasible alternatives)} in the {\sl PPP} scenario, and comment on the basic properties of our approach.

In Fig.~\ref{fig:MC_metric}, we present a utility comparison between the \texttt{Ground} scheme, optima results from \texttt{MC} and our proposal PADD with the three main metrics studied in this paper (\texttt{MaxThr}, \texttt{PropFair} and \texttt{MaxMin}).
Note that different values of $\alpha$ make utility values very different, so we cannot compare utilities across fairness metrics. So, we will compare PADD and the existing schemes on a per-metric basis.

We observe that the difference between utilities achieved with PADD and \texttt{MC} is negligible, below $1\%$ in all cases. This shows that our proposal PADD is able to achieve close-to-optimal results with a much lower complexity.
Furthermore, PADD matches very well also the throughput achievable in the optimal case, as illustrated in Fig.~\ref{fig:MC_thr}.
This shows that PADD is able to pursue the optimal system configuration, not just an operational configuration that is near-optimal according to the chosen utility metric.

The figures show that, in the PPP scenario, utility and throughput increase with fleet size.
This is because user's density is homogeneously spread over the entire area, so that drones are spaced apart, and inter-drone interference constraints do not kick in. 
Note also that the gain in terms of utility and throughput with respect to the {\tt Ground} configuration is remarkable under all considered metrics, which confirms that coordinated drone relays have huge potential.

\subsection{Robustness of PADD}

\begin{figure*}
\centering
\hspace{-0cm}
\vspace{3.5mm}

\hspace{-0cm}
\vspace{3mm}
\begin{minipage}[T]{.3\textwidth}
\hspace{-3mm}
\includegraphics[width=6.15cm]{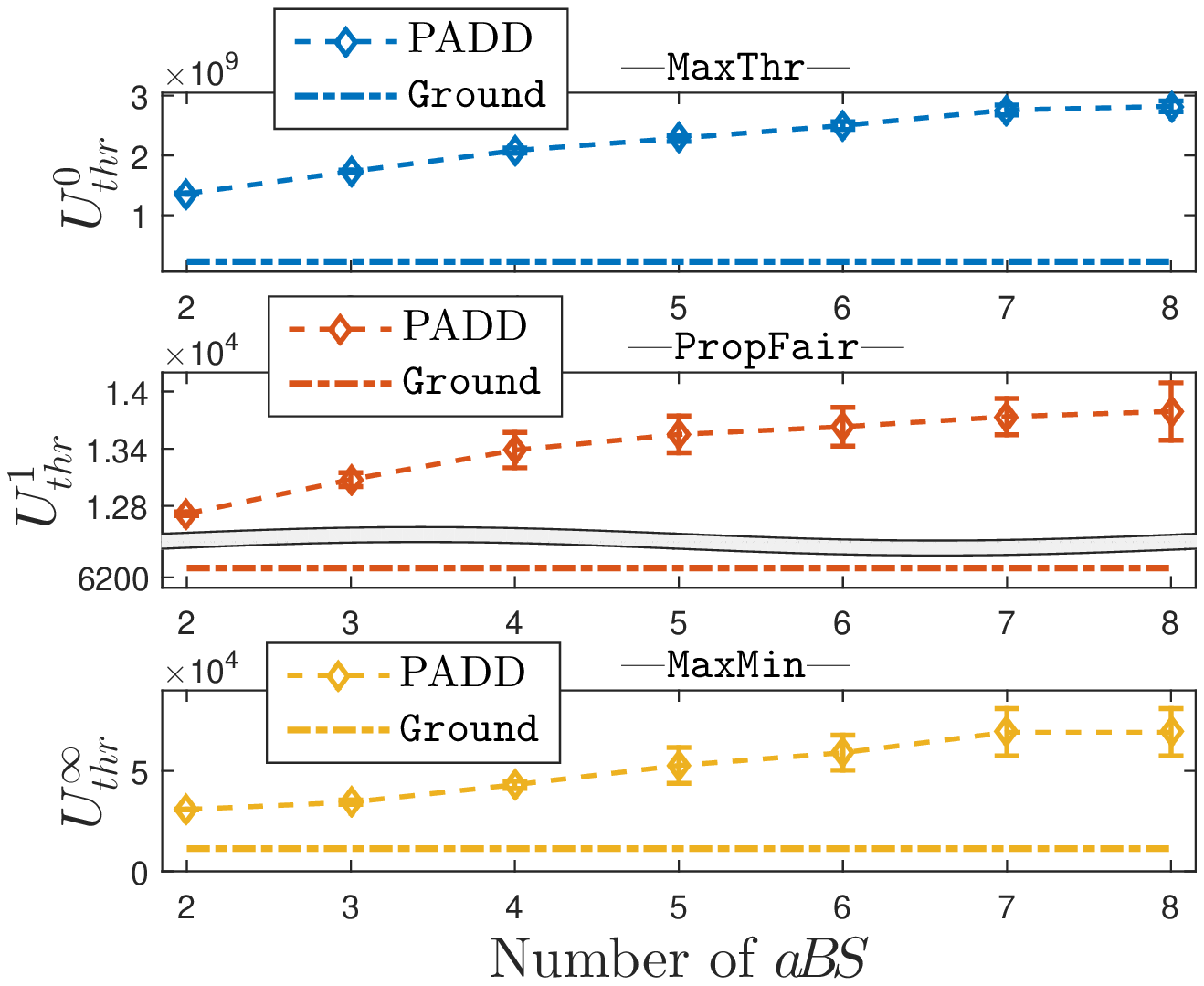}
    \vspace{-5.5mm}
    \caption{Utility of all users for $\alpha\in\{0, 1, \infty\}$. $G = 10$, $U = 1000$. Scenario: {\sl Stadium} with $U_d = 600$.}
\label{fig:metric_net}
\end{minipage}
%
\hspace{5mm}
\begin{minipage}[T]{.3\textwidth}
    \includegraphics[width=6.15cm]{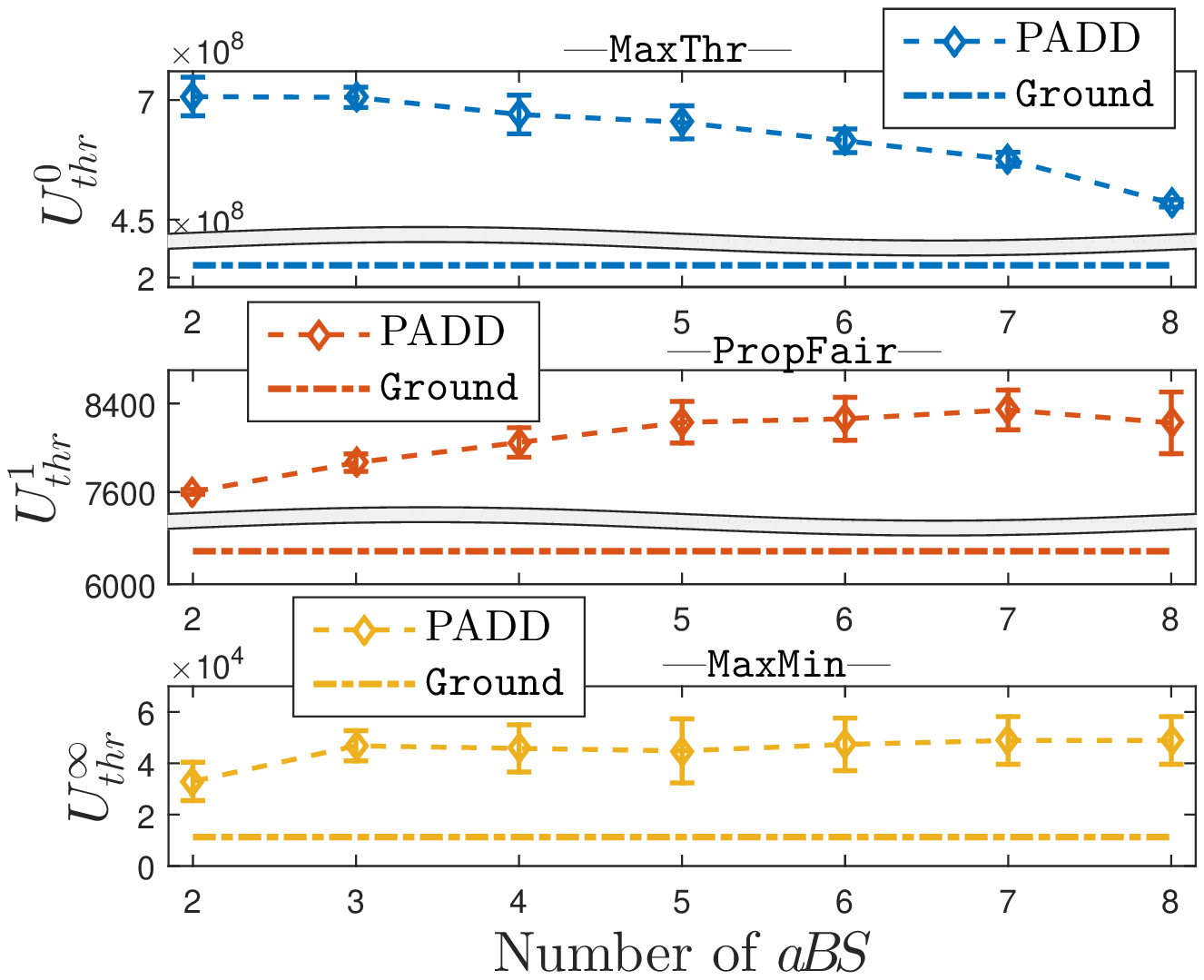}
    \vspace{-5.5mm}
    \caption{Utility of stadium users for $\alpha\in\{0, 1, \infty\}$. $G = 10$, $U = 1000$. Scenario: {\sl Stadium} with $U_d = 600$.}
\label{fig:metric_sta}
\end{minipage}
%
%
\hspace{5mm}
\begin{minipage}[T]{.3\textwidth}
\vspace{-1mm}
\includegraphics[width=6.15cm]{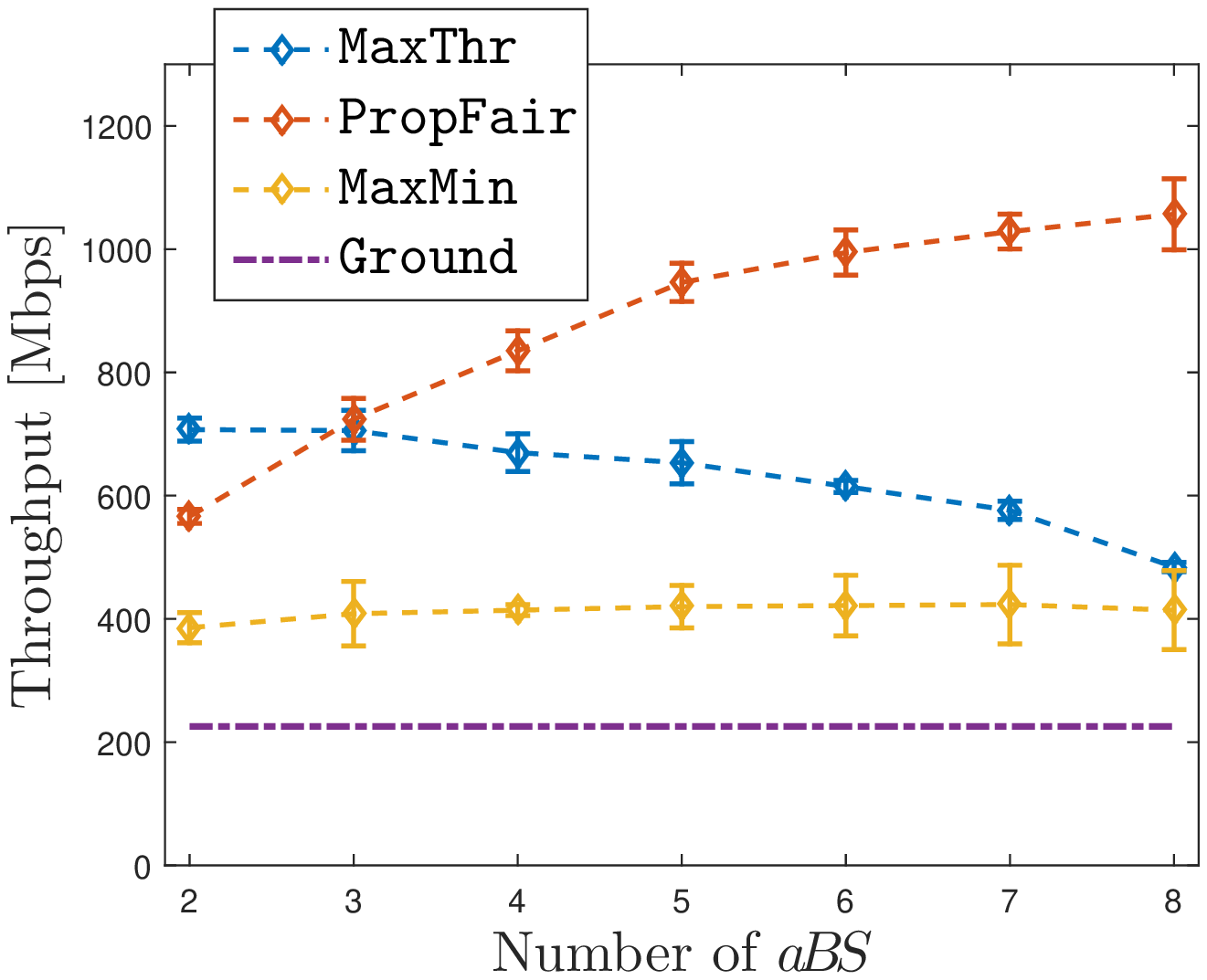}
    \vspace{-5.5mm}
    \caption{Throughput of stadium users for $\alpha\in\{0, 1, \infty\}$. $G = 10$, $U = 1000$. Scenario: {\sl Stadium} with $U_d = 600$.}
\label{fig:thr_sta}
\end{minipage}
\vspace{-7mm}
\end{figure*}

In the system model described in Section~\ref{s:model}, we have assumed that the positions of users are known. However, it might be not realistic to estimate user locations with negligible error unless GPS is enabled or many drones are available~\cite{petrolo2018astro}. For instance, using a trilateration on the signal strength at the base station, the error on the position is normally below $50$~m~\cite{del2017survey}. Hence, here we consider the PPP scenario to numerically analyze the robustness of PADD by introducing uncertainty in the position of the ground users, uniformly at random, within $50$~m.


In Fig.~\ref{fig:Rob_pos}, we observe that the position of drones selected by PADD is similar with and without localization errors. Indeed, we have quantified an average utility relative loss below $5\%$ (see Section~\ref{app:robustness} in the online supplemental material for further details).
The reason of such robust behavior stands in the fact that the optimization of drone positions is done based on many users and in relatively large areas, so that multiple errors in user positions are not so important, whereas the presence of a distributed mass of users in a given area is what actually catalyzes the presence of a drone.

Next, we analyze non-homogeneous user topologies in static and dynamic cases.

\subsection{Performance Evaluation in the Static \textsl{Stadium} Case}

The \textsl{Stadium} scenario allows to study network performance when the ground network cannot sustain the UE's demand.

\begin{figure*}
\centering
\begin{minipage}[T]{.32\textwidth}
\hspace{-3mm}
\includegraphics[width=6.15cm]{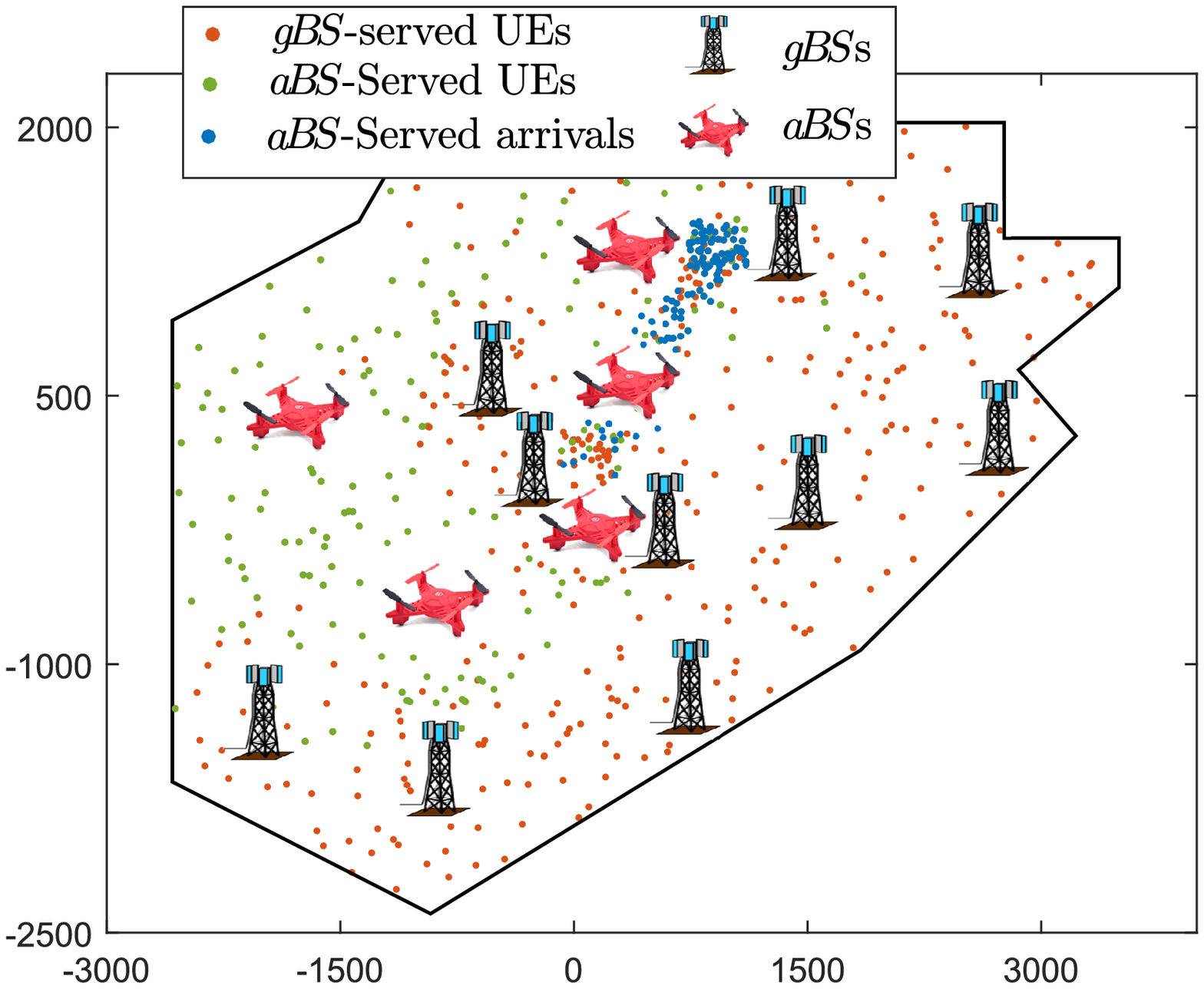} 
    \vspace{-5.5mm}
    \caption{Network state at $t=25$~min. $G=10$, $A=5$, $U=640$. Scenario: {\tt PropFair}, {\sl Event}.}
\label{fig:dyn_map1}
\end{minipage}
\hspace{0.1cm}
\begin{minipage}[T]{.32\textwidth}
    \includegraphics[width=6.15cm]{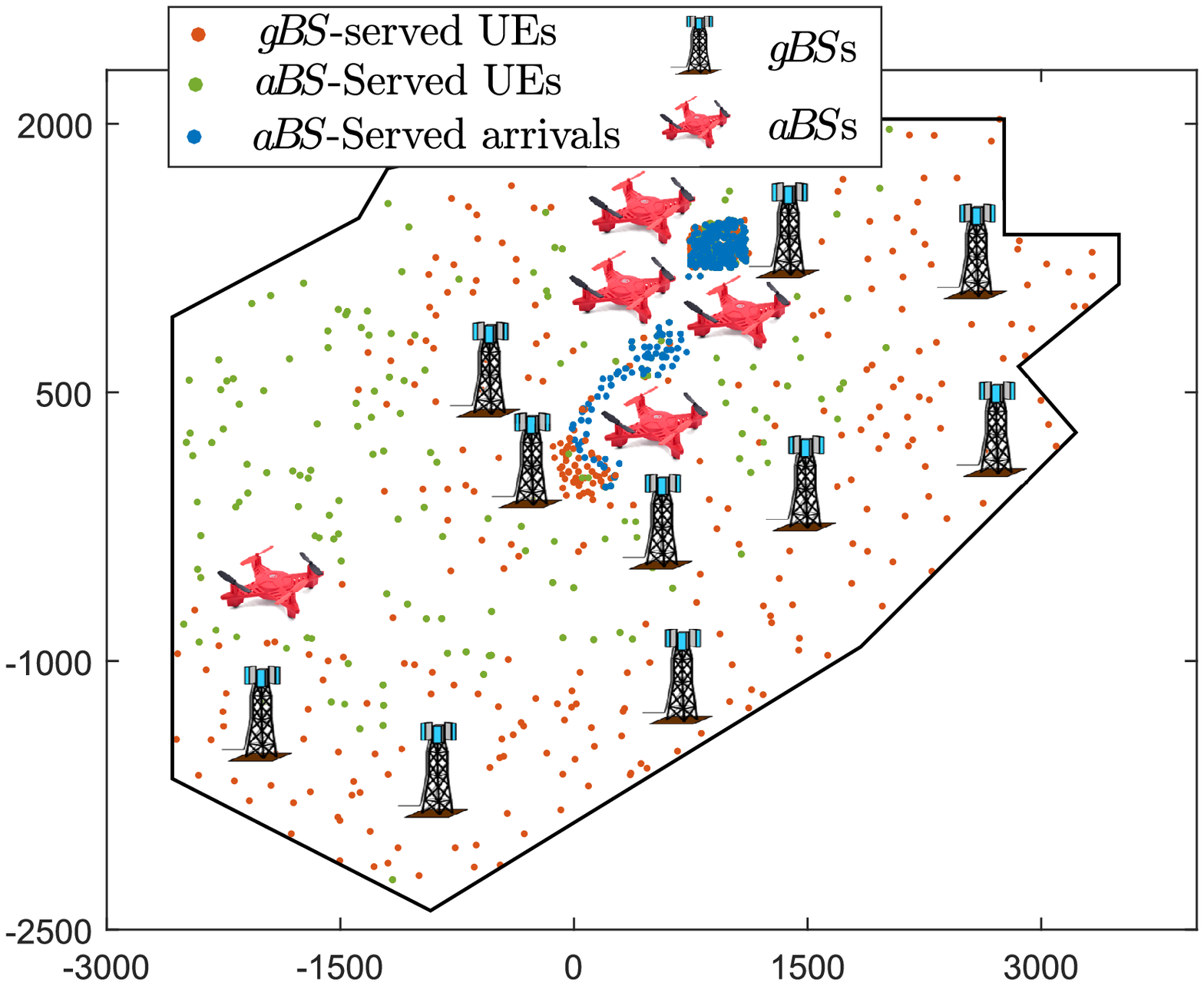}
    \vspace{-5.5mm}
    \caption{Network state at $t=60$~min. $G=10$, $A=5$, $U=880$. Scenario: {\tt PropFair}, {\sl Event}.}
\label{fig:dyn_map2}
\end{minipage}
\hspace{2mm}
\begin{minipage}[T]{.32\textwidth}
\vspace{0mm}
\includegraphics[width=6.15cm]{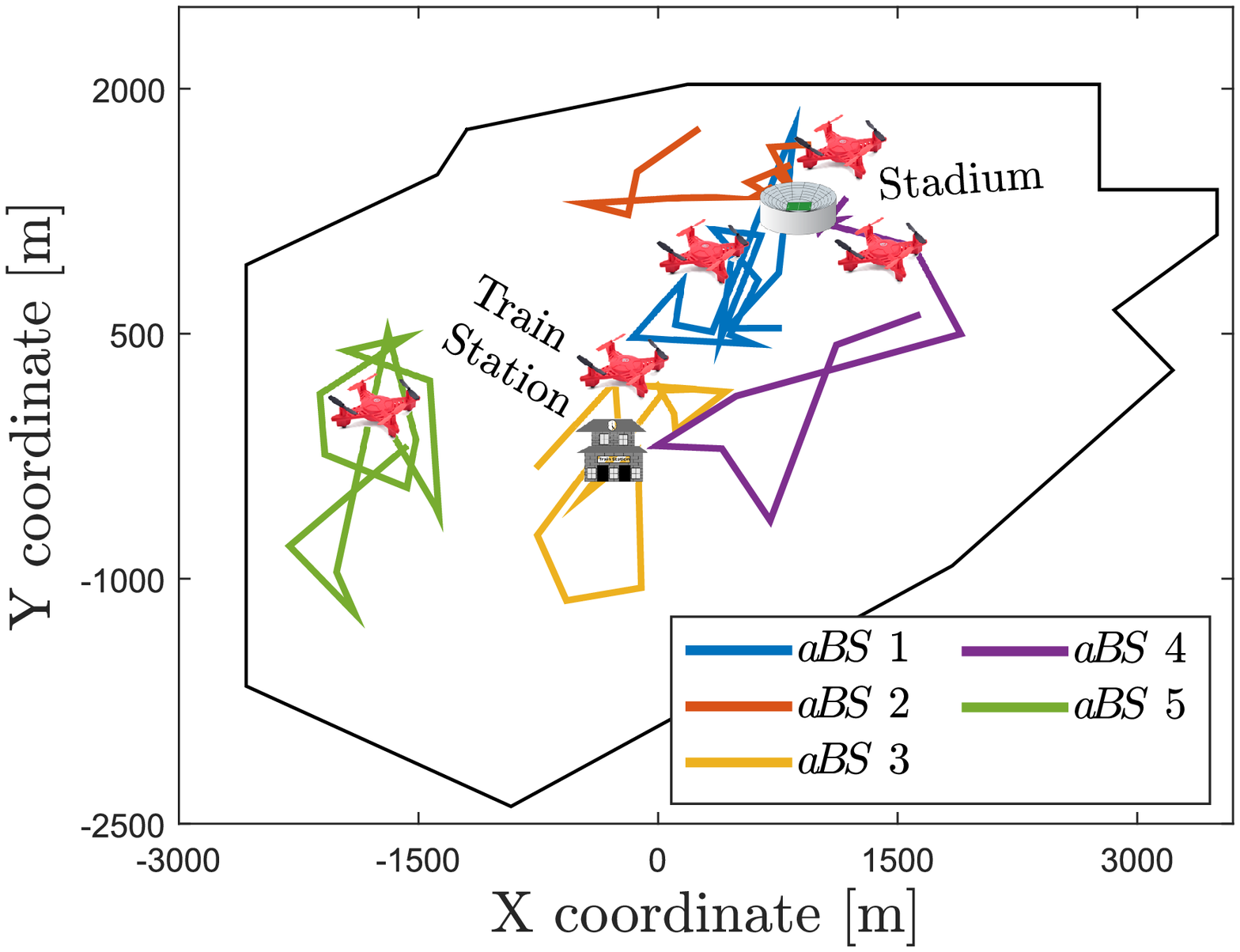}
    \vspace{-5.5mm}
    \caption{Drone trajectories during 75~minutes. $G\!=\!10$, $A\!=\!5$, $U\!=\![400,$ $ \hdots, 1000]$. Scenario: {\tt PropFair}, {\sl Event}.}
\label{fig:dyn_trajectories}
\end{minipage}
\vspace{-4mm}
\end{figure*}

\begin{figure}
\centering
\hspace{-9mm}
\vspace{-5mm}
\hspace{1mm}
\begin{minipage}[T]{.4\textwidth}
\vspace{-4mm}
\includegraphics[width=7.4cm]{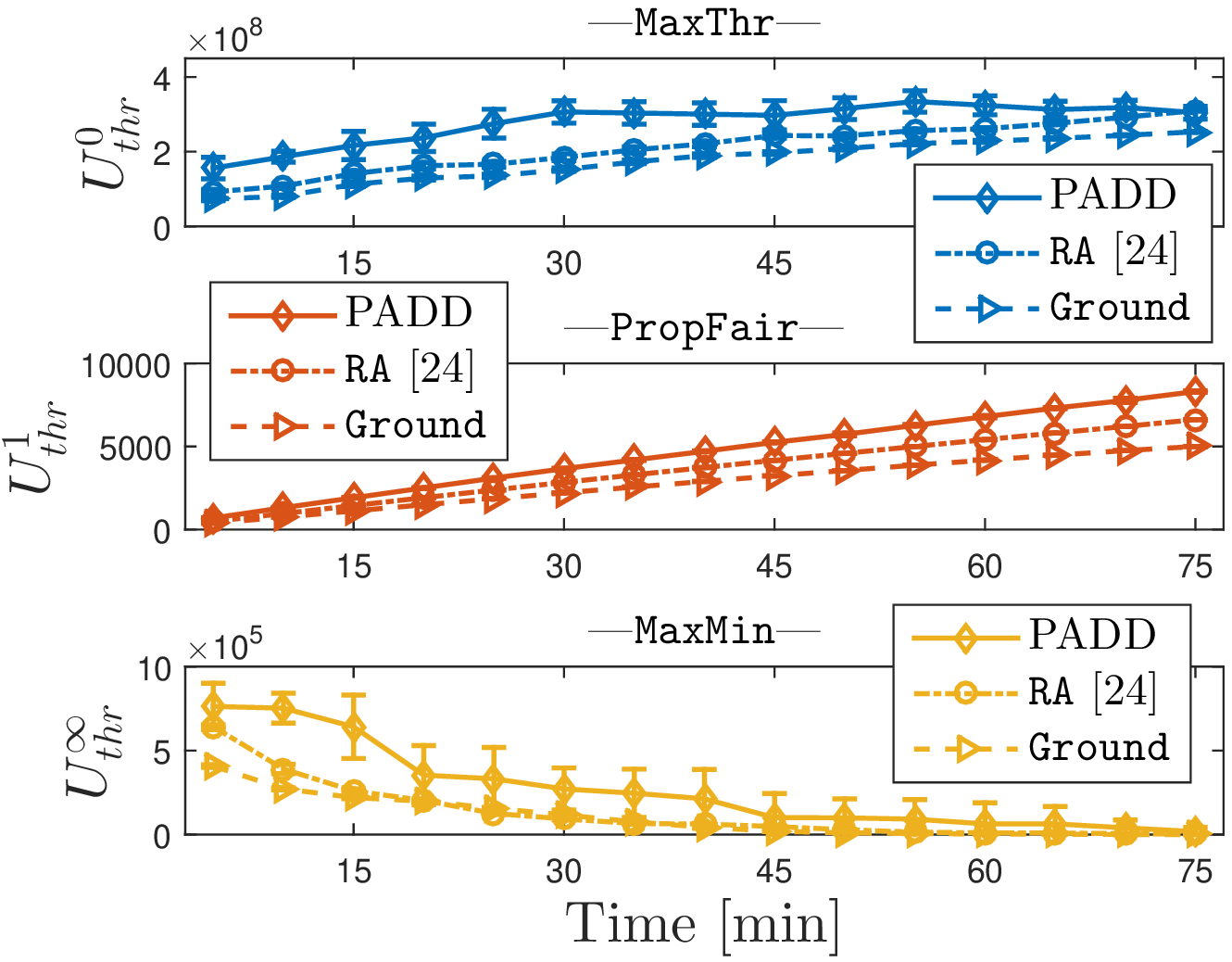}
    \vspace{-5.5mm}
    \caption{Attendance utility for $\alpha\in\{0, 1, \infty\}$. $G = 10$, $A = 5$, $U = [400,\hdots, 1000]$. Scenario: {\sl Event}.}
\label{fig:dyn_utility}
\end{minipage}
%
\vspace{-0.75mm}
\end{figure}

In Fig.~\ref{fig:metric_net} we show the average performance experienced by all users in the scenario, while in Fig.~\ref{fig:metric_sta} we focus on the performance of users by the stadium.
We observe that PADD with {\tt MaxThr} benefits of the presence of drones (see Fig.~\ref{fig:metric_net}, top), although adding drones is negative for users by the stadium (see Fig.~\ref{fig:metric_sta}, top). In fact, adding \aBSs introduces more capacity and connectivity opportunities, hence the benefit at aggregate city level. However, PADD with {\tt MaxThr} seeks for drone positions where they are less impaired by interference, so to generate a few good-quality channels. This behavior results into having only one or two \aBSs at most located near the stadium. More \aBSs would interfere too much. Hence, as long as \aBSs are added, PADD with {\tt MaxThr} positions them apart from the stadium, yet they generate some interference, which progressively worsens the performance of users by the stadium.

PADD with {\tt PropFair} behaves completely different.
It brings drones where it favors users otherwise served below the average, so to increase the system's log utility. This results in deploying \aBSs by the stadium, where the density of users in presence of limited radio resources hinders performance more than radio channel quality issues.

With  {\tt MaxMin},
PADD  positions drones where users would otherwise suffer the worst connection quality, irrespective of their closeness to densely populated areas. Therefore the aggregated throughput is lower than with {\tt PropFair}, which in turn is much lower than with  {\tt MaxThr}. 

If we now consider only users by the stadium, Fig.~\ref{fig:thr_sta} illustrates how using {\tt PropFair} can clearly outperform {\tt MaxThr} in terms of throughput. This is due to the fact that the optimization of throughput requires positioning drones where they interfere less, which is not necessarily by the stadium. Indeed, due to interference, users by the stadium could incur a loss by increasing the number of drones even in the case of using {\tt MaxMin}, as shown in the figure.

As a result of the previous numerical analysis, we observe that using \aBSs can be beneficial to help the ground cellular network in dense spots, except it cannot help in purely maximizing throughput (e.g., with {\tt MaxThr}). Drones cannot either ``rescue'' all users with bad channel conditions, as PADD would seek with {\tt MaxMin} and {\tt PropFair}. However, PADD can always provide fairness and large throughput gain with respect to the {\tt Ground} case. Besides, the version of PADD with {\tt PropFair} results to be quite effective in case of dense masses of users.

\subsection{Performance Evaluation in the Dynamic \textsl{Event}  Case}
\label{ss:Event}

The last scenario we consider is dynamic and allows us to study the evolution of network performance {\sl while} the density of users increases. Moreover, it shows the importance of designing a fast and reactive algorithm to re-position drones as the user topology changes.

In the Event scenario, small masses of 40 users arrive periodically to the train station of the city every 5 minutes.
The initial population is 400 users and keeps growing during 75 minutes up to 1000 users. There are 5 \aBSs hovering the area in this example.
Upon a train arrival, the new mobile users walk towards the stadium, located 1.5~km away from the train station.
The fleet is repositioned every 5 simulated minutes, using as initial condition the positions of the 5 drones in the previous optimization epoch.

We first illustrate how the network evolves over time in Figs.~\ref{fig:dyn_map1} and~\ref{fig:dyn_map2}, using PADD with {\tt PropFair}.
After 25 minutes (Fig.~\ref{fig:dyn_map1}) some people are already at the stadium, while many others keep arriving and are walk towards it. At that point in time, 3 \aBSs 
are getting prepared to serve the users nearby the stadium and also the smaller groups on their way from the train station to the stadium.
We see how drones adapt their positions after other 35 minutes, in Fig.~\ref{fig:dyn_map2}, when much more users have reached the stadium. By that time, one more drone has been dispatched as well, to serve the stadium.
The trajectories of the drones for a 75-minute simulation instance are shown in Fig.~\ref{fig:dyn_trajectories}, where the red drones represent the source position of \aBSs.
In the figure, we can see that \aBS~5 is not required to assist the dense stadium spot, while \aBSs~1 and~2 are always hovering between the train station and the stadium. Also, \aBSs~3 and~4 keep moving back and forth within different regions, to fairly supply the demand of users.


As concerns performance,
Fig.~\ref{fig:dyn_utility} illustrates
utility 
as they evolve over time.
Here, in addition to PADD with the three selected 
$\alpha$'s, we also compare the {\it Repulsion-Attraction} (\texttt{RA}) scheme.
With {\tt RA}, \aBSs are attracted by UE's inverse SNR, and repulsed by proximity to \gBSs to avoid interference. \texttt{RA} does not target any fairness metric, so we quantify its impact with the same utility functions used for PADD 
computed for
users arriving over time (the {\it attendance}).

As time passes by, and more people reach the stadium,
Fig.~\ref{fig:dyn_utility} shows a significant utility raise under the adoption of either {\tt MaxThr} or {\tt PropFair} schemes.
With {\tt MaxThr}, the gain of PADD over \texttt{RA} and {\tt Ground} schemes is high, although it saturates quickly. Instead, with {\tt PropFair}, PADD  exhibits a smoother behavior, as its gain keeps increasing.

Under the {\tt MaxMin} scheme, PADD performs better than {\tt RA}, although now we observe a decay of performance over time, for all schemes.
This is due to the fact that, with more attendance, the minimum per-user achieved rate will have  decrease, unless more drones where deployed.

Clearly, the {\tt RA} scheme is not able to opportunistically take advantage of user's diversity and improve utility because it does not target a throughput-based metric, unlike PADD. This permits PADD to optimize the network with much better guarantees in terms of throughput and fairness.

\vspace{-1mm}
\section{Lessons learnt and discussion}
\label{s:lessonslearnt}

The performance assessment carried in this paper shows the importance of integrating a fleet of drone relay stations in a cellular network.
It also unveils that optimizing drone positions to maximize throughput, without taking into account fairness, has little relevance in the presence of dense spots or ground users. Instead, a fair metric like {\tt PropFair}  provides notable throughput and utility improvements.

The fact that PADD is fast, allows to design an almost continuous reconfiguration of positions in realistic networks, as shown in this paper. In turn, our scheme is fast because we have designed it by segmenting the problem into a few phases: we use EO for optimizing drone positions, while wireless backhaul attachment, BS selection and distributed resource allocation are sequentially solved optimally and analytically for each drone deployment topology.

The actual implementation of PADD requires the exchange of signaling information between drones and a centralized orchestrator, as well as the implementation of a mechanism to track users.
Signaling incurs some limited overhead to instruct drones and to gather user positions and interference reports (which are however already collected by current BSs), depending on the frequency of reconfigurations. However, we have seen that PADD is robust to imprecise tracking of user's positions, which can be therefore strongly simplified, so that the additional overhead due to PADD will mostly be due to controlling drones.

Finally, we comment on drone mechanicals. Current commercial drones can carry small BSs and access points, although they cannot fly for long time due to battery limitations (around 30 minutes at most). They can easily move at the reasonable speed of 15~m/s. Therefore, it is possible to derive a repositioning scheme that accounts for replacing drones that go back to the charging station if the drones do not fly too far from it. For the case considered in Section~\ref{ss:Event}, the routes flown by drones in 5 minutes are short enough, and it is easy to hover a city district in a few minutes. Thus, notwithstanding the intricacies of the analysis, the performance evaluation discussed in this paper is quite relevant for realistic systems.

\section{Conclusions}
\label{s:conclusions}

In this paper we have proposed an analytic framework to optimize drone-aided cellular networks in terms of an $\alpha$-fair throughput utility function under realistic stochastic models.
Specifically, we have studied the integration of a coordinated fleet of {\it aerial base stations} carried by drones and relaying traffic for ground base stations by means of 3D-beamforming wireless backhaul connections.
Due to the complexity of the studied problem, we have proposed PADD, an optimization algorithm based on extremal optimization that works in parallel threads.
PADD provides near-optimal solutions in
low-degree polynomial time, with a linear dependency on all parameters but for the number of base stations, which causes a sub-quadratic dependency. 
This makes PADD suitable for implementation in dynamically changing environments.
The performance evaluation presented in the paper shows that PADD brings significant gain and outperforms existing approaches. It also unveils that using fairness is key to get benefit from coordinated yet interfering drone relay stations.



\ifCLASSOPTIONcaptionsoff
  \newpage
\fi

\bibliographystyle{IEEEtran}
\balance
\bibliography{biblio}

\vspace{-11.5mm}
\begin{IEEEbiography}[{\vspace{0mm}\includegraphics[width=0.8in,height=1.2in]{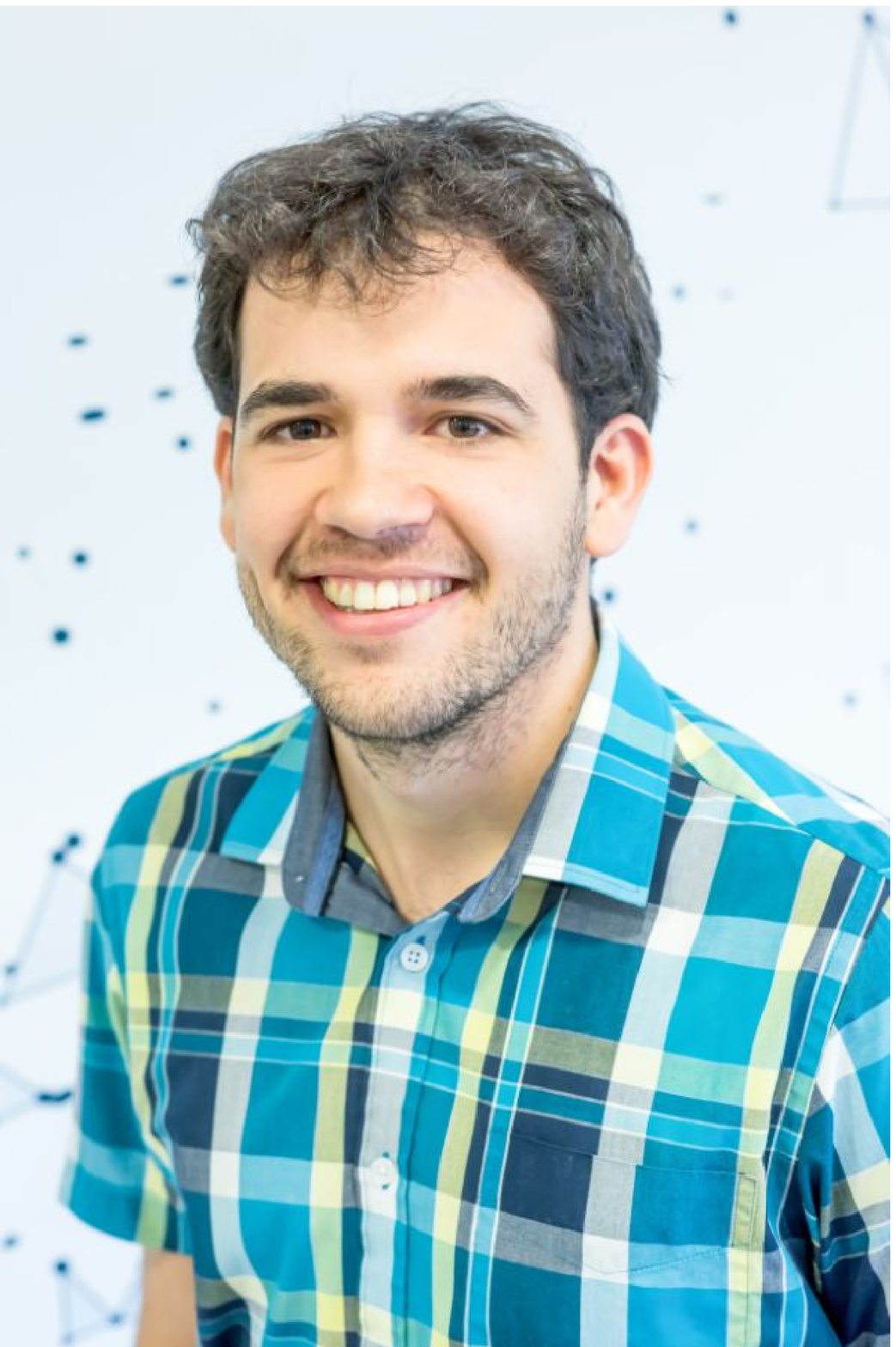}}]{Edgar~{Arribas}}
obtained his PhD in Telematic Engineering in 2020 at IMDEA Networks Institute and Universidad Carlos III de Madrid (UC3M), funded by the MECD FPU15/02051 grant. He is currently a lecturer and researcher at the Applied Mathematics and Statistics Department of Universidad CEU San Pablo (Spain).
He works on optimization of dynamic relay in wireless networks.
\end{IEEEbiography}

\vspace{-11.5mm}
\begin{IEEEbiography}[{\includegraphics[width=1in,height=1.1in,keepaspectratio]{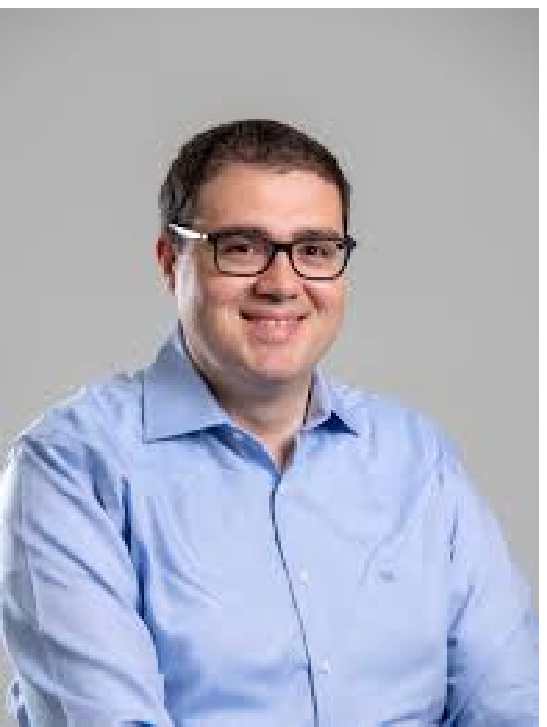}}]{Vincenzo~{Mancuso}}
is Research Associate Professor at IMDEA Networks, Madrid, Spain, and recipient of a Ramon y Cajal research grant
of the Spanish Ministry of Science and Innovation.
Previously, he was with INRIA (France), Rice University (USA) and University of Palermo (Italy), from where he obtained his Ph.D. in 2005.
His research focus is on analysis, design, and experimental evaluation of
opportunistic wireless architectures and mobile broadband services.
\end{IEEEbiography}

\vspace{-11.5mm}
\begin{IEEEbiography}[{\vspace{-11mm}\includegraphics[width=0.95in,height=1.25in,keepaspectratio]{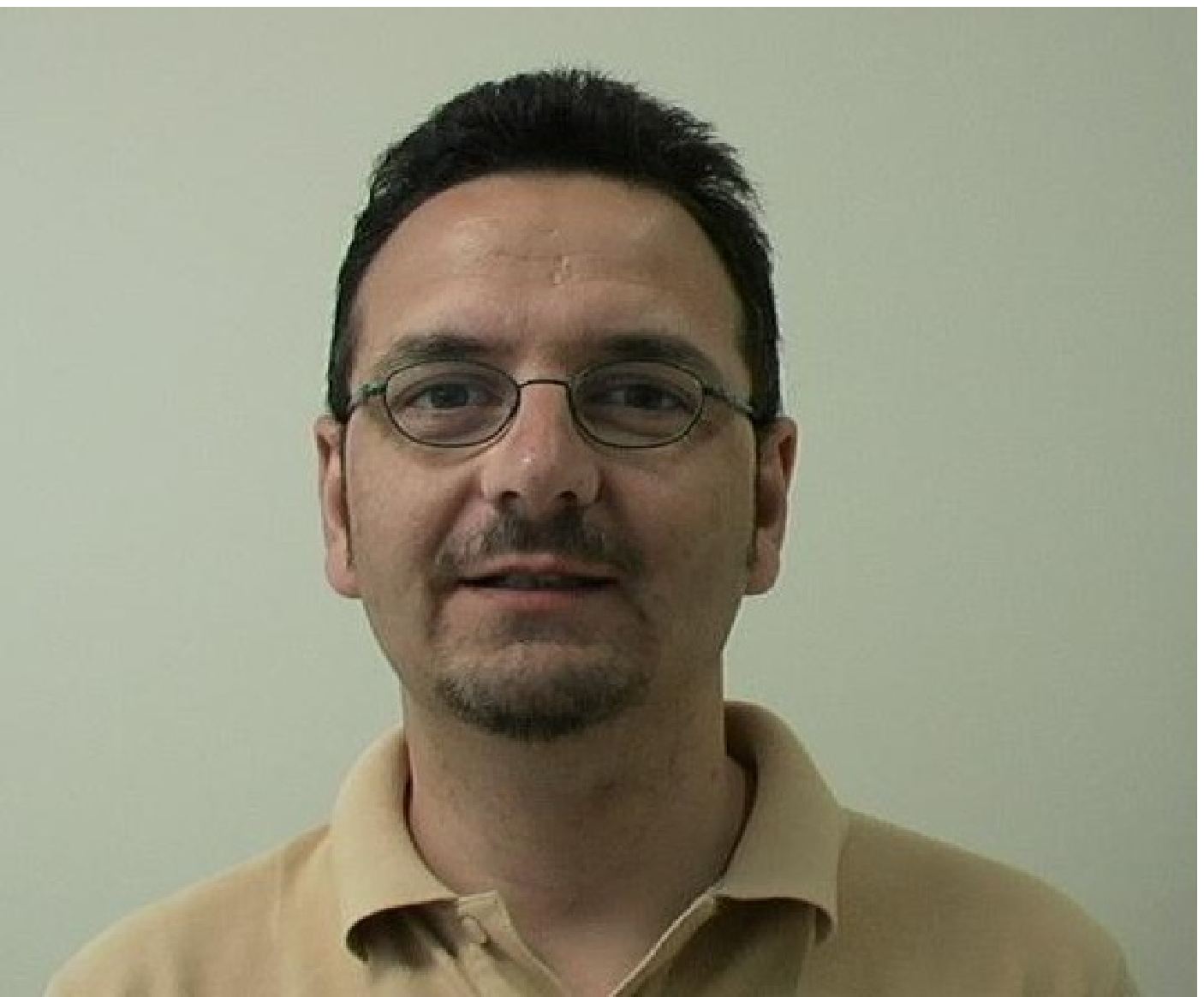}}]{Vicent~{Cholvi}}
graduated in Physics from the University of Valencia, Spain and received his doctorate in Computer Science in 1994 from the Polytechnic University of Valencia. In 1995, he joined the Jaume~I University in Castell\'on, Spain where he is currently a Professor. His interests are in distributed and communication systems.
\end{IEEEbiography}





\clearpage

\renewcommand{\baselinestretch}{1}
\nobalance

\section*{Online Supplemental Material}
\renewcommand\thesection{\Alph{section}}
\setcounter{section}{0}
\renewcommand{\baselinestretch}{0.96}
\section{Channel Modelling}
\label{a:ss:pathloss}
We assume that the network operator disposes of two orthogonal frequency bands. One band is assigned to \gBSs to provide access service to ground users as well as aerial backhaul service to \aBSs. 
The other band is assigned to \aBSs for aerial user access.
Hence, we model three different channel types: (i) air-to-ground and (ii) ground-to-ground channels in the access network, and (iii) ground-to-air channels in the backhaul network.

Indeed, the access network communication channels between BSs and UEs differ much depending on whether users connect to a \gBS or to an \aBS.
While the ground-to-ground channel attenuation for \gBS--UE links follows conventional path-loss modeling based on slow and fast fading, air-to-ground channels (\aBS--UE links) suffer additional attenuation depending on the LoS---or NLoS---state of the channel.
Such additional attenuation is referred to in the literature as an \textsl{excess attenuation}~\cite{al2014modeling}. Moreover, antennas used for the access network differ from backhaul network antennas performing 3D-beamforming, which directly affects the interference suffered in each case. In the following sections, we detail these features for each type of modelled channel.

\vspace{-1mm}
\subsection{Air-to-Ground Channels}
\label{a:ss:A2G}

Depending on whether links between \aBSs and UEs are free of obstacles (e.g., buildings, traffic, etc.), the attenuation differs notably~\cite{al2014modeling}.
The LoS-likelihood is a complex function of the elevation angle between UE $u\in\mathcal{U}$ and \aBS $a\in\mathcal{A}$:

\vspace{-3mm}
\small
\begin{eqnarray}
    P_{LoS}(a, u) = \frac{1}{1+\beta_1\cdot \exp\left(-\beta_2\left(\frac{180}{\pi}\arctan\left(\frac{h_a}{r_{a,u}}\right)-\beta_1\right)\right)},
\label{eq:LoS}
\end{eqnarray}
\normalsize
where the elevation of $a$ is $h_a$, while $\beta_1$ and $\beta_2$ are parameters depending on the number of large signal obstructions per unit area, building's height distribution, ratio of built-up area and clean surfaces, etc., as derived in~\cite{al2014optimal}, based on ITU recommendations~\cite{recommendations2015propagation}.
In Eq.~\eqref{eq:LoS}, $\theta_{a, u}\!=\!\arctan(h_a / r_{a, u})$ is the elevation angle.
$\theta_{a, u}$ approaches $\frac{\pi}{2}$ when the \aBS~$a$ hovers just above the user~$u$, i.e., when the LoS likelihood reaches its maximum.
The elevation angle $\theta_{a, u}$ is characterized by the \aBS height and the ground distance between the user and the \aBS, that is
$r_{a,u} = \|(X_a, Y_a) - (x_u, y_u)\|$.
Since safety, legislation and technology constraints allow drones to fly only at low altitudes---a few hundreds of meters at most---the ground distance can be of the same order of the drone's elevation, so that the argument of the $\arctan$ function in Eq.~\eqref{eq:LoS} can vary sensibly with user and drone positions, and so can LoS likelihood. This means that drone positions, including their elevation, play an important role on air-to-ground channel conditions.

In particular, the average attenuation (in dB units) of an air-to-ground channel between drone $a$ and user $u$ depends on the LoS likelihood, with the following expression~\cite{al2014optimal}:

\small
\vspace{-3mm}
\begin{eqnarray}
    L_\mathcal{A}(a,u) &=& 20\log_{10}\left(\frac{4\pi f_\mathcal{A}}{c} \cdot \sqrt{h_a^2 + r_{a,u}^2} \right) + \nonumber \\
    &+& \hspace{-2mm}P_{LoS}(a, u) \cdot \left(\xi_{LoS} - \xi_{N\!LoS}\right) + \xi_{N\!LoS},
    \label{eq:PLdrone}
\end{eqnarray}
\normalsize
where $\xi_{LoS}, \xi_{N\!LoS}$ are the {\it excess attenuation} components in LoS/NLoS conditions; $f_{\!\mathcal{A}}$ is the carrier frequency in~Hz; and $c$ is the speed of light in~m/s.

Since \gBSs and \aBSs operate onto orthogonal bands, there is no interference between drone-served users and cellular users, which is commonly the main limiting factor in aided cellular networks, as, e.g., {\it inband} D2D networks \cite{arribas2017multi}.
With the above, the experienced SINR for air-to-ground access links $(a, u)$ is:

\small
\vspace{-4.5mm}
\begin{eqnarray}
  \gamma^\mathcal{A}_{a, u} = \frac{P^a_{Tx} \!\cdot\! 10^{-L_\mathcal{A}(a, u)/10}}{N_{a, u} + I_{a, u}^\mathcal{A}},
\end{eqnarray}
\vspace{-3mm}
\normalsize

\noindent
where $P^a_{Tx}$ is the transmission power of an omnidirectional antenna in the \aBSs $a\in\mathcal{A}$;
$N_{a, u}$ is thermal noise according to the allocated bandwidth; and $I_{a,u}^\mathcal{A}$ is the interference level that user~$u$ suffers from other \aBSs. However, note that the 3D position of an \aBS is a decision parameter that directly affects interfering signals received by user $u$, i.e.:

\small
\vspace{-3.5mm}
\begin{eqnarray}
  I_{a, u}^\mathcal{A} = \sum\limits_{a'\in\mathcal{A}\setminus\{a\}} P^a_{Tx} \!\cdot\! 10^{-L_\mathcal{A}(a', u)/10}, \quad \forall a\in\mathcal{A},
\end{eqnarray}
\vspace{-3mm}
\normalsize

\noindent
where $L_\mathcal{A}(a', u)$ depends on the 3D position of \aBSs $a'\!\in\!\mathcal{A}$, as shown in Eq.~\eqref{eq:PLdrone}.

\vspace{-0.5mm}
\subsection{Ground-to-Ground Channels.}
\label{a:ss:G2G}
Connections in the access network between \gBSs and users experience an attenuation based on a well known path-loss model with slow fading (in dB units):

\vspace{-3mm}
\small
\begin{eqnarray}
    L_\mathcal{G}(g, u) = 10\eta_\mathcal{G}\log_{10}\left(\frac{4\pi f_\mathcal{G}}{c_{l}} \cdot  \|\Pi^g - \pi_u\|\right)  + \mathcal{N}(0, \sigma_\mathcal{G}^2),
  \label{eq:PLeNB}
\end{eqnarray}
\vspace{-3mm}
\normalsize

\noindent
where $\eta_\mathcal{G}>2$ is the path-loss exponent in ground communications; $f_\mathcal{G}$ is the operating carrier frequency of the \gBSs; and $\sigma_{\mathcal{G}}$ is the standard deviation of the gaussian random variable $\mathcal{N}(0, \sigma^2_\mathcal{G})$, modelling the effects of shadowing.

As mentioned above, since there is no interference between cellular users and drone-served users, the SINR for access links $(g, u)$ is:

\small
\vspace{-4.5mm}
\begin{eqnarray}
  \gamma^\mathcal{G}_{g, u} = \frac{P^g_{Tx} \!\cdot\! 10^{-L_\mathcal{G}(g, u)/10}}{N_{g, u} + I_{g, u}^\mathcal{G}},
\end{eqnarray}
\vspace{-2mm}
\normalsize

\noindent
where $P^g_{Tx}$ is the transmission power of an omnidirectional antenna integrated in the \gBSs $g\!\in\!\mathcal{G}$;
$N_{g, u}$ represents thermal noise according to the allocated bandwidth; and most importantly, $I_{g,u}^\mathcal{G}$ is the interference level that user~$u$ suffers from other \gBSs.

\vspace{-0.5mm}
\subsection{Ground-to-Air Channels}
\label{a:ss:G2A}

The aerial network relays traffic from the \gBSs by means of LoS backhaul wireless links. Hence, the attenuation of a \gBS--\aBS link $(g, a)$ is the following:

\vspace{-3.5mm}
\small
\begin{eqnarray}
L_\mathcal{B}(g, a) = 10\eta_\mathcal{B} \log_{10} \left(\frac{4\pi f_\mathcal{B}}{c}\!\cdot\!  \left\|\Pi^g \!-\! \Pi_a\right\|\right) \!+\! \mathcal{N}\left(0, \sigma_\mathcal{B}^2\right),
  \label{eq:PL_G2A}
\end{eqnarray}
\vspace{-2mm}
\normalsize

\noindent
where $\eta_\mathcal{B}\approx 2$ is the path-loss exponent in LoS; $f_\mathcal{B}$ is the operating carrier frequency of the backhaul wireless links; and $\sigma_\mathcal{B}^2$ is the standard deviation of the gaussian random variable $\mathcal{N}\left(0, \sigma_\mathcal{B}^2\right)$, modeling the effects of shadowing.

Backhaul links operate on the bandwidth shared with user access to \gBSs.
However, as backhaul links perform 3D-beamforming pointing to the air (where \aBSs hover), the interference between \gBS-served users and backhaul-served \aBSs is very limited. 
Although the majority of the \gBS radiating power is focused in one direction towards the air thanks to the adoption of 3D-beamforming,
non-ideal beam-patterns also radiate energy in other directions.
Therefore, the SINR experienced by an \aBS $a\!\in\!\mathcal{A}$ depends also on the direction in which other \gBSs transmit to other \aBSs. The SINR experienced by a \gBS--\aBS link $(g, a)$ is:

\small
\vspace{-3.5mm}
\begin{eqnarray}
  \gamma^\mathcal{B}_{g, a} = \frac{P_{Tx}^g\!\cdot\! G_g\!\cdot\!  10^{-L_\mathcal{B}(g, a)/10}}{N^{g,a} + I^\mathcal{B}_{g, a}},
  \label{eq:SINRbackhaul}
\end{eqnarray}
\vspace{-2mm}
\normalsize

\noindent
where $P_{Tx}^g$ is the transmission power of the \gBS $g$; $G_g$ is the antenna gain over the main lobe of the beam-pattern of \gBS $g$; $N^{g,a}$ is the thermal noise; and $I_{g, a}^\mathcal{B}$ is the interference coming from the remaining backhaul links of the network.

Backhaul links reuse the spectrum used for ground cellular connections, although using beam-patterns pointing to the air, while antennas that provide service to ground users are pointing mainly to the ground. Hence, we assume that the interference suffered by a backhaul link $(g, a)$ is dominated by the interference from other backhaul links.
Hence, the interference suffered by a backhaul link $(g, a)$ is:

\small
\vspace{-2mm}
\begin{eqnarray}
I_{g, a}^\mathcal{B} = \sum\limits_{g'\in\mathcal{G}\setminus\{g\}} P_{Tx}^{g'} \!\cdot\!  G_{g'}(\phi_{g', a}) \!\cdot\!  10^{-L_\mathcal{B}(g', a)/10},
  \label{eq:Interf_Backhaul}
\end{eqnarray}
\vspace{-1mm}
\normalsize

\noindent
where $\phi_{g', a}$ is the angle between the main lobe direction of the antenna of $g'$ and the position of \aBS $a$.
In case a \gBS $g'$ does no set any backhaul wireless link, this \gBS will not affect interference, and $P_{Tx}^{g'}$ will be considered as zero.

\renewcommand{\baselinestretch}{1}
\section{Overall Complexity of PADD}
\label{a:Complexity}

Our proposed algorithm consists of sequential steps, some of which involve operations that can be parallelized and can run on a centralized or distributed network orchestrator.
Specifically, at each iteration, i.e., for fixed positions of drones and having identified the least fit drone, we have:
\begin{enumerate}[label={\bfseries Step~\arabic*},, wide = 0.5em, leftmargin = 0em,  rightmargin = 0em]
\setlength\itemsep{0.1cm}
\item {\it User association}.
\label{i:association}
This can be implemented on parallel threads: one thread per UE 
ranks the candidate list of BSs to attach to, then a separate thread computes the association in at most as many rounds as the number of BSs, as seen in Section~\ref{ss:userassociation}.
With $| \mathcal{U} |$ parallel threads, the complexity required for this step to rank the list of BSs is $\mathcal{O}\!\left(\frac{1}{2\log\!2}|\mathcal{B}|\log|\mathcal{B}|\right)$ sums and comparisons. 
\item {\it Backhaul association}.
\label{i:backhaul}
The next step consists in solving the {\tt GAP-knap} problem for backhaul association, which must be done with a single thread, as shown in Section~\ref{ss:backhaulassociation}. 
The complexity of combining dynamic programming and the knapsack problem to solve this step is analyzed in~\cite{fisher1986multiplier} and it is $\mathcal{O}\left(2A_g |\mathcal{G}| |\mathcal{A}| \right)$ sums or comparisons.
\item {\it Resource allocation}.
\label{i:resource_alloc}
This requires a thread per each \gBS for which we need to solve the convex problem 
of Fig.~\ref{fig:CP}, as shown in Section~\ref{ss:bandwidthallocation}.
The time needed to complete this step is therefore the time needed to solve a single problem. 
The complexity of the costliest case is $\mathcal{O}\left(4A_gU_{\max}\right)$ simple operations (e.g., sums), as discussed in Section~\ref{a:OptimalCP}.
\item {\it Utility evaluation}.
\label{i:utility}
The current configuration is evaluated in terms of $\alpha$-fair utility, which has the cost of $| \mathcal{U} | | \mathcal{B} |$ sums or comparisons (for $\alpha\!\rightarrow\!\infty$) plus $| \mathcal{B} |$ power operations (for $\alpha \neq 0,1$) or  logarithms (for $\alpha = 1$).
As discussed at the beginning of this section and depicted in Fig.~\ref{fig:eoa}, the current configuration can be discarded and a new position is probed
for the current least fit drone (back to \ref{i:association}).%
\footnote{When the optimization problem is initialized, there is no least fit drone yet; thus, the first iteration executes \ref{i:association}-\ref{i:utility} only once, then it moves to \ref{i:least} to make the first least fit selection.}
\item {\it Least fit selection}.
\label{i:least}
Eventually, the algorithm computes the least fit drone with a single thread, with complexity $\mathcal{O}(4|\mathcal{A}| U_{\max})$ operations, as two kinds of additional different utilities need to be computed in Eq.~\eqref{eq:leastfit} of Section~\ref{ss:leastfit}.
%
\end{enumerate}
With discretized drone positions, and indicating $N$ the number of points that can be probed 
at each iteration for each drone, the complexity of one iteration is $N$ times the complexity of \ref{i:association}-\ref{i:utility}, %
%
%
plus the complexity of \ref{i:least}.
The overall complexity is therefore 
$\mathcal{O} \left( N \left(\frac{1}{2\log\!2}|\mathcal{B}| \log |\mathcal{B}| + 2A_g |\mathcal{G}| |\mathcal{A}| + 4A_g  U_{\max} + | \mathcal{U} | | \mathcal{B} |\right) \right) $ sums or comparisons and $\mathcal{O} \left(N | \mathcal{B}|  + 4|\mathcal{A}| U_{\max}\right)$ powers (or logarithms) in the worst case.
This complexity is 
low-degree polynomial, with small constant factors, and linear in most of the parameters.
The number of iterations required by EO is not bounded unless a hard limit is 
imposed, however, the approach is designed to quickly approach a local optimum.
\\
\indent
In our experiments 
we have observed no more than a hundred of iterations, without imposing any limit to the number of iterations.
The number of possible testing positions for the least fit drone is $N=400$ but we have observed numbers below $5$.
Indeed, on average, the initial position can be improved with probability 0.5, which means that, if we assume that positions are selected at random with no memory, the iteration stops after testing, on average, 2 positions in the ball around the least fit drone (it would be a Bernoulli process). If we consider memory, 
the Bernoulli approximation upper bounds the average number of probing attempts, because 
the probability that the next position will be better than the current one, and the iteration will hence stop, will grow attempt after attempt.
\\
\indent
Besides, our algorithm implemented on \MATLAB and running on a linux machine with an Intel Xeon E5-2670 processor at $2.6$~GHz took
10~s in the worst case (i.e., it required about 800 floating-point operations, at 32 flops/cycle involving 64-bit operations), consuming only $1$~GB of RAM, to optimize drone positions.

\section{Robustness of PADD}
\label{app:robustness}


Fig.~\ref{fig:Rob_CDF} depicts the CDF of the relative loss due to erroneous user position estimation, i.e., the relative loss of utility due to optimizing drone positions according to erroneous user positions. The figure also shows the error in terms of throughput and fairness separately.
The loss is below $10\%$ in all cases, and below $3\%$ for utility and throughput in more than half of the cases, while the average loss is below $5\%$.

\begin{figure}[H]
\hspace{-0mm}
\vspace{3.5mm}
\begin{minipage}[H]{0.99\linewidth}
\centering
\vspace{-4.5mm}
\hspace{-5mm}
\includegraphics[width=7.55cm]{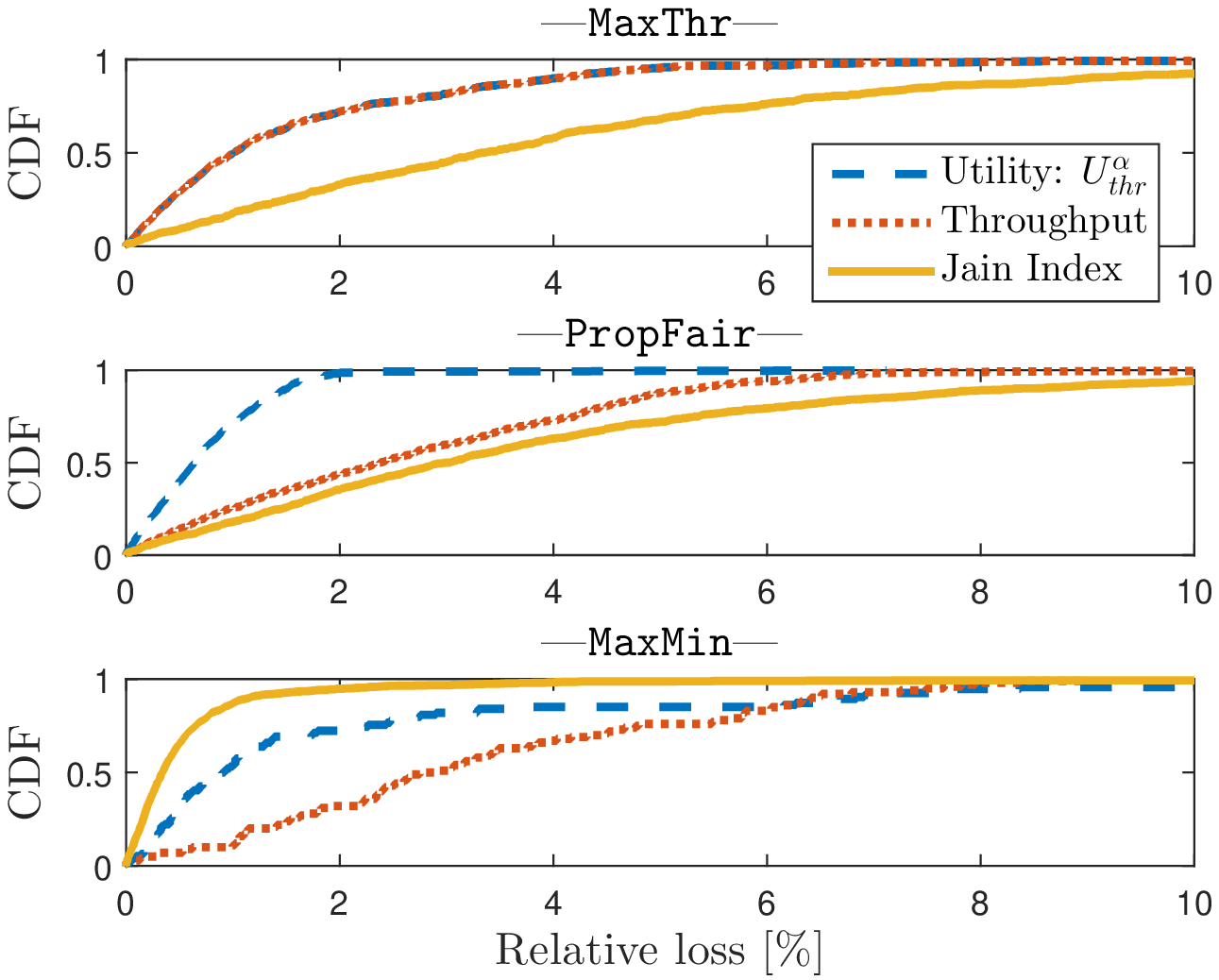}
    \vspace{-1mm}
    \caption{Robustness validation for $\alpha\in\{0, 1, \infty\}$. CDF of the relative loss. $G = 10$, $A = 5$, $U = 1000$. Scenario: {\sl PPP}.}
\label{fig:Rob_CDF}
\end{minipage}
\vspace{-2mm}
\end{figure}

\renewcommand{\baselinestretch}{1}
\section{Optimal Solution for Bandwidth Allocation}
\label{a:OptimalCP}

Bandwidth allocation is done by solving the convex optimization program presented in Fig.~\ref{fig:CP}.
We apply KKT conditions~\cite{kuhn2014nonlinear} so to find solutions without having to resort to a solver.
To simplify the notation, we consider that, given a set of generic entities $\mathcal{E}$, there is a bijective function $\sigma\!:\!\{1,\hdots,|\mathcal{E}|\}\!\longrightarrow\! \mathcal{E}$ that maps entities onto integer numbers. With that, we can denote, e.g., the throughput of a user $u$ indistinctly as $T_u$ (with $u$ being a user) or $T_i$ (with $i$ being an integer number), where $\sigma(i) = u$.

We next show how to solve the bandwidth allocation problem by considering two separate cases: with and without drones.

\subsection{Without Drones}
\label{ss:BHfree}

In this scenario, there are no drone. CP
of Fig.~\ref{fig:CP}
simplifies considerably since \gBS~$g$ only needs to manage resources to be split among \gBS-served users, i.e., the set $\mathcal{U}_g$. Hence, we solve the following CP:

\small
\begin{equation}
\hspace{-3.6cm}
\eag{
  \begin{cases}
    \qquad\qquad\qquad\qquad\qquad\mathclap{\max\limits_{\sepia{w_u,} \redd{T_u}} U_{\mathrm{cvx}, g}^\alpha =  \begin{cases}\sum\limits_{u\in\mathcal{U}_g} \left( T_u \right)^{1-\alpha} \cdot  \frac{1}{1-\alpha}, & \alpha \neq 1; \\
    \sum\limits_{u\in\mathcal{U}_g} \log\left( T_u \right), &  \alpha = 1;
    \end{cases}}\\
    \text{s.t.:} \\
    \sepia{w_u \geq W^{\min}_\mathcal{G},} & \sepia{\!\!\forall  u\!\in\!\mathcal{U}_g;} \\
    \sepia{\sum\limits_{u\in\mathcal{U}_g} \! w_u = W_\mathcal{G};} & \!\!\\
    \redd{T_u \!\leq\!  w_u \log_2\!\left(1 \!+\! \gamma_{b, u}\right)\!,} & \redd{\!\!\forall u\!\in\!\mathcal{U}_g;} \\
    \blue{\sum\limits_{u\in\mathcal{U}_g} \! T_u \! \leq  \tau_g.} &\!\!
  \end{cases}}
  \label{eq:BHfree}
\end{equation}
\normalsize

Denote as $x=[\{w_u\}_{u\in\mathcal{U}_g}, \{T_u\}_{u\in\mathcal{U}_g}]\in\mathds{R}^{2|\mathcal{U}_g|}$ the vector of variables. Then, depending on the value of $\alpha$, we define the KKT functions as follows:

\vspace{-2mm}
\small
\begin{eqnarray}
  \!\!\!\!\!\!\!\!f^\alpha(x) =& \!\!\!\!\!\!\!\!\!\!\!\!\!\!\!\!\!\!\!\!\!\!\!\!\!\!\!\!\!\!\!\! -\sum\limits_{u\in\mathcal{U}_g} T_u^{1-\alpha} & \text{if } \alpha \neq 1; \label{eq:KKTfunc1} \\
  \!\!\!\!\!\!\!\!f^{\log}(x) =& \!\!\!\!\!\!\!\!\!\!\!\!\!\!\!\!\!\!\!\!\!\!\!\!\!\!\!\!\!\!\!\! -\sum\limits_{u\in\mathcal{U}_g}\log T_u & \text{if } \alpha = 1; \label{eq:KKTfunc2} \\
  \!\!\!\!\!\!\!\!g_u(x)  =& \!\!\!\!\!\!\!\!\!\!\!\!\!\!\!\!\!\!\!\!\!\!\!\!\!\!\!\!\!\!\! W_\mathcal{G}^{\min} - w_u, & \forall  u\in\mathcal{U}_g; \label{eq:KKTfunc3} \\
  \!\!\!\!\!\!\!\!g_{|\mathcal{U}_g|+u}(x)  =& T_u - w_u \log_2\!\left(1\!+\!\gamma_{g,u}\right), & \forall u\in\mathcal{U}_g;\label{eq:KKTfunc4} \\
  \!\!\!\!\!\!\!\!g_{2|\mathcal{U}_g|+1}(x) =& \!\!\!\!\!\!\!\!\!\!\!\!\!\!\!\!\!\!\!\!\!\!\!\!\!\!\!\!\!\!\!\! \sum\limits_{u\in\mathcal{U}_g}T_u-\tau_g; & \label{eq:KKTfunc5} \\
  \!\!\!\!\!\!\!\!h(x) =& \!\!\!\!\!\!\!\!\!\!\!\!\!\!\!\!\!\!\!\!\!\!\!\!\!\!\!\!\!\!\!\! \sum\limits_{u\in\mathcal{U}_g} w_u - W_\mathcal{G}. &\label{eq:KKTfunc6}
\end{eqnarray}
\normalsize

The corresponding KKT gradients are as follows:

\vspace{-2mm}
\small
\begin{eqnarray}
  \!\!\nabla\! f^\alpha(\!x\!) =& \hspace{-3.2cm} \left[0,\hdots ,\left\{(\alpha\!-\!1)T_u^{-\alpha}\right\}_{u\in\mathcal{U}_g}\right]\!; & \label{eq:KKTgrad1} \\
  \!\!\nabla\! f^{\log}(\!x\!) =& \hspace{-3.9cm} \left[0, \hdots , 0, \left\{\frac{-1}{T_u}\right\}_{u\in\mathcal{U}_g}\right]\!; & \label{eq:KKTgrad2} \\
  \!\!\nabla\! g_u(\!x\!) =& \hspace{-3.6cm} \left[0,\hdots,0, -1|_u,0,\hdots,0\right]\!,  & \nonumber \\
  \forall u\!\in\!\mathcal{U}_g;  \label{eq:KKTgrad3} && \\
  \!\!\nabla\! g_{|\mathcal{U}_g|\!+\!u}(\!x\!)  =& \hspace{-0.3cm} \left[0,...,0,-\!\log_2\!(\!1\!\!+\!\!\gamma_{g, u})|_u,0,...,0,1|_{|\mathcal{U}_g|\!+\!u},0,...,0\right]\!, &\nonumber \\
    \forall  u\!\in\!\mathcal{U}_g; \label{eq:KKTgrad4} && \\
  \!\!\nabla\! g_{2|\mathcal{U}_g|\!+\!1}(\!x\!) =& \hspace{-3.3cm} \left[0, \hdots, 0, 1|_{|\mathcal{U}_g|+1}, 1, \hdots, 1\right]\!; & \label{eq:KKTgrad5} \\
  \!\!\nabla\! h(\!x\!) =& \hspace{-3.9cm} \left[1, \hdots, 1|_{|\mathcal{U}_g|}, 0, \hdots, 0\right]\!. & \label{eq:KKTgrad6}
\end{eqnarray}
\normalsize

Now, the KKT conditions state that if we find a vector $x^{*}\in\mathds{R}^{2|\mathcal{U}_g|}$ that is feasible (i.e., it satisfy the original constraints of the problem), and multipliers $\mu_i\geq0$ $\forall 1\leq i\leq 2|\mathcal{U}_g|+1$ and $\nu$, and not all of them null, then the following equations hold and $x^{*}$ is the optimal solution:
%

\vspace{-2mm}
\small
\begin{eqnarray}
  \overrightarrow{0} \hspace{-3mm}&=&\hspace{-3mm} \nabla f(x^{*}) + \sum\limits_{i=1}^{2|\mathcal{U}_g|+1} \mu_i \nabla g_i(x^{*}) + \nu\nabla h(x^{*}); \label{eq:KKTcondition1} \\
  \mu_i g_i(x^{*}) \hspace{-3mm}&=&\hspace{-3mm} 0, \qquad \forall  1\leq i\leq 2|\mathcal{U}_g|+1; \label{eq:KKTcondition2} \\
  h(x^{*}) \hspace{-3mm}&=&\hspace{-3mm} 0, \label{eq:KKTcondition3}
\end{eqnarray}
\normalsize
\vspace{-1mm}

\noindent
where $f\in\{\{f^\alpha\}_{\alpha\in[0, 1[}, f^{\log}\}$.

Hence, in what follows we describe how we have found the optimal solution for each particular case. We analyze separately the \texttt{MaxThr} case (when $\alpha=0$), the \texttt{$\alpha$Fair} case (when $\alpha\in]0, 1[$), the \texttt{PropFair} case (when $\alpha = 1$) and the \texttt{MaxMin} case (when $\alpha\rightarrow \infty$).

\subsubsection{\texttt{$\alpha$Fair} optimum ($\alpha \in]0, 1[$)}

In this case, $f(x) = f^{\alpha}(x) = -\!\!\sum\limits_{u\in\mathcal{U}_g}\!T_u^{1-\alpha}$, and the KKT conditions are as follows:

\vspace{-4mm}
\small
\begin{eqnarray}
  -\mu_i - \log_2\!\left(1\!+\!\gamma_{g, i}\right)\!\cdot\! \mu_{|\mathcal{U}_g|+i} \!+\! \nu = 0, & \forall  1\leq  i \leq  |\mathcal{U}_g|; \label{eq:KKTalphaFair1}\\
  \mu_{|\mathcal{U}_g|+i} + \mu_{2|\mathcal{U}_g|+1} - (1-\alpha)T_i^{-\alpha} = 0, & \forall  1\leq  i \leq  |\mathcal{U}_g|; \label{eq:KKTalphaFair2}\\
  \mu_i \cdot  \left(W_\mathcal{G}^{\min} - w_i\right) = 0, & \forall  1\leq  i \leq  |\mathcal{U}_g|; \label{eq:KKTalphaFair3}\\
  \mu_{|\mathcal{U}_g|+i} \!\cdot\!  \left(T_i - w_i  \log_2\!\left(1\!+\!\gamma_{g, i}\right)\right) = 0, &  \forall  1\leq  i \leq  |\mathcal{U}_g|; \label{eq:KKTalphaFair4}\\
  \mu_{2|\mathcal{U}_g|+1} \cdot  \Big( \sum\limits_{u\in\mathcal{U}_g} T_u - \tau_g\Big) = 0;  & \label{eq:KKTalphaFair5}\\
  \sum\limits_{u\in\mathcal{U}_g} w_u - W_\mathcal{G} = 0. & \label{eq:KKTalphaFair6}
\end{eqnarray}
\normalsize

\textbf{With unbounded backbone capacity $\pmb{\tau_g}$.}
Assuming that $\tau_g \rightarrow + \infty$, then $\sum\limits_{u\in\mathcal{U}_g} T_u < \tau_g$ for all feasible $x$. Hence, according to Eq.~\eqref{eq:KKTalphaFair5}, $\mu_{2|\mathcal{U}_g|+1} = 0$. Therefore, for $1\leq i\leq |\mathcal{U}_g|$, Eq.~\eqref{eq:KKTalphaFair2} becomes $\mu_{|\mathcal{U}_g|+i} = (1-\alpha)T_{i}^{-\alpha}$, and
Eq.~\eqref{eq:KKTalphaFair1} becomes $\nu=\mu_i+(1-\alpha)T_i^{-\alpha}\cdot \log_2\left(1+\gamma_{g,i}\right)$. As $\nu$ is a constant, we search for expressions of $\mu_i$ and $T_i$ that make $\nu$ independent of index $i$.
For convenience, we define set $\mathcal{J}$ as those indices $i$ for which $w_i$ is the minimum guaranteed value, i.e.:
%
\vspace{-4mm}
\begin{eqnarray}
  \mathcal{J} = \left\{i \suchthat w_i = W_\mathcal{G}^{\min}\right\}.
\end{eqnarray}
\vspace{-4mm}
%

We start by assuming that $\mu_i=0$, $\forall i\notin\mathcal{J}$, so that~\eqref{eq:KKTalphaFair3} is satisfied $\forall i\notin\mathcal{J}$. Given the SINR values, this assumption makes $T_i$ become a function of $\nu$ only, $\forall i\notin\mathcal{J}$, i.e., $T_i = \left(\frac{1-\alpha}{\nu} \log_2\left(1+\gamma_{g,i}\right)\right)^{\frac{1}{\alpha}}$, $\forall i\notin\mathcal{J}$.
In addition, as $\tau_g$ is unbounded, we can assume that 
$T_i$ must take the highest possible value, which is the Shannon capacity $w_i \log_2\left(1+\gamma_{g,i}\right)$, $\forall 1\leq i\leq |\mathcal{U}_g|$.
Hence, $w_i = \left(\frac{1-\alpha}{\nu}\right)^{\frac{1}{\alpha}} \left(\log_2\left(1+\gamma_{g,i}\right)\right)^{\frac{1-\alpha}{\alpha}}$ for all cases in which $w_i > W_\mathcal{G}^{\min}$ (i.e., $\forall i\notin\mathcal{J}$) and $w_i = W_\mathcal{G}^{\min}, \forall i \in \mathcal{J}$.
Replacing the above expression for $w_i$ in Eq.~\eqref{eq:KKTalphaFair6}, we obtain an equation in which $\nu$ is the only unknown to be derived.
We therefore have obtained the following solution:
\begin{algorithm}[t]
  \caption{Derivation of set $\mathcal{J}$ in \texttt{$\alpha$Fair} (unbounded $\tau_g$).}
  \label{alg:SetJforAlphaFair}
  \begin{algorithmic}[1]
    \REQUIRE{$\log_2\left(1+\gamma_{g,i}\right)$, $\forall 1\leq i\leq |\mathcal{U}_g|$.}
    \STATE{Initialize: $\mathcal{J} \leftarrow \emptyset$; $w_i \leftarrow \frac{W_\mathcal{G} - |\mathcal{J}|\cdot W_\mathcal{G}^{\min}}{\sum\limits_{j\notin\mathcal{J}}\left(\frac{\log_2\left(1+\gamma_{g,j}\right)}{\log_2\left(1+\gamma_{g,i}\right)}\right)^{\frac{1-\alpha}{\alpha}}}$, $\forall i\notin\mathcal{J}$.}
    \WHILE{$\exists i_1\notin\mathcal{J}\suchthat w_{i_1} < W_\mathcal{G}^{\min}$}
    \STATE $i_0\leftarrow\arg\max_{i\notin\mathcal{J}} \log_2\left(1+\gamma_{g,i}\right)$.
    \STATE $\mathcal{J} \leftarrow \mathcal{J} \cup \{i_0\}$.
    \STATE $w_i \leftarrow \frac{W_\mathcal{G} - |\mathcal{J}|\cdot W_\mathcal{G}^{\min}}{\sum\limits_{j\notin\mathcal{J}}\left(\frac{\log_2\left(1+\gamma_{g,j}\right)}{\log_2\left(1+\gamma_{g,i}\right)}\right)^{\frac{1-\alpha}{\alpha}}}$, $\forall i\notin\mathcal{J}$.
    \ENDWHILE
  \end{algorithmic}
\end{algorithm}
\vspace{-4mm}
\small
\begin{align}
  &w_i = W_\mathcal{G}^{\min}, & \forall i\in\mathcal{J}; \label{eq:BHfreewialphaFair}\\
  &w_i = \frac{W_\mathcal{G} - |\mathcal{J}|\cdot W_\mathcal{G}^{\min}}{\sum\limits_{j\notin\mathcal{J}}\!\!\left(\frac{\log_2\!\left(1\!+\!\gamma_{g,j}\right)}{\log_2\!\left(1\!+\!\gamma_{g,i}\right)}\right)^{\frac{1\!-\!\alpha}{\alpha}}}\!, & \forall i\notin\mathcal{J}; \label{eq:BHfreewialphaFair2}\\
  &T_i = w_i\log_2(1\!+\!\gamma_{g,i}) & \forall 1\leq  i \leq |\mathcal{U}_g|; \label{eq:BHfreeTialphaFair}\\
  &\mu_i = \nu-\log_2\left(1+\gamma_{g,i}\right)(1-\alpha)T_i^{-\alpha}, &\forall  i\in\mathcal{J}; \label{eq:BHfreemuiJalphaFair}\\
  &\mu_i = 0, &\forall  i\notin\mathcal{J}; \\
  &\mu_{|\mathcal{U}_g|+i}= (1\!-\!\alpha)\,T_i^{-\alpha}, & \forall 1\leq  i \leq |\mathcal{U}_g|; \\
  &\mu_{2|\mathcal{U}_g|+1} = 0; & \\
  &\nu = \hspace{2.5cm}\mathclap{ (1\!-\!\alpha)\!\!\left(\frac{\sum\limits_{j\notin \mathcal{J}}  \left(\log_2\left(1+\gamma_{g,j}\right)\right)^{\frac{1-\alpha}\alpha}} {W_\mathcal{G}-|\mathcal{J}|\cdot W_\mathcal{G}^{\min}}\!\right)^{\!\!\!\alpha}\!\!\!. } \label{eq:BHfreeNualphaFair}
\end{align}
\normalsize
\\
\indent
The solution we have built accomplishes all KKT equations and works also for the case $\sum\limits_{u\in\mathcal{U}_g} T_u  = \tau_g$. In particular, note in Eq.~\eqref{eq:BHfreemuiJalphaFair} that $\mu_i \geq 0$, $\forall i\in\mathcal{J}$, which can be seen by replacing the expressions for $T_i$, $w_i$ and $\nu$ in
the equation and consider that,   $\forall i\in\mathcal{J}$,
$W_\mathcal{G}^{\min} \ge \frac{W_\mathcal{G} - |\mathcal{J}|\cdot W_\mathcal{G}^{\min}}{\sum\limits_{j\notin\mathcal{J}}\!\!\left(\frac{\log_2\!\left(1\!+\!\gamma_{g,j}\right)}{\log_2\!\left(1\!+\!\gamma_{g,i}\right)}\right)^{\frac{1\!-\!\alpha}{\alpha}}}$.

A valid set $\mathcal{J}$ always exists and it is built as shown in Algorithm~\ref{alg:SetJforAlphaFair}.
The algorithm starts with an empty set and, while it is not true that $w_i\geq W_\mathcal{G}^{\min}$, $\forall i\notin\mathcal{J}$,
keeps adding to $\mathcal{J}$ one user at a time, the one with the highest SINR, which is the user that consumes less resources.
The algorithm stops for sure, eventually after moving all users, with a set $\mathcal{J}$ for which KKT and feasibility conditions are satisfied,
therefore the built solution is optimal.


\textbf{With limited backbone capacity $\pmb{\tau_g}$.}
In order to find the optimal solution in the general case, we start from the solution for  $\tau_g \rightarrow + \infty$.
In case this procedure yields $\sum\limits_{u\in\mathcal{U}_g} T_u \le  \tau_g$, we have the optimal solution described above.
Otherwise, we have that $\sum\limits_{u\in\mathcal{U}_g} T_u  > \tau_g$, which is unfeasible. However, we can build a feasible solution by decreasing some values of $T_i$, motivated by the consideration that the utility function is strictly concave, so that the least utility reduction is obtained by reducing the highest throughput value, as shown in Algorithm~\ref{alg:CPxi} (for $n=|\mathcal{U}_g|$, $\{x_i\}_{i=1}^n = \{T_u\}_{u\in\mathcal{U}_g}$ and $X=\tau_g$).

Algorithm~\ref{alg:CPxi} is iterative, and it is based on the following principle: Because of concavity, if only one user has maximal throughput $T^{\prime} \!=\! \max_i \{T_i\}$, and second best throughput is $T^{\prime\prime}$, any throughput reduction from $T^{\prime}$ to $T^{\prime}-y \ge T^{\prime\prime}$ maximizes utility for a reduction of $y$ in $\sum_i T_i$;
if we have two users with maximal throughput, and we have to reduce the sum of throughputs by $y$, then strict concavity assures that $2 \left(T^{\prime}-y/2 \right)^{1-\alpha} > \left(T^{\prime}-(y/2-\epsilon)\right)^{1-\alpha}+\left(T^{\prime}-(y/2+\epsilon)\right)^{1-\alpha}, \forall \epsilon \in ]0; y/2], y/2 \in ]0,T^{\prime}-T^{\prime\prime}]$, so the best aggregate utility is obtained by decreasing the two highest throughput values both to $T^{\prime}-y/2$. As it is easy to see, in case of tie between $n$ users, the best choice consists in reducing their throughputs to $T^{\prime}-y/n, \forall y/n \in ]0, T^{\prime}-T^{\prime\prime}]$.

\subsubsection{\texttt{MaxThr} optimum ($\alpha = 0$)}
In this case, $f(x) = f^0(x) = -\!\!\sum\limits_{u\in\mathcal{U}_g}T_u$, and the  KKT conditions are like for the case $\alpha \in ]0,1[$, except $\alpha=0$.
%

\begin{algorithm}[t]
  \caption{Optimum for strictly concave increasing utility}
  \label{alg:CPxi}
  \begin{algorithmic}[1]
    \REQUIRE{$\{x_i\}_{i=1}^n$, $X$.}
    \WHILE{$X-\sum\limits_{i=1}^n x_i < 0$}
    \STATE $X_{M} \leftarrow \{x_i\suchthat x_i=\max_{1\leq j\leq n}x_j\}$.
    \STATE $x_{M_2}\leftarrow \arg\max\{x_i\suchthat x_i\notin X_M\}$ (assume $|x_{M_2}|=1$).
    \STATE $R\leftarrow -X+\sum\limits_{i=1}^n x_i > 0$.
    \STATE $ R_j \leftarrow \frac{R}{|X_M|}$, $\forall  1\leq j\leq n \text{ with } x_j\in X_M$.
    \STATE $x_j \leftarrow x_j \!-\! \min\!\left\{R_j, x_j\!-\!x_{M_2}\right\}$, $\!\forall 1\!\leq\! j\!\leq\! n \text{ with } x_j\!\in\! X_M$.
    \ENDWHILE
  \end{algorithmic}
\end{algorithm}

\textbf{With unbounded backbone capacity $\pmb{\tau_g}$.}
First, we look for a solution assuming that $\tau_g$ is unbounded, i.e., assuming that $\sum\limits_{u\in\mathcal{U}_g} T_u <  \tau_g$ for all feasible $x$. Hence, according to Eq.~\eqref{eq:KKTalphaFair5}, $\mu_{2|\mathcal{U}_g|+1} = 0$. Also, according to Eq.~\eqref{eq:KKTalphaFair2}, $\mu_{|\mathcal{U}_g|+i} = 1$, $\forall  1\leq i \leq |\mathcal{U}_g|$, and according to Eq.~\eqref{eq:KKTalphaFair4}, $T_i=w_i \log_2\left(1+\gamma_{g,i}\right)$, $\forall 1\leq i\leq |\mathcal{U}_g|$. Moreover, according to Eq.~\eqref{eq:KKTalphaFair1}, $\nu-\mu_i=\log_2\left(1+\gamma_{g, i}\right)$, $\forall 1\leq i\leq  |\mathcal{U}_g|$.

Now, let $i_0=\arg\max_{1\leq i\leq |\mathcal{U}_g|} \log_2\left(1+\gamma_{g,i}\right)$ and let $w_i=W_\mathcal{G}^{\min}$, $\forall 1\leq i\neq i_0\leq |\mathcal{U}_g|$. Hence, according to Eq.~\eqref{eq:KKTalphaFair6}, $$w_{i_0} = W_\mathcal{G} - \sum\limits_{\substack{i=1 \\i\neq i_0 }}^{|\mathcal{U}_g|} w_i = W_\mathcal{G} - \sum\limits_{\substack{i=1 \\i\neq i_0 }}^{|\mathcal{U}_g|}W_\mathcal{G}^{\min} = W_\mathcal{G}-(|\mathcal{U}_g|-1)W_{\mathcal{G}}^{\min}.$$

Now, let $\mu_{i_0} = 0$. Hence, according to Eq.~\eqref{eq:KKTalphaFair1}, $\nu = \log_2\left(1+\gamma_{g, i_0}\right)$ and $\mu_i=\log_2\left(1+\gamma_{g, i_0}\right)-\log_2\left(1+\gamma_{g,i}\right)$, $\forall 1\leq i\leq  |\mathcal{U}_g|$ (note that $\mu_i\geq 0$ $\forall 1\leq  i \leq |\mathcal{U}_g|$ by definition of $i_0$).

Since the found solution $x$ is feasible and accomplishes all KKT conditions, this is necessary the optimal solution.

\textbf{With limited backbone capacity $\pmb{\tau_g}$.}
We can observe that the optimal solution when $\sum\limits_{u\in\mathcal{U}_g} T_u <  \tau_g$ corresponds to assigning the maximum possible amount of resources to the users with highest SINR.
If we now consider a bounded value for $\tau_g$, we can build the solution following the same scheme to build Algorithm~\ref{alg:BHfreeMaxThr}, which finds the optimal solution, as explained in what follows.

Let $i_0=\arg\max_{1\leq i\leq |\mathcal{U}_g|}\log_2\left(1+\gamma_{g,i}\right)$. Let $w_i=W_\mathcal{G}^{\min}$, $\forall 1\leq i\neq i_0\leq |\mathcal{U}_g|$ and $w_{i_0}=W_\mathcal{G}-\sum\limits_{\substack{i=1 \\i\neq i_0 }}W_\mathcal{G}^{\min} = W_\mathcal{G}-(|\mathcal{U}_g|-1)W_\mathcal{G}^{\min}$.
In the algorithm, we initially assign $T_{i_0}\!=\!\min\left(\tau_g, w_{i_0} \log_2\!\left(1\!+\!\gamma_{g, i_0}\right)\right)$, $T_i=0$, $\forall 1\!\leq\! i\!\neq\! i_0 \!\leq\! |\mathcal{U}_g|$, and define the set $\overline{\mathcal{U}_g}=\mathcal{U}_g\setminus\{i_0\}$ (step~\ref{alg:BHfreeMaxThrStep1}).
Next, while $\sum\limits_{i=1}^{|\mathcal{U}_g|} T_i < \tau_g$ and $\overline{\mathcal{U}_g}\neq \emptyset$, we do the following three assignments: (i)~$j_0=\arg\max_{j\in\overline{\mathcal{U}_g}}\log_2\left(1+\gamma_{g,j}\right)$ (step~\ref{alg:BHfreeMaxThrStep3}), (ii)~$T_{j_0} = \min\left(\tau_g-\sum\limits_{i\in\mathcal{U}_g} T_i, w_{j_0} \log_2\left(1+\gamma_{g, j_0}\right)\right)$ (step~\ref{alg:BHfreeMaxThrStep4}), and (iii)~$\overline{\mathcal{U}_g}=\overline{\mathcal{U}_g}\setminus\{j_0\}$ (step~\ref{alg:BHfreeMaxThrStep5}). This procedure finds values of $\{T_i\}_{i=1}^{|\mathcal{U}_g|}$ such that $\sum\limits_{i=1}^{|\mathcal{U}_g|} T_i \leq  \tau_g$. Besides, if $\overline{\mathcal{U}_g}\neq \emptyset$ when the algorithm concludes, we have that $\sum\limits_{i=1}^{|\mathcal{U}_g|} T_i =  \tau_g$.

Note that in case $\sum\limits_{i=1}^{|\mathcal{U}_g|} T_i <  \tau_g$, the solution of Algorithm~\ref{alg:BHfreeMaxThr}
corresponds with the solution for unbounded $\tau_g$.
Moreover, in case the output of the algorithm is such that $\sum\limits_{i=1}^{|\mathcal{U}_g|} T_i =  \tau_g$, the solution is also optimal because, with $\alpha=0$, $\sum\limits_{i=1}^{|\mathcal{U}_g|} T_i$ is the utility function, and $\tau_g$ is the maximum that that sum can achieve, according to the last constraint in~\eqref{eq:BHfree}. Therefore, the solution computed with Algorithm~\ref{alg:BHfreeMaxThr} is always optimal for $\alpha=0$.

\begin{algorithm}[t]
  \caption{Optimum \texttt{MaxThr} in Backhaul-free scenario}
  \label{alg:BHfreeMaxThr}
  \begin{algorithmic}[1]
    %
    \STATE Initialize: $T_{i_0}=\min\left(\tau_g,\; w_{i_0} \log_2\left(1+\gamma_{g, i_0}\right)\right)$, $T_i=0$, $\forall 1\leq i\neq i_0 \leq |\mathcal{U}_g|$, $\overline{\mathcal{U}_g}=\mathcal{U}_g\setminus\{i_0\}$. \label{alg:BHfreeMaxThrStep1}
    \WHILE{$\sum\limits_{i=1}^{|\mathcal{U}_g|} T_i < \tau_g$ \AND $\overline{\mathcal{U}_g}\neq \emptyset$} \label{alg:BHfreeMaxThrStep2}
    \STATE $j_0=\arg\max_{j\in\overline{\mathcal{U}_g}}\log_2\left(1+\gamma_{g,j}\right)$. \label{alg:BHfreeMaxThrStep3}
    \STATE $T_{j_0} = \min\left(\tau_g-\sum\limits_{i\in\mathcal{U}_g} T_i, w_{j_0} \log_2\left(1+\gamma_{g, j_0}\right)\right)$. \label{alg:BHfreeMaxThrStep4}
    \STATE $\overline{\mathcal{U}_g}=\overline{\mathcal{U}_g}\setminus\{j_0\}$. \label{alg:BHfreeMaxThrStep5}
    \ENDWHILE \label{alg:BHfreeMaxThrStep6}
  \end{algorithmic}
\end{algorithm}

\subsubsection{\texttt{PropFair} optimum ($\alpha = 1$)}
This case aims to maximize the network proportional fairness metric, $f(x) = f^{\log}(x) = -\sum\limits_{u\in\mathcal{U}_g}\log\left(T_u\right)$. The KKT conditions for this case are like for the case $\alpha \in ]0,1[$, except condition~\eqref{eq:KKTalphaFair2} is replaced by the following one:
\begin{eqnarray}
  \mu_{|\mathcal{U}_g|+i} + \mu_{2|\mathcal{U}_g|+1} - \frac{1}{T_i} = 0, & \forall  1\leq  i \leq  |\mathcal{U}_g|; \label{eq:KKTPropFair2}
\end{eqnarray}

\textbf{With unbounded backbone capacity $\pmb{\tau_g}$.}
If $\tau_g$ is unbounded, Eq.~\eqref{eq:KKTalphaFair5} requires that $\mu_{2|\mathcal{U}_g|+1} = 0$. Hence, according to Eq.~\eqref{eq:KKTPropFair2}, $\mu_{|\mathcal{U}_g|+i} = \frac{1}{T_i}$, $\forall 1\leq i\leq |\mathcal{U}_g|$.

Now, $\forall 1\leq i\leq |\mathcal{U}_g|$, let $\mu_i=0$ as in the previous cases, and $w_i=\frac{W_\mathcal{G}}{|\mathcal{U}_g|}$, $T_i=\frac{W_\mathcal{G}}{|\mathcal{U}_g|} \log_2\left(1+\gamma_{g,i}\right)$.
Hence, according to Eq.~\eqref{eq:KKTalphaFair1}, $\nu=\frac{|\mathcal{U}_g|}{W_\mathcal{G}}$.

Since the found feasible solution $x$ accomplishes all KKT conditions, this is the optimal solution. Note that, as expected,  in a \texttt{PropFair} without backbone constraints, all users would obtain the same amount of resources.

\textbf{With limited backbone capacity $\pmb{\tau_g}$.}
To find the optimal solution for the generic case, in which $\tau_g$ is limited, we start from the solution for unbounded backbone capacity.
If this results in $\sum\limits_{u\in\mathcal{U}_g} T_u > \tau_g$, being $\log(x)$ is a strictly concave increasing function, we can apply Algorithm~\ref{alg:CPxi}  with $n=|\mathcal{U}_g|$, $\{x_i\}_{i=1}^n = \{T_u\}_{u\in\mathcal{U}_g}$ and $X=\tau_g$ in order to find the optimal solution of the \texttt{PropFair} case.

\subsubsection{\texttt{MaxMin} optimum ($\alpha \rightarrow \infty$)}
The utility function $f(x) = \min_{u\in\mathcal{U}_g} T_u$ is not differentiable, so we cannot directly apply KKT conditions.
However, following the spirit of \texttt{MaxMin} optimization, we can reformulate the problem by adding a new decision variable $T$, changing the utility function and adding one extra set of constraints. This results in the following convex program:

\small
\begin{equation}
\hspace{-3.6cm}
  \begin{cases}
    \max T; & \\
    \text{s.t.:} \\
    T_u \geq T, & \!\!\forall u\!\in\!\mathcal{U}_g; \\
    \sepia{w_u \geq W^{\min}_\mathcal{G},} & \sepia{\!\!\forall  u\!\in\!\mathcal{U}_g;} \\
    \sepia{\sum\limits_{u\in\mathcal{U}_g} \! w_u = W_\mathcal{G};} & \!\!\\
    \redd{T_u \!\leq\!  w_u \log_2\!\left(1 \!+\! \gamma_{b, u}\right)\!,} & \redd{\!\!\forall u\!\in\!\mathcal{U}_g;} \\
    \blue{\sum\limits_{u\in\mathcal{U}_g} \! T_u \! \leq  \tau_g.} &\!\!
  \end{cases}
  \label{eq:CPMaxMinEquiv}
\end{equation}
\normalsize

Since we have added a new decision variable $T$, we denote the solution as $x = \left[\{w_u\}_{u\in\mathcal{U}_g}\}, \{T_u\}_{u\in\mathcal{U}_g}, T\right]\in\mathds{R}^{2|\mathcal{U}_g|+1}$. Hence, we add to the KKT functions of Eqs.~\eqref{eq:KKTfunc1}--\eqref{eq:KKTfunc6} the following KKT functions:
\begin{eqnarray}
  f^{\min}(x) =& \hspace{-1cm}-T; & \\
  g_{2|\mathcal{U}_g|+1+u}(x) =& \hspace{-0.2cm} T- T_u, & \forall u\in\mathcal{U}_g.
\end{eqnarray}

Also, we have to add to KKT gradients of Eqs.~\eqref{eq:KKTgrad1}--\eqref{eq:KKTgrad6} the following KKT gradients:
\begin{eqnarray}
  \hspace{-5mm} \nabla\!f^{\min}(x) =& \hspace{-2.3cm} [0,\hdots,0,-1]; & \\
  \hspace{-5mm} \nabla\!g_{2|\mathcal{U}_g\!|\!+\!1\!+\!u} (x) =& \hspace{-2mm} [0,...,0,-\!1|_{|\mathcal{U}_g\!|\!+\!u}, 0,...,0,1], & \!\!\forall u\!\in\!\mathcal{U}_g.
\end{eqnarray}

In this case, applying KKT conditions, i.e., finding $\nu\in\mathds{R}$ and positive constants $\{\mu_i\}_{i=1}^{3|\mathcal{U}_g|+1}$ not all of them null, and $x\in\mathds{R}^{2|\mathcal{U}_g|+1}$ such that $\overrightarrow{0} = \nabla f^{\min}(x) + \sum\limits_{i=1}^{3|\mathcal{U}_g|+1} \mu_i \nabla g_i(x) + \nu\nabla h(x)$ results into the following set of KKT equations:
\begin{eqnarray}
  -\mu_i - \log_2\!\left(1\!+\!\gamma_{g, i}\right)\!\cdot\! \mu_{|\mathcal{U}_g|+i} \!+\! \nu = 0, & \!\forall  1\!\leq\!  i \!\leq\!  |\mathcal{U}_g|; \label{eq:KKTMaxMin1}\\
  -\mu_{|\mathcal{U}_g|+i} + \mu_{2|\mathcal{U}_g|+1} - \mu_{2|\mathcal{U}_g|+1+i} = 0, & \!\forall  1\!\leq\!  i \!\leq\!  |\mathcal{U}_g|; \label{eq:KKTMaxMin2}\\
  -1 + \sum\limits_{i=1}^{|\mathcal{U}_g|} \mu_{2|\mathcal{U}_g|+1+i} = 0; & \label{eq:KKTMaxMin3}\\
  \mu_i \cdot  \left(W_\mathcal{G}^{\min} - w_i\right) = 0, & \!\forall  1\!\leq\!  i \!\leq\!  |\mathcal{U}_g|; \label{eq:KKTMaxMin4}\\
  \mu_{|\mathcal{U}_g|+i} \!\cdot\!  \left(T_i - w_i  \log_2\!\left(1\!+\!\gamma_{g, i}\right)\right) = 0, &  \!\forall  1\!\leq\!  i \!\leq\!  |\mathcal{U}_g|; \label{eq:KKTMaxMin5}\\
  \mu_{2|\mathcal{U}_g|+1} \cdot  \Big( \sum\limits_{u\in\mathcal{U}_g} T_u - \tau_g\Big) = 0;  & \label{eq:KKTMaxMin6}\\
  \mu_{2|\mathcal{U}_g|+1+i}\cdot \left(T-T_i\right) = 0, & \!\forall  1\!\leq\!  i \!\leq\!  |\mathcal{U}_g|; \label{eq:KKTMaxMin7}\\
  \sum\limits_{u\in\mathcal{U}_g} w_u - W_\mathcal{G} = 0. & \label{eq:KKTMaxMin8}
\end{eqnarray}

\textbf{With unbounded backbone capacity $\pmb{\tau_g}$.}
First, we look for a solution assuming that $\tau_g$ is unbounded.
Hence, according to Eq.~\eqref{eq:KKTMaxMin6}, $\mu_{2|\mathcal{U}_g|+1}=0$.
As $\{T_i\}_{i=1}^{|\mathcal{U}_g|}$ is not limited by $\tau_g$, we take $T_i=w_i \log_2\left(1+\gamma_{g,i}\right)$, $\forall 1\leq i\leq |\mathcal{U}_g|$.

We initially assume that $w_i=W_\mathcal{G}^{\min}$, $\forall 1\leq i\leq |\mathcal{U}_g|$ and build an algorithm that finds the optimal solution of CP~\eqref{eq:CPMaxMinEquiv} (see Algorithm~\ref{alg:CPMaxMinEquiv}).
Please note that now $T_i = W_\mathcal{G}^{\min}\cdot \log_2\left(1+\gamma_{g,i}\right)$, $\forall 1\leq i\leq |\mathcal{U}_g|$ (step~\ref{alg:CPMaxMinEquiv:Step1}).
We assume, without loss of generality, that $T_{i-1} \leq T_i$, $\forall 2\leq i\leq |\mathcal{U}_g|$ (step~\ref{alg:CPMaxMinEquiv:Step2}), and define the set of indices $\mathcal{J}$ as those indices $i$ such that $T_i$ is minimum (step~\ref{alg:CPMaxMinEquiv:Step3}), i.e.,
\begin{eqnarray}
  \mathcal{J}=\left\{i\in\{1,\hdots,|\mathcal{U}_g|\}\suchthat T_i=\min\limits_{1\leq j\leq |\mathcal{U}_g|} T_j\right\}.
  \label{eq:JPF}
\end{eqnarray}

Algorithm~\ref{alg:CPMaxMinEquiv} works in the following way: while there are still bandwidth resources to be allocated,
i.e., $\sum\limits_{i=1}^{|\mathcal{U}_g|} w_i < W_{\mathcal{G}}$, and $|\mathcal{J}|\neq |\mathcal{U}_g|$,
we select the index $j_0=\arg\min\{T_i\suchthat i\notin \mathcal{J}\}$ such that $T_{j_0}$ is the lowest user throughput rate not equal to the minimum of the throughput rates (step~\ref{alg:CPMaxMinEquiv:Step5}).
Now, we aim to increase $w_i$, $\forall i\in\mathcal{J}$ as much as possible in a way that is $\max$--$\min$ fair and $T_i\leq T_{j_0}$, $\forall i\in\mathcal{J}$. Hence, we have to find $\{k_i\}_{i\in\mathcal{J}}$ such that $\{w_i\}_{i\in\mathcal{J}}$ are increased by $k_i$ each. The optimal way of doing this is by first assigning $k_i=\frac{T_{j_0}}{\log_2\left(1+\gamma_{g,i}\right)}-w_i$, $\forall i\in\mathcal{J}$ (step~\ref{alg:CPMaxMinEquiv:Step6}) and checking that $\sum\limits_{i\in\mathcal{J}}k_i\leq W_{\mathcal{G}}-\sum\limits_{i=1}^{|\mathcal{U}_g|} w_i$. In case such an inequality is not satisfied, then assign $k_i=\frac{W_\mathcal{G}-\sum\limits_{i\notin\mathcal{J}}w_i}{\log_2\left(1+\gamma_{g,i}\right)\sum\limits_{i\in\mathcal{J}}\frac{1}{\log_2\left(1+\gamma_{g,i}\right)}}-w_i$, $\forall i\in\mathcal{J}$ (step~\ref{alg:CPMaxMinEquiv:Step8}).

Once $k_i$ is derived, we assign $w_i \leftarrow w_i + k_i$, $\forall i\in\mathcal{J}$ (step~\ref{alg:CPMaxMinEquiv:Step10}). Now, in case that we have assigned $k_i=\frac{T_{j_0}}{\log_2\left(1+\gamma_{g,i}\right)}-w_i$, $\forall i\in\mathcal{J}$, we reassign set $\mathcal{J}$ as $\mathcal{J}\leftarrow \mathcal{J}\cup\{j_0\}$ (step~\ref{alg:CPMaxMinEquiv:Step12}), and start all over. Otherwise, the algorithm stops and the optimal solution is found.

\begin{algorithm}[t]
  \caption{Optimal solution of CP~\eqref{eq:CPMaxMinEquiv} (unbounded $\tau_g$).}
  \label{alg:CPMaxMinEquiv}
  \begin{algorithmic}[1]
    \STATE Initialize: $w_i\!\leftarrow\! W_\mathcal{G}^{\min}$, $T_i\!\leftarrow\! w_i \log_2\!\left(1\!+\!\gamma_{g,i}\!\right)$, $\forall 1\!\leq\!i\!\leq\! |\mathcal{U}_g|$.\label{alg:CPMaxMinEquiv:Step1}
    \STATE Assume, WLOG, $T_{i-1}\leq T_i$, $\forall 2\leq i\leq |\mathcal{U}_g|$.\label{alg:CPMaxMinEquiv:Step2}
    \STATE $\mathcal{J} \leftarrow  \left\{i\in\{1,\hdots,|\mathcal{U}_g|\} \suchthat T_i = \min\limits_{1\leq j\leq |\mathcal{U}_g|} T_j\right\}$.\label{alg:CPMaxMinEquiv:Step3}
    \WHILE{$\sum\limits_{i=1}^{|\mathcal{U}_g|} w_i < W_\mathcal{G}$ \AND $|\mathcal{J}|\neq |\mathcal{U}_g|$}\label{alg:CPMaxMinEquiv:Step4}
    \STATE $j_0 \leftarrow \arg\min\{T_i\suchthat i\notin\mathcal{J}\}$ \AND $K_b \leftarrow 1$.\label{alg:CPMaxMinEquiv:Step5}
    \STATE $k_i \leftarrow \frac{T_{j_0}}{\log_2\left(1+\gamma_{g,i}\right)}-w_i$, $\forall i\in\mathcal{J}$.\label{alg:CPMaxMinEquiv:Step6}
    \IF{$\sum\limits_{i\in\mathcal{J}}k_i > W_\mathcal{G} - \sum\limits_{i=1}^{|\mathcal{U}_g|} w_i$}\label{alg:CPMaxMinEquiv:Step7} \STATE{$\!\!k_i \!\leftarrow\! \frac{W_\mathcal{G}-\sum\limits_{i\notin\mathcal{J}}w_i}{\log_2\!\left(1+\gamma_{g,i}\right)\!\sum\limits_{i\!\in\!\mathcal{J}} \!\frac{1}{\log_2\left(1+\gamma_{g,i}\right)}}\!-\!w_i$, $\forall i\!\in\!\mathcal{J}$ \AND $K_b \!\leftarrow\! 0$.}\label{alg:CPMaxMinEquiv:Step8} \ENDIF\label{alg:CPMaxMinEquiv:Step9}
    \STATE $w_i \leftarrow w_i \!+\! k_i$, $\forall i\!\in\!\mathcal{J}$ \AND $T_i \leftarrow w_i \log_2\!\left(1\!+\!\gamma_{g,i}\right)$, $\forall i\!\in\!\mathcal{J}$.\label{alg:CPMaxMinEquiv:Step10}
    \IF{$K_b = 1$}\label{alg:CPMaxMinEquiv:Step11} \STATE{$\!\!\mathcal{J} \leftarrow \mathcal{J} \cup \{j_0\}$.}\label{alg:CPMaxMinEquiv:Step12} \ENDIF\label{alg:CPMaxMinEquiv:Step13}
    \ENDWHILE\label{alg:CPMaxMinEquiv:Step14}
    \STATE $T \leftarrow \min\limits_{1\leq i\leq |\mathcal{U}_g|} T_i$.\label{alg:CPMaxMinEquiv:Step15}
  \end{algorithmic}
\end{algorithm}

In Note~\ref{note:CPMaxMinEquiv} we prove that the assigned $\{k_i\}_{i\in\mathcal{J}}$ at each iteration of Algorithm~\ref{alg:CPMaxMinEquiv} provides the optimal $\max$--$\min$ fair distribution of resources.

\begin{note}
  Given a distribution of resources $\{w_i\}_{i=1}^{|\mathcal{U}_g|}$ and users throughput rates $\{T_i\}_{i=1}^{|\mathcal{U}_g|}$ such that $T_i=w_i \log_2\left(1+\gamma_{g,i}\right)$, $\forall 1\leq i\leq |\mathcal{U}_g|$, we define the set $\mathcal{J}$ as in Eq.~\eqref{eq:JPF}.

  Hence, we have that $w_i \log_2\!\left(1\!+\!\gamma_{g,i}\right) = w_k \log_2\!\left(1\!+\!\gamma_{g,k}\right)$, $\forall i,k\in\mathcal{J}$.

  Now, given $j_0 = \arg\min\{T_i\suchthat i\notin\mathcal{J}\}$, we want to increase $\{w_i\}_{i\in\mathcal{J}}$ as much as possible by $k_i$ each in a $\max$--$\min$ fair way so that $(w_i+k_i) \log_2\left(1+\gamma_{g,i}\right) \leq  T_{j_0}$, $\forall i\in\mathcal{J}$. Hence, we must solve the following convex program:

    \begin{equation}
    \hspace{-2cm}
    \begin{cases}
        \max \min\limits_{i\in\mathcal{J}}\left(w_i+k_i\right) \log_2\left(1+\gamma_{g,i}\right); & \\
        \text{s.t.:} \\
        (w_i+k_i) \log_2\left(1+\gamma_{g,i}\right)\leq T_{j_0}, & \!\!\forall i\!\in\!\mathcal{J}; \\
        \sum\limits_{i=1}^{|\mathcal{U}_g|} w_i + \sum\limits_{i\in\mathcal{J}} k_i \leq  W_\mathcal{G}; &
    \end{cases}
    \label{eq:CPMaxMinki}
    \end{equation}

  Since CP~\eqref{eq:CPMaxMinki} has a $\max$--$\min$ utility function, we reformulate this CP into an equivalent CP to which we can directly apply KKT conditions:

    \begin{equation}
    \hspace{-2cm}
    \begin{cases}
        \max L; & \\
        \text{s.t.:} \\
        (w_i+k_i) \log_2\left(1+\gamma_{g,i}\right) \geq L, & \forall  i\!\in\!\mathcal{J}; \\
        k_i \leq  \frac{T_{j_0}}{\log_2\left(1+\gamma_{g,i}\right)} - w_i, & \forall i\!\in\!\mathcal{J}; \\
        \sum\limits_{i\in\mathcal{J}} k_i \leq  W_\mathcal{G} - \sum\limits_{i=1}^{|\mathcal{U}_g|} w_i; &
    \end{cases}
    \label{eq:CPMaxMinkiEquiv}
    \end{equation}

  Please, note that the first constraint of CP~\eqref{eq:CPMaxMinkiEquiv} can be rearranged as $k_i\geq\frac{L}{\log_2\left(1+\gamma_{g,i}\right)}- w_i$, $\forall i\in\mathcal{J}$. Also, the second and third constraints of CP~\eqref{eq:CPMaxMinkiEquiv} are equivalent to the constraints of the $\max$--$\min$ CP~\eqref{eq:CPMaxMinki}.

  Now, we make use of KKT conditions to solve CP~\eqref{eq:CPMaxMinkiEquiv}. The decision variables are gathered in vector $x = \left[\{k_i\}_{i\in\mathcal{J}}, L\right]\in\mathds{R}^{|\mathcal{J}|+1}$. We define the following KKT functions:
  \begin{eqnarray}
    f(x) =& \hspace{-2.7cm} -L; \\
    g_i(x) =& \hspace{-2mm} \frac{L}{\log_2\left(1+\gamma_{g,i}\right)}-k_i-w_i, & \forall i\in\mathcal{J}; \\
    g_{|\mathcal{J}|+i}(x) =& \hspace{-2mm} k_i - \frac{T_{j_0}}{\log_2\left(1+\gamma_{g,i}\right)}-w_i, & \forall i\in\mathcal{J};\\
    g_{2|\mathcal{J}|+1}(x) =& \hspace{-2mm} \sum\limits_{i\in\mathcal{J}}k_i - W_\mathcal{G} - \sum\limits_{i=1}^{|\mathcal{U}_g|} w_i, &\forall i\in\mathcal{J}.
  \end{eqnarray}

  Hence, the KKT gradients are the following:
  \begin{eqnarray}
    \!\!\!\!\!\!\!\!\nabla \!f(x) =& \hspace{-3.1cm} [0, \hdots, 0, 1]; \\
    \!\!\!\!\!\!\!\!\nabla \!g_i(x) =& \hspace{-3mm}\left[\!0,..., 0, -1\!|_i, 0, ..., 0, \frac{1}{\log_2\!\left(1\!+\!\gamma_{g,i}\!\right)}\!\right]\!\!, & \!\!\forall i\!\in\!\mathcal{J}\!; \\
    \!\!\!\!\!\!\!\!\nabla \!g_{|\mathcal{J}|\!+\!i}(x) =& \hspace{-1.6cm} \left[0,\hdots,0,1\!|_i,0,\hdots,0\right], & \!\!\forall i\!\in\!\mathcal{J}\!; \\
    \!\!\!\!\!\!\!\!\nabla \!g_{2|\mathcal{J}|\!+\!1}(x) =&  \hspace{-3.05cm} \left[1,\hdots,1,0\right]. &
  \end{eqnarray}

%
 We derive the following KKT conditions:
  \begin{eqnarray}
    -\mu_i + \mu_{|\mathcal{J}|+i} + \mu_{2|\mathcal{J}|+1} = 0, & \forall  i\in\mathcal{J}; \label{eq:KKTCPMaxMinkiEquiv1}\\
    -1 + \sum\limits_{i=1}^{|\mathcal{J}|} \frac{\mu_i}{\log_2\left(1+\gamma_{g,i}\right)} = 0; & \label{eq:KKTCPMaxMinkiEquiv2}\\
    \mu_i \cdot \left(\frac{L}{\log_2\left(1+\gamma_{g,i}\right)}-k_i-w_i\right) = 0, & \forall i\in\mathcal{J}; \label{eq:KKTCPMaxMinkiEquiv3}\\
    \mu_{|\mathcal{J}|+i}\cdot  \left(k_i - \frac{T_{j_0}}{\log_2\left(1+\gamma_{g,i}\right)}+w_i\right) = 0, & \forall i\in\mathcal{J}; \label{eq:KKTCPMaxMinkiEquiv4}\\
    \mu_{2|\mathcal{J}|+1} \cdot  \left(\sum\limits_{i\in\mathcal{J}} k_i - W_\mathcal{G} - \sum\limits_{i=1} ^{|\mathcal{U}_g|} w_i \right) = 0. &\label{eq:KKTCPMaxMinkiEquiv5}
  \end{eqnarray}

  First, we note that if assigning $k_i = \frac{T_{j_0}}{\log_2\left(1+\gamma_{g,i}\right)} - w_i$, $\forall i\in\mathcal{J}$ accomplishes that $\sum\limits_{i\in\mathcal{J}} k_i \leq  W_\mathcal{G} - \sum\limits_{i=1}^{|\mathcal{U}_g|} w_i$, we have the optimal solution, as each $k_i$ receives the maximum possible value and constraints are satisfied.

  Hence, we can assume that $\exists i_0\in\mathcal{J} \suchthat k_{i_0} < \frac{T_{j_0}}{\log_2\left(1+\gamma_{g,i_0}\right)}-w_{i_0}$ and hence, according to Eq.~\eqref{eq:KKTCPMaxMinkiEquiv4}, $\mu_{|\mathcal{J}|+i_0} = 0$.

  The optimal $\{k_i\}_{i\in\mathcal{J}}$ and $L$ that solve the KKT conditions are:
  \begin{eqnarray}
    k_i =& \hspace{-2mm} \frac{W_\mathcal{G}-\sum\limits_{i\notin\mathcal{J}}w_i}{\log_2\left(1+\gamma_{g,i}\right) \sum\limits_{i\in\mathcal{J}} \frac{1}{\log_2\left(1+\gamma_{g,i}\right)}}-w_i, & \forall  i\in\mathcal{J}; \\
    L =& \hspace{-2.5cm} \frac{W_\mathcal{G} - \sum\limits_{i\notin\mathcal{J}}w_i}{\sum\limits_{i\in\mathcal{J}}\frac{1}{\log_2\left(1+\gamma_{g,i}\right)}}. &
  \end{eqnarray}

  It is simple to check that $\sum\limits_{i\in\mathcal{J}} k_i = W_\mathcal{G} - \sum\limits_{i=1}^{|\mathcal{U}_g|} w_i$, so that Eq.~\eqref{eq:KKTCPMaxMinkiEquiv5} is satisfied.

  Now, we assign $\mu_i=\frac{w_i}{\sum\limits_{i\in\mathcal{J}}w_i}\log_2\left(1+\gamma_{g, i}\right)$, $\forall i\in\mathcal{J}$, and hence, since $\mu_{|\mathcal{J}|+i_0} = 0$, and according to Eq.~\eqref{eq:KKTCPMaxMinkiEquiv1}, we have that $\mu_{2|\mathcal{J}| + 1} = \mu_{i_0} = \frac{w_{i_0}}{\sum\limits_{i\in\mathcal{J}}w_i}\log_2\left(1+\gamma_{g,i_0}\right)$. Hence, according also to Eq.~\eqref{eq:KKTCPMaxMinkiEquiv1}, $\mu_{|\mathcal{J}|+i} = \mu_i - \mu_{2|\mathcal{J}|+1} = \frac{w_i}{\sum\limits_{i\in\mathcal{J}}w_i}\log_2\left(1+\gamma_{g,i}\right) - \frac{w_{i_0}}{\sum\limits_{i\in\mathcal{J}}w_i}\log_2\left(1+\gamma_{g,i_0}\right)$, $\forall i\in\mathcal{J}$. Since $i,i_0\in\mathcal{J}$, we have that, by definition on $\mathcal{J}$, $\mu_{|\mathcal{J}|+i} = 0$, $\forall i\in\mathcal{J}$.

  As a result, since with such a solution all KKT conditions,
  are satisfied, we have the optimal solution to CP~\eqref{eq:CPMaxMinkiEquiv}.


  \label{note:CPMaxMinEquiv}
\end{note}

\textbf{With limited backbone capacity $\pmb{\tau_g}$.} Now, we assume that $\tau_g$ is bounded. In order to find the optimal solution under this assumption, we first solve the problem assuming unbounded $\tau_g$, as detailed above. If such an output provides a feasible solution, i.e., $\sum\limits_{u\in\mathcal{U}_g} T_u \leq \tau_g$, we have the optimal solution. Hence, we assume that such an output provides an unfeasible solution, i.e., $\sum\limits_{u\in\mathcal{U}_g} T_u > \tau_g$.

Let $\{w_u\}_{u\in\mathcal{U}_g}$, $\{T_u\}_{u\in\mathcal{U}_g}$ be the unfeasible solution provided by the $\max$--$\min$ optimization with unbounded $\tau_g$, and let $T=\min\limits_{u\in\mathcal{U}_g}\{T_u\}$ be the minimum achieved throughput by users so far (note that $T$ is the value of the utility). Due to the $\max$--$\min$ fairness nature, every user~$u$ such that $T_u > T$ disposes of the minimum amount of resources, $W_\mathcal{G}^{\min}$ and $T_u = W_\mathcal{G}^{\min}  \log_2\left(1+\gamma_{g,u}\right)$, $\forall u\in\mathcal{U}_g \suchthat T_u > T$ (otherwise, if such users disposed of more than $W_\mathcal{G}^{\min}$ resources, such exceeded resources could be reallocated to those users with minimum throughput to increase the utility function, which is not possible from the output $\max$--$\min$ fairness optimization with unbounded $\tau_g$).

Hence, as no resources can be removed from any user $u$ such that $T_u>T$, and $\sum\limits_{u\in\mathcal{U}_g} T_u > \tau_g$, we can reduce the rate of these users to, for instance, $T_u = T$ (hence, now $T_u = T$, $\forall u\in\mathcal{U}_g$). If now $\sum\limits_{u\in\mathcal{U}_g} T_u \leq \tau_g$, we have found the optimal solution. Otherwise, we need to reduce more individual throughput rates. In order to be $\max$--$\min$ fair, we have to reduce every individual rate the same amount until the $\tau_g$--constraint is satisfied. Hence, we define $R=\sum\limits_{u\in\mathcal{U}_g} T_u - \tau_g > 0$ and decrease every individual rate by $\frac{R}{|\mathcal{U}_g|}$, i.e., $T_u = T - \frac{R}{|\mathcal{U}_g|}$, $\forall  u\in\mathcal{U}_g$. Hence, the $\max$--$\min$ fairness optimization of CP~\eqref{eq:CPMaxMinEquiv} is solved also under the assumption of bounded~$\tau_g$.

\subsection{With One Drone}
\label{ss:BHmanaged1aBS}

In this scenario, we consider that \gBS~$g$ serves its users $\mathcal{U}_g$ and also one \aBS~$a$, i.e., $\exists a\in\mathcal{A} \suchthat \mathcal{A}_g = \{a\}$. Hence, CP
of Fig.~\ref{fig:CP}
simplifies since backhaul resources do not need to be split over multiple \aBSs, but all of them are assigned to only one \aBS, i.e., $w^a = W_\mathcal{B}$. Hence, we solve the following~CP:

\vspace{-1.5mm}
\small
\begin{equation}
\hspace{-3cm}
  \eag{
  \begin{cases}
    \,\,\,\qquad\qquad\qquad\qquad\qquad\mathclap{\max\limits_{\sepia{w_u,} \redd{T_u}} U_{\mathrm{cvx}, g}^\alpha =  \begin{cases}\sum\limits_{u\in\mathcal{U}_g \cup\, \mathcal{U}_a} \!\left( T_u \right)^{1-\alpha} \cdot  \frac{1}{1-\alpha}, & \alpha \neq 1; \\
    \sum\limits_{u\in\mathcal{U}_g \cup\, \mathcal{U}_a} \!\log\left( T_u \right), &  \alpha = 1;
    \end{cases}}\\
    \text{s.t.:} \\
    \blue{T^a \!\leq\! W_\mathcal{B} \log_2\!\left(1 \!+\! \gamma_{g, a}^\mathcal{B}\right)\!;} & \\
    \sepia{w_u \geq W^{\min}_\mathcal{G},} & \sepia{\forall  u\!\in\!\mathcal{U}_g;} \\
    \sepia{\sum\limits_{u\in\mathcal{U}_g} \! w_u = W_\mathcal{G};} & \!\!\\
    \redd{w_u \geq W^{\min}_\mathcal{A},} & \redd{\forall  u\!\in\!\mathcal{U}_a;} \\
    \redd{\sum\limits_{u\in\mathcal{U}_a} \! w_u = W_{\!\mathcal{A}};} & \\
    \redd{T_u \!\leq\!  w_u \log_2\!\left(1 \!+\! \gamma_{g, u}\right)\!,} & \redd{\forall u\in\mathcal{U}_g;} \\
    \blue{T_u \!\leq\!  w_u \log_2\!\left(1 \!+\! \gamma_{a, u}\right)\!,} & \blue{\forall u\in\mathcal{U}_a;} \\
    \redd{\sum\limits_{u\in\mathcal{U}_a} \! T_u \!\leq  T^a;} & \\
    \blue{\sum\limits_{u\in\mathcal{U}_g} \! T_u \!+\! T^a \leq  \tau_g.} &\!\!
  \end{cases}}
  \label{eq:BHmanaged1aBS}
\end{equation}
\normalsize

In order to solve CP~\eqref{eq:BHmanaged1aBS}, we do not derive KKT conditions, as we will make use of the analysis conducted in Subection~\ref{ss:BHfree}.

\textbf{With unbounded backbone capacity $\pmb{\tau_g}$.} First, we assume that $\tau_g$ is unbounded, and hence users throughput rates are only limited by their Shannon capacity, according to the allocated bandwidth resources. As \gBS~$g$ disposes of two independent baskets of resources, $W_\mathcal{G}$ for \gBS-served users, and $W_\mathcal{B}$ for the backhaul-served \aBS, the distribution of \gBS-served users resources disregards from \aBS-served users resources (since $\tau_g$ is unbounded). Hence, the backhaul rate for \aBS~$a$, $T^a$, is limited only by the Shannon capacity, so that we can assume that $T^a = W_\mathcal{B} \log_2\left(1 + \gamma_{g,a}^{\mathcal{B}}\right)$.

Hence, we solve optimal resource allocation for \gBS-served users assuming unbounded $\tau_g$, and solve also optimal resource allocation for \aBS-served users assuming that their aggregated throughput is limited by the backhaul capacity $T^a$, i.e., $\sum\limits_{u\in\mathcal{U}_a} T_u \leq  T^a$. Both cases can be solved as detailed in Subsection~\ref{ss:BHfree}. As a result, CP~\eqref{eq:BHmanaged1aBS} is solved under the assumption of unbounded $\tau_g$, for any value of $\alpha$, including $\alpha\rightarrow \infty$.

The solution provided is optimal although not necessarily unique, as the backhaul capacity $T^a$ might be higher than the aggregated served throughputs. Hence, to ease the understanding of upcoming sections, we assume that the provided optimal solution satisfies that $T^a = \sum\limits_{u\in\mathcal{U}_a} T_u$, which remains feasible and optimal.

\textbf{With limited backbone capacity $\pmb{\tau_g}$.} Now, we do not assume that $\tau_g$ is unbounded. In order to solve this case, we first search the optimal solution assuming that $\tau_g$ is indeed unbounded and checking if $\sum\limits_{u\in\mathcal{U}_g} T_u + T^a = \sum\limits_{u\in\mathcal{U}_g} T_u + \sum\limits_{u\in\mathcal{U}_a} T_u \leq  \tau_g$. In case the inequality is satisfied, we have found the optimal solution. Otherwise, users throughputs must be decreased in order to satisfy the $\tau_g$--constraint.

Hence, according to what proven and described in Subsection~\ref{ss:BHfree}, we can decrease the convenient users throughputs $T_u$ until we get to satisfy that $\sum\limits_{u\in\mathcal{U}_g} T_u + T^a = \sum\limits_{u\in\mathcal{U}_g} T_u + \sum\limits_{u\in\mathcal{U}_a} T_u = \tau_g$ and hence provide the optimal solution.

\subsection{Generic Case}
In this scenario, we consider the generic case, formulated in Fig.~\ref{fig:CP}.
As in the previous cases, we distinguish between unbounded and bounded $\tau_g$ assumptions. In Algorithm~\ref{alg:BHmanaged} we show how to find the optimal solution, as detailed in what follows.

\textbf{With unbounded backbone capacity $\pmb{\tau_g}$.}
Here, we assume that $\tau_g$ is unbounded, so that distribution of user resources at \gBSs is not affected by the presence of \aBSs and their associated users.
However, now we cannot know a priori how many resources each \aBS will get, as they share a common bandwidth of $W_{\!\mathcal{B}}$. Each \aBS must receive $w^a$ resources according to the optimization output. We find the optimal solution as detailed as follows.

First, the optimal resource allocation for \gBS-served users is as previously detailed in Section~\ref{ss:BHfree}.
Then, for each \aBS $a\in\mathcal{A}_g$ we solve also optimal resource allocation assuming unbounded backhaul capacity $T^a$, i.e., \aBS-served users throughput rates are limited only by their Shannon capacity, according to the bandwidth allocated, using the scheme detailed in Section~\ref{ss:BHfree}.
For convenience, we now denote by $T_{a,u}$ the throughput of access link $(a, u)$, $\forall a\in\mathcal{A}_g$ $\forall u\in\mathcal{U}_a$.
For each $a\in\mathcal{A}_g$, we need a backhaul capacity of $T^a = \sum\limits_{u\in\mathcal{U}_a} T_{a, u}$. Hence, \aBS $a$ needs $w^a = \max\left(W_{\!\mathcal{B}}^{\min}, \sum\limits_{u\in\mathcal{U}_a} \frac{T_{a, u}}{\log_2\left(1 + \gamma_{g, a}^{\mathcal{B}}\right)}\right)$.

Now, in case $\sum\limits_{a\in\mathcal{A}_g} w^a \leq  W_{\!\mathcal{B}}$, we choose $a_0\in\mathcal{A}_g$ arbitrarily and add resources to $w^{a_0}$ so that we get to $\sum\limits_{a\in\mathcal{A}_g} w^a = W_{\!\mathcal{B}}$ (i.e., $w^{a_0} = w^{a_0} + W_{\!\mathcal{B}} - \sum\limits_{a\in\mathcal{A}_g} w^a$). Hence, in this case the optimization is solved.

Conversely, in case $\sum\limits_{a\in\mathcal{A}_g} w^a >  W_{\!\mathcal{B}}$, we are assigning to backhaul \aBSs more resources than what available at the \gBS. Hence, some resources should be removed. In this case, we need to build the optimal solution from scratch.
First, we assign to each \aBS $a\in\mathcal{A}_g$ the minimum bandwidth: $w^a = W_{\!\mathcal{B}}^{\min}$, $\forall a\in\mathcal{A}_g$ (step~\ref{alg:BHmanagedStep1} in Algorithm~\ref{alg:BHmanaged}).
Second, we assign to each \aBS $a\in\mathcal{A}_g$ the highest achievable throughput: $T^a = w^a \log_2\left(1 + \gamma_{g, a}^\mathcal{B}\right)$, $\forall  a\in\mathcal{A}_g$ (step~\ref{alg:BHmanagedStep2}).
Third, we solve optimal resource allocation at each $a\in\mathcal{A}_g$ and also at \gBS~$g$.
Hence, we dispose of $\{T_{a, u}\}_{u\in\mathcal{U}_a}$, $\forall a\in\mathcal{A}_g$, and of $\{T_u\}_{u\in\mathcal{U}_g}$ (step~\ref{alg:BHmanagedStep3}). Please note that $\forall a\in\mathcal{A}_g$, $\forall u\in\mathcal{U}_a$ it might happen that $T_{a,u} < w_u \log_2\left(1 + \gamma_{a,u}\right)$ because of the backhaul limitation $T^a$. Hence, if $T^a$ increases, such $T_{a, u}$'s can also increase.
The contribution of $\{T_{a,u}\}_{u\in\mathcal{U}_a}$, $\forall a\in\mathcal{A}_g$ to utility $\mathcal{U}_{\mathrm{cvx}, g}^\alpha$ is either $\left(T_{a, u}\right)^{1-\alpha}$ or $\log\left(T_{a, u}\right)$. Hence, the distribution of throughput contribution follows an increasing and concave function. Since $\xi(x) = x^{1-\alpha}$ and $\xi(x) = \log\left(x\right)$ are both increasing and concave functions, the way of increasing most utility $\mathcal{U}_{\mathrm{cvx}, g}^\alpha$ is by raising the lowest values of $\{T_{a, u}\}_{u\in\mathcal{U}_a}$, $\forall a\in\mathcal{A}_g$ as much as possible as long as constraints are not violated. Also in the $\max$--$\min$ case: the way of increasing as much as possible the utility is by equally raising the lowest values of $\{T_{a,u}\}_{u\in\mathcal{U}_a}$, $\forall a\in\mathcal{A}_g$, as long as constraints are not violated, due to the well-known $\max$--$\min$ fairness nature.
Let \begin{eqnarray}T_m \!=\! \min \!\left\{T_{a,u} \!\!\suchthat\! T_{a,u} \!<\! w_u \log_2\!\left(1 \!+\! \gamma_{a, u}\right), a\!\in\!\mathcal{A}_g, u\!\in\!\mathcal{U}_a\!\right\}\end{eqnarray} (step~\ref{alg:BHmanagedStep6}) be the minimum throughput rate that has not reached the corresponding to the Shannon capacity (note that if such a $T_m$ does not exist, we are done). Let \begin{eqnarray}
\mathcal{L}_m = \left\{(a,u)\in\mathcal{A}_g\times \bigcup\limits_{a\in\mathcal{A}_g} \mathcal{U}_a \suchthat T_{a,u} = T_m\right\}
\end{eqnarray} be the set of those \aBS--UE links such that the link rate is the same as the minimum $T_m$ (step~\ref{alg:BHmanagedStep7}). Let \begin{eqnarray}
T_M = \min\left\{T_{a,u} \suchthat T_{a,u} > T_m, a\in\mathcal{A}_g, u\in\mathcal{U}_a\right\}
\end{eqnarray} be the minimum throughput rate among those rates that are not as the minimum $T_m$ (step~\ref{alg:BHmanagedStep7.5}). Let \begin{eqnarray}
T_{M_2} = \min\left(T_M, \min\limits_{(a,u)\in\mathcal{L}_m} w_u \log_2\left(1 + \gamma_{a,u}\right)\right)
\end{eqnarray} (step~\ref{alg:BHmanagedStep8}). The goal now is to increase $\{T_{a,u}\}_{(a,u)\in\mathcal{L}_m}$ as much as possible not exceeding $T_{M_2}$, as long as those involved $a\in\mathcal{A}_g$ can request more resources to increase $T^a$. Let $\beta\in[0,1]$ be an undetermined parameter, $\{T_{a,u}\}_{(a,u)\in\mathcal{L}_m}$ will be increased by $\beta\!\cdot\! (T_{M_2} - T_m)$, i.e., at most, by $T_{M_2} - T_m$ (step~\ref{alg:BHmanagedStep13}). Parameter $\beta$ will be defined later. Let \begin{eqnarray}\overline{\mathcal{U}}_a = \{u\in\mathcal{U}_a\suchthat (a,u)\in\mathcal{L}_m\}, \forall  a\in\mathcal{A}_g\end{eqnarray} (step~\ref{alg:BHmanagedStep9}). Now, we set $T_{a,u}^{'} = T_{a,u} + \beta\!\cdot\! (T_{M_2}-T_m)$, $\forall (a,u)\in\mathcal{L}_m$ to increase the involved throughput rates. Hence, we set $\forall a\in\mathcal{A}_g$ \begin{eqnarray}T^a &\!\!\!\!=\!\!\!\!& \sum\limits_{u\notin\overline{\mathcal{U}}_a}T_{a,u} + \sum\limits_{u\in\overline{\mathcal{U}}_a}T_{a,u}^{'} = \nonumber\\ &\!\!\!\!=\!\!\!\!& \sum\limits_{u\notin\overline{\mathcal{U}}_a}T_{a,u} + \sum\limits_{u\in\overline{\mathcal{U}}_a}\left(T_{a,u} + \beta\!\cdot\! (T_{M_2}-T_m)\right) = \nonumber\\ &\!\!\!\!=\!\!\!\!& \sum\limits_{u\notin\overline{\mathcal{U}}_a}T_{a,u} + \sum\limits_{u\in\overline{\mathcal{U}}_a}T_{a,u} + |\overline{\mathcal{U}}_a|\beta\!\cdot\! (T_{M_2}-T_m).\end{eqnarray} Hence, we set $\forall a\in\mathcal{A}_g$ \begin{eqnarray}w^a_{new} &\!\!\!\!=\!\!\!\!& \frac{T_a}{\log_2\!\left(1\!+\!\gamma_{g,a}^{\mathcal{B}}\!\right)} =  \frac{\sum\limits_{u\in\mathcal{U}_a} \!\!T_{a,u} \!+\! |\overline{\mathcal{U}}_a|\beta\!\cdot\! (\!T_{M_2}\!\!-\!T_m\!)}{\log_2\!\left(1\!+\!\gamma_{g,a}^{\mathcal{B}}\!\right)} = \nonumber\\ &\!\!\!\!=\!\!\!\!& w^a + \frac{|\overline{\mathcal{U}}_a|\beta\!\cdot\! (T_{M_2}\!-\!T_m)}{\log_2\left(1 + \gamma_{g,a}^{\mathcal{B}}\right)}\end{eqnarray} (step~\ref{alg:BHmanagedStep11}). Now, the aggregation of the new backhaul bandwidth allocation has to be lower than the total bandwidth, i.e., \begin{eqnarray}\!\!\!\!\!\!\!\!\!\!\sum\limits_{a\in\mathcal{A}_g} \!\!w^a_{new} &\!\!\!\!=\!\!\!\!& \!\sum\limits_{a\in\mathcal{A}_g}\!\!\left(w^a + \frac{|\overline{\mathcal{U}}_a|\beta\!\cdot\! (T_{M_2}-T_m)}{\log_2\left(1+\gamma_{g,a}^{\mathcal{B}}\right)}\right) = \nonumber\\ &\!\!\!\!=\!\!\!\!& \!\sum\limits_{a\in\mathcal{A}_g} \!\!\!w^a \!+\! \beta\!\cdot\! (T_{M_2}\!-\!T_m)\!\!\sum\limits_{a\in\mathcal{A}_g} \!\!\!\frac{|\overline{\mathcal{U}}_a|}{\log_2\!\left(1\!\!+\!\gamma_{g,a}^{\mathcal{B}}\!\right)}\end{eqnarray} has to be lower or equal than $W_{\!\mathcal{B}}$. Hence, isolating $\beta$ we get that necessarily, \begin{eqnarray}\beta \leq \frac{W_{\!\mathcal{B}} - \sum\limits_{a\in\mathcal{A}_g} w^a}{(T_{M_2}-T_m) \sum\limits_{a\in\mathcal{A}_g}\frac{|\overline{\mathcal{U}}_a|}{\log_2\left(1+\gamma_{g,a}^{\mathcal{B}}\right)}}.\end{eqnarray} Hence, we define $\beta$ as \begin{eqnarray}\beta = \min\left(1, \frac{W_{\!\mathcal{B}} - \sum\limits_{a\in\mathcal{A}_g} w^a}{(T_{M_2}-T_m) \sum\limits_{a\in\mathcal{A}_g}\frac{|\overline{\mathcal{U}}_a|}{\log_2\left(1+\gamma_{g,a}^{\mathcal{B}}\right)}}\right)\end{eqnarray} (step~\ref{alg:BHmanagedStep10}). Once the parameter $\beta$ is derived, we assign $w^a = w^a_{new}$ and $T^a = w^a \log_2\left(1 + \gamma_{g,a}^{\mathcal{B}}\right)$, $\forall  a\in\mathcal{A}_g$ (step~\ref{alg:BHmanagedStep12}). In case that $\beta = 1$ (step~\ref{alg:BHmanagedStep5}), we repeat the process defining $T_m$ again and increasing the corresponding throughput rates. This yields the optimal solution.

\textbf{With limited backbone capacity $\pmb{\tau_g}$.} Now, we do not assume that $\tau_g$ is unbounded. Hence, we now find the optimal solution by solving the problem as above, assuming that $\tau_g$ is indeed unbounded, and hence progressively decreasing highest individual throughput rates $T_u$, $\forall  u\in\mathcal{U}_g\cup \bigcup\limits_{a\in\mathcal{A}_g} \mathcal{U}_a$ until $\sum\limits_{u\in\mathcal{U}_g} T_u + \sum\limits_{a\in\mathcal{A}_g}\sum\limits_{u\in\mathcal{U}_a}T_u = \tau_g$, as done also in the analyzed cases in Subsections~\ref{ss:BHfree} and~\ref{ss:BHmanaged1aBS} (step~\ref{alg:BHmanagedStep17} of Algorithm~\ref{alg:BHmanaged}).

\balance
\begin{algorithm}[H]
  \caption{Optimal solution to CP of Fig.~\ref{fig:CP}. 
  Generic case}
  \label{alg:BHmanaged}
  \begin{algorithmic}[1]
    %
    \STATE $w^a \leftarrow W_\mathcal{B}^{\min}$, $\forall a\in\mathcal{A}_g$.  \label{alg:BHmanagedStep1}
    \STATE $T^a \leftarrow w^a \log_2\left(1 + \gamma_{g,a}^{\mathcal{B}}\right)$, $\forall a\in\mathcal{A}_g$. \label{alg:BHmanagedStep2}
    \STATE Derive $\{T_u\}_{u\in\mathcal{U}_g}$ and $\{T_{a,u}\}_{u\in\mathcal{U}_a}$, $\forall a\in\mathcal{A}_g$ by solving optimal resource allocation for \gBS~$g$ assuming unbounded $\tau_g$ and also $\forall a\in\mathcal{A}_g$ assuming a backhaul limitation of $T^a$, as detailed in Subsection~\ref{ss:BHfree}. \label{alg:BHmanagedStep3}
    \STATE $\beta \leftarrow 1$. \label{alg:BHmanagedStep4}
    \WHILE{$\beta = 1$} \label{alg:BHmanagedStep5}
    \STATE $T_m \!\leftarrow\! \min\!\left\{T_{a,u} \!\!\suchthat\!\! T_{a,u} \!\!<\! w_u \log_2\!\left(\!1\!\!+\!\gamma_{g,a}\!\right)\!, a\!\in\!\mathcal{A}_g, u\!\in\!\mathcal{U}_a\right\}\!$. \label{alg:BHmanagedStep6}
    \STATE $\mathcal{L}_m \!\leftarrow\! \left\{(a,u)\in\mathcal{A}_g\times  \bigcup\limits_{a\in\mathcal{A}_g}\mathcal{U}_a \suchthat T_{a,u} = T_m\right\}$. \label{alg:BHmanagedStep7}
    \STATE $T_M \!\leftarrow\! \min\left\{T_{a,u}\suchthat T_{a,u} > T_m, a\in\mathcal{A}_g, u\in\mathcal{U}_a\right\}$. \label{alg:BHmanagedStep7.5}
    \STATE $T_{M_2} \!\leftarrow\! \min\left(T_M, \min\limits_{(a,u)\in\mathcal{L}_m} w_u \log_2\left(1+\gamma_{a,u}\right)\right)$. \label{alg:BHmanagedStep8}
    \STATE $ \overline{\mathcal{U}}_a \!\leftarrow\! \left\{u\in\mathcal{U}_a \suchthat (a,u)\in\mathcal{L}_m\right\}$, $\forall a\in\mathcal{A}_g$. \label{alg:BHmanagedStep9}
    \STATE $\beta \!\leftarrow\! \min\left(1, \frac{W_\mathcal{B} - \sum\limits_{a\in\mathcal{A}_g}w^a}{\left(T_{M_2} - T_m\right)\cdot \sum\limits_{a\in\mathcal{A}_g}\frac{|\overline{\mathcal{U}}_a|}{\log_2\left(1+\gamma_{g,a}^{\mathcal{B}}\right)}}\right)$. \label{alg:BHmanagedStep10}
    \STATE $w^a \!\leftarrow\! w^a + \frac{|\overline{\mathcal{U}}_a| \cdot \beta\cdot \left(T_{M_2}-T_m\right)}{\log_2\left(1+\gamma_{g,a}^{\mathcal{B}}\right)}$, $\forall a\in\mathcal{A}_g$. \label{alg:BHmanagedStep11}
    \STATE $T^a \!\leftarrow\! w^a  \log_2\left(1+\gamma_{g,a}^{\mathcal{B}}\right)$, $\forall a\in\mathcal{A}_g$. \label{alg:BHmanagedStep12}
    \STATE $T_{a,u} \!\leftarrow\! T_{a,u} + \beta\cdot \left(T_{M_2}-T_m\right)$, $\forall (a,u)\in\mathcal{L}_m$. \label{alg:BHmanagedStep13}
    \ENDWHILE \label{alg:BHmanagedStep14}
    \STATE $T^a \!\leftarrow\! \sum\limits_{u\in\mathcal{U}_a} T_{a,u}$, $\forall a\in\mathcal{A}_g$. \label{alg:BHmanagedStep15}
    \IF{$\sum\limits_{u\in\mathcal{U}_g} T_u + \sum\limits_{a\in\mathcal{A}_g}T^a > \tau_g$} \label{alg:BHmanagedStep16}
    \STATE Apply Algorithm~\ref{alg:CPxi} to $n \!=\! |\mathcal{U}_g| +\!\! \sum\limits_{a\in\mathcal{A}_g}\!\!|\mathcal{U}_a|$, \phantom{get new line} $\{x_i\}_{i=1}^n \!=\! \{T_u\}_{u\in\mathcal{U}_g} \!\cup\! \bigcup\limits_{a\in\mathcal{A}_g}\!\!\{T_{a,u}\}_{u\in\mathcal{U}_a}$, $X \!=\! \tau_g$. \label{alg:BHmanagedStep17}
    \ENDIF \label{alg:BHmanagedStep18}
  \end{algorithmic}
\end{algorithm}

\end{document}